\documentclass[longauth]{aa}
\usepackage[varg]{txfonts}
\usepackage{fancyhdr}
\usepackage[english]{babel}
\usepackage{graphicx}
\usepackage{amssymb}
\usepackage{geometry}
\usepackage{amsmath}
\usepackage{footnote}
\usepackage{natbib}
\usepackage{pdflscape}
\usepackage{longtable}
\bibpunct{(}{)}{;}{a}{}{,}
\usepackage[pdftex,colorlinks]{hyperref}
\begin{document}
\title{SUPER I. Toward an unbiased study of ionized outflows \\
in $z \sim 2$ active galactic nuclei:\\ 
survey overview and sample characterization\thanks{Table \ref{tab:photo_summary} is only available in electronic form at the CDS via anonymous ftp to cdsarc.u-strasbg.fr (130.79.128.5) or via \url{http://cdsweb.u-strasbg.fr/cgi-bin/qcat?J/A+A/}}}
\titlerunning{SUPER I. Toward an unbiased study of ionized outflows in $z \sim 2$ AGN}

\author{C.~Circosta\inst{\ref*{ESO},\ref*{LMU}}\thanks{\email{ccircost@eso.org}}
 \and V.~Mainieri\inst{\ref*{ESO}}
 \and P.~Padovani\inst{\ref*{ESO}}
 \and G.~Lanzuisi\inst{\ref*{DIFA},\ref*{OABO}}
 \and M.~Salvato\inst{\ref*{MPE},\ref*{cluster}}
 \and C.~M.~Harrison\inst{\ref*{ESO}}
 \and D.~Kakkad\inst{\ref*{ESO},\ref*{LMU},\ref*{ESOchile}}
 \and A.~Puglisi\inst{\ref*{CEA},\ref*{pd}}
 \and G.~Vietri\inst{\ref*{cluster},\ref*{ESO}}
 \and G.~Zamorani\inst{\ref*{OABO}}
 \and C.~Cicone\inst{\ref*{brera}}
 \and B.~Husemann\inst{\ref*{MPIA}}
 \and C.~Vignali\inst{\ref*{DIFA},\ref*{OABO}}
 \and B.~Balmaverde\inst{\ref*{brera}}
 \and M.~Bischetti\inst{\ref*{OAroma},\ref*{TorVergata}}
 \and M.~Brusa\inst{\ref*{DIFA},\ref*{OABO}}
 \and A.~Bongiorno\inst{\ref*{OAroma}}
 \and S.~Carniani\inst{\ref*{cambridge},\ref*{cambridge2}}
 \and F.~Civano\inst{\ref*{cfa}}
 \and A.~Comastri\inst{\ref*{OABO}}
 \and G.~Cresci\inst{\ref*{OAarcetri}}
 \and C.~Feruglio\inst{\ref*{OAtrieste}}
 \and F.~Fiore\inst{\ref*{OAtrieste}}
 \and S.~Fotopoulou\inst{\ref*{durham}}
 \and A.~Karim\inst{\ref*{bonn}}
 \and A.~Lamastra\inst{\ref*{OAroma},\ref*{ASI}}
 \and B.~Magnelli\inst{\ref*{bonn}}
 \and F.~Mannucci\inst{\ref*{OAarcetri}}
 \and A.~Marconi\inst{\ref*{uniFI},\ref*{OAarcetri}}
 \and A.~Merloni\inst{\ref*{MPE}}
 \and H.~Netzer\inst{\ref*{telAviv}}
 \and M.~Perna\inst{\ref*{OAarcetri}}
 \and E.~Piconcelli\inst{\ref*{OAroma}}
 \and G.~Rodighiero\inst{\ref*{uniPD}}
 \and E.~Schinnerer\inst{\ref*{MPIA}}
 \and M.~Schramm\inst{\ref*{NAO}}
 \and A.~Schulze\inst{\ref*{NAO}}
 \and J.~Silverman\inst{\ref*{japan}}
 \and L.~Zappacosta\inst{\ref*{OAroma}}
 }
 
 \institute{European Southern Observatory, Karl-Schwarzschild-Str. 2, 85748 Garching bei M\"{u}nchen, Germany\label{ESO}
\and Ludwig Maximilian Universit\"{a}t, Professor-Huber-Platz 2, 80539 M\"{u}nchen, Germany\label{LMU}
\and Dipartimento di Fisica e Astronomia dell’Universit\`a degli Studi di Bologna, via P. Gobetti 93/2, 40129 Bologna, Italy\label{DIFA}
\and INAF/OAS, Osservatorio di Astrofisica e Scienza dello Spazio di Bologna, via P. Gobetti 93/3, 40129 Bologna, Italy\label{OABO}
\and MPE, Giessenbach-Str. 1, 85748 Garching bei M\"{u}nchen, Germany\label{MPE}
\and Cluster of Excellence, Boltzmann-Str. 2, 85748 Garching bei M\"{u}nchen, Germany\label{cluster}
\and European Southern Observatory, Alonso de Cordova 3107, Casilla 19, Santiago 19001, Chile\label{ESOchile}
\and CEA, IRFU, DAp, AIM, Université Paris-Saclay, Université Paris Diderot, Sorbonne Paris Cité, CNRS, 91191 Gif-sur-Yvette, France\label{CEA}
\and INAF - Osservatorio Astronomico di Padova, Vicolo dell’Osservatorio 5, I-35122, Padova, Italy\label{pd}
\and INAF - Osservatorio Astronomico di Brera, Via Brera 28, 20121 Milano, Italy\label{brera}
\and Max-Planck-Institut f\"{u}r Astronomie, K\"{o}nigstuhl 17, 69117 Heidelberg, Germany\label{MPIA}
\and INAF - Osservatorio Astronomico di Roma, Via Frascati 33, 00078 Monte Porzio Catone (Roma), Italy\label{OAroma}
\and Università degli Studi di Roma ``Tor Vergata'', Via Orazio Raimondo 18, 00173 Roma, Italy\label{TorVergata}
\and Cavendish Laboratory, University of Cambridge, 19 J. J. Thomson Avenue, Cambridge CB3 0HE, UK\label{cambridge}
\and Kavli Institute for Cosmology, University of Cambridge, Madingley Road, Cambridge CB3 0HA, UK\label{cambridge2}
\and Harvard-Smithsonian Center for Astrophysics, 60 Garden Street, Cambridge, MA 02138, USA\label{cfa}
\and INAF - Osservatorio Astrofisico di Arcetri, Largo E. Fermi 5, 50125 Firenze, Italy\label{OAarcetri}
\and INAF - Osservatorio Astronomico di Trieste, via G.B. Tiepolo 11, 34143 Trieste, Italy\label{OAtrieste}
\and Centre for Extragalactic Astronomy, Department of Physics, Durham University, South Road, Durham, DH1 3LE, UK\label{durham}
\and Argelander-Institut f\"{u}r Astronomie, Universit\"{a}t Bonn, Auf dem H\"{u}gel 71, 53121 Bonn, Germany\label{bonn}
\and Space Science Data Center - ASI, via del Politecnico SNC, 00133 Roma, Italy\label{ASI}
\and Dipartimento di Fisica e Astronomia, Università di Firenze, Via G. Sansone 1, 50019 Sesto Fiorentino (Firenze), Italy\label{uniFI}
\and School of Physics and Astronomy, Tel Aviv University, Tel Aviv 69978, Israel\label{telAviv}
\and Dipartimento di Fisica e Astronomia, Università di Padova, vicolo Osservatorio 3, 35122 Padova, Italy\label{uniPD}
\and National Astronomical Observatory of Japan, Mitaka, Tokyo 181-8588, Japan\label{NAO}
\and Kavli Institute for the Physics and Mathematics of the Universe, The University of Tokyo, Kashiwa, Japan 277-8583\label{japan}
}


\abstract{Theoretical models of galaxy formation suggest that the presence of an active galactic nucleus (AGN) is required to regulate the growth of its host galaxy through feedback mechanisms, produced by, for example, AGN-driven outflows. Although many observational studies have revealed that such outflows are common both at low and high redshift, a comprehensive picture is still missing. In particular, the peak epoch of galaxy assembly ($1<z<3$) has been poorly explored so far, and current observations in this redshift range are mostly limited to targets with high chances to be in an outflowing phase. This paper introduces SUPER (a SINFONI Survey for Unveiling the Physics and Effect of Radiative feedback), an ongoing ESO's VLT/SINFONI Large Programme. SUPER will perform the first systematic investigation of ionized outflows in a sizeable and blindly-selected sample of 39 X-ray AGN at $z \sim 2$, which reaches high spatial resolutions ($\sim 2$ kpc) thanks to the adaptive optics-assisted IFU observations. The outflow morphology and star formation in the host galaxy will be mapped through the broad component of [\ion{O}{iii}]$\lambda 5007$ and the narrow component of H$\alpha$ emission lines. The main aim of our survey is to infer the impact of outflows on the on-going star formation and to link the outflow properties to a number of AGN and host galaxy properties. We describe here the survey characteristics and goals, as well as the selection of the target sample. Moreover, we present a full characterization of its multi-wavelength properties: we measure, via spectral energy distribution fitting of UV-to-FIR photometry, stellar masses ($4\times10^{9} - 2\times10^{11}$ $M_{\odot}$), star formation rates ($25 - 680$ $M_{\odot}$ yr$^{-1}$) and AGN bolometric luminosities ($2\times10^{44} - 8\times10^{47}$ erg s$^{-1}$), along with obscuring column densities (up to $2\times10^{24}$ cm$^{-2}$) and luminosities in the hard $2-10$ keV band ($2\times10^{43} - 6\times10^{45}$ erg s$^{-1}$) derived through X-ray spectral analysis. Finally, we classify our AGN as jetted or non-jetted according to their radio and FIR emission.  
}
\maketitle
\section{Introduction}

Supermassive black holes (SMBHs) at the center of galaxies undergo periods of gas accretion becoming visible as active galactic nuclei (AGN). The enormous amount of energy released during these growth episodes is thought to shape the evolutionary path of AGN host galaxies. It may play a significant role in regulating and even quenching star formation in the galaxy by expelling gas out of the galaxy itself or preventing gas cooling. The process by which the energy is injected by the AGN and coupled to the surrounding medium is the so-called AGN feedback \citep{fabian12,king15,harrison17}. It can be particularly crucial at $z \sim 2$, since this redshift corresponds to the peak of star formation and SMBH accretion in the Universe \citep[e.g.,][]{madau14} and therefore the energy injected by the central engine into the host galaxy may be maximized. However, the full details of the specific effects this may have on the host galaxy's life are still not clear. 

Feedback of AGN is invoked from a theoretical perspective \citep[e.g.,][]{ciotti97,silk98,dimatteo05,king05,somerville08} to explain key observations of the galaxy population, such as the tight correlation between black hole masses and bulge masses as well as velocity dispersions of the host galaxies \citep{kormendy13}, the bimodal color distribution of galaxies \citep{strateva01}, and the lack of very massive galaxies in the most massive galaxy haloes \citep{somerville08,behroozi13}. 
According to some models \citep[e.g.,][]{king05,springel05,debuhr12,costa14}, fast winds are launched by the accretion disk surrounding the SMBH and driven by radiative and mechanical energy during its active and bright phase. 
These winds propagate into the host galaxy coupling to the interstellar medium (ISM) and drive fast outflows out to large scales (up to $\sim 1000$ km s$^{-1}$ on kpc scales), potentially removing the gas which fuels star formation. It is important to test the models with observations by measuring key outflow properties such as kinetic energy and momentum injection rates \citep{fiore17,harrison18}.

AGN-driven outflows can therefore be a manifestation of AGN feedback. The presence of outflows in AGN host galaxies is now quite well established: they have been detected at different physical scales \citep[e.g.,][]{feruglio10,tombesi15,veilleux17} and in different gas phases \citep[e.g.,][]{cano_diaz12,cicone14,rupke17}, both in the nearby \citep[e.g.,][]{rupke13,perna17} and distant Universe \citep[e.g.,][]{nesvadba11,carniani15,cicone15}. An important property shown by outflows is their multi-phase nature so to fully characterize them we need to trace all the gas phases, neutral and ionized, atomic and molecular \citep{cicone18}. The ionized phase has been studied through absorption and emission lines in rest-frame optical \citep[e.g.,][]{bae17,concas17,perna17}, ultraviolet (UV) \citep[e.g.,][]{liu15} and X-ray \citep[e.g.,][]{tombesi10}. When the velocity shift of these lines with respect to the rest-frame velocity is not representative of ordered motion in the galaxy as traced by stellar kinematics, it can be considered as evidence for the presence of non-gravitational kinematic components, such as outflowing gas \citep{karouzos16,woo16}. To understand the impact of AGN outflows on the gas and star formation in the host galaxy, it is necessary to explore large galactic scales ($\approx 1-10$ kpc). A commonly used diagnostic for this kind of studies is the forbidden emission line doublet [\ion{O}{iii}]$\lambda$5007,4959 \AA. It traces the kinematics of ionized gas on galaxy-wide scales, in the narrow line region (NLR), since being a forbidden line it cannot be produced in the high-density environment of the broad line region (BLR) on sub-parsec scales. Therefore  asymmetric [\ion{O}{iii}]$\lambda$5007 profiles, showing a broad and blue-shifted wing, are used to trace outflowing kinematic components.

\begin{figure*}[h!]
\centering
  \includegraphics[width=9cm]{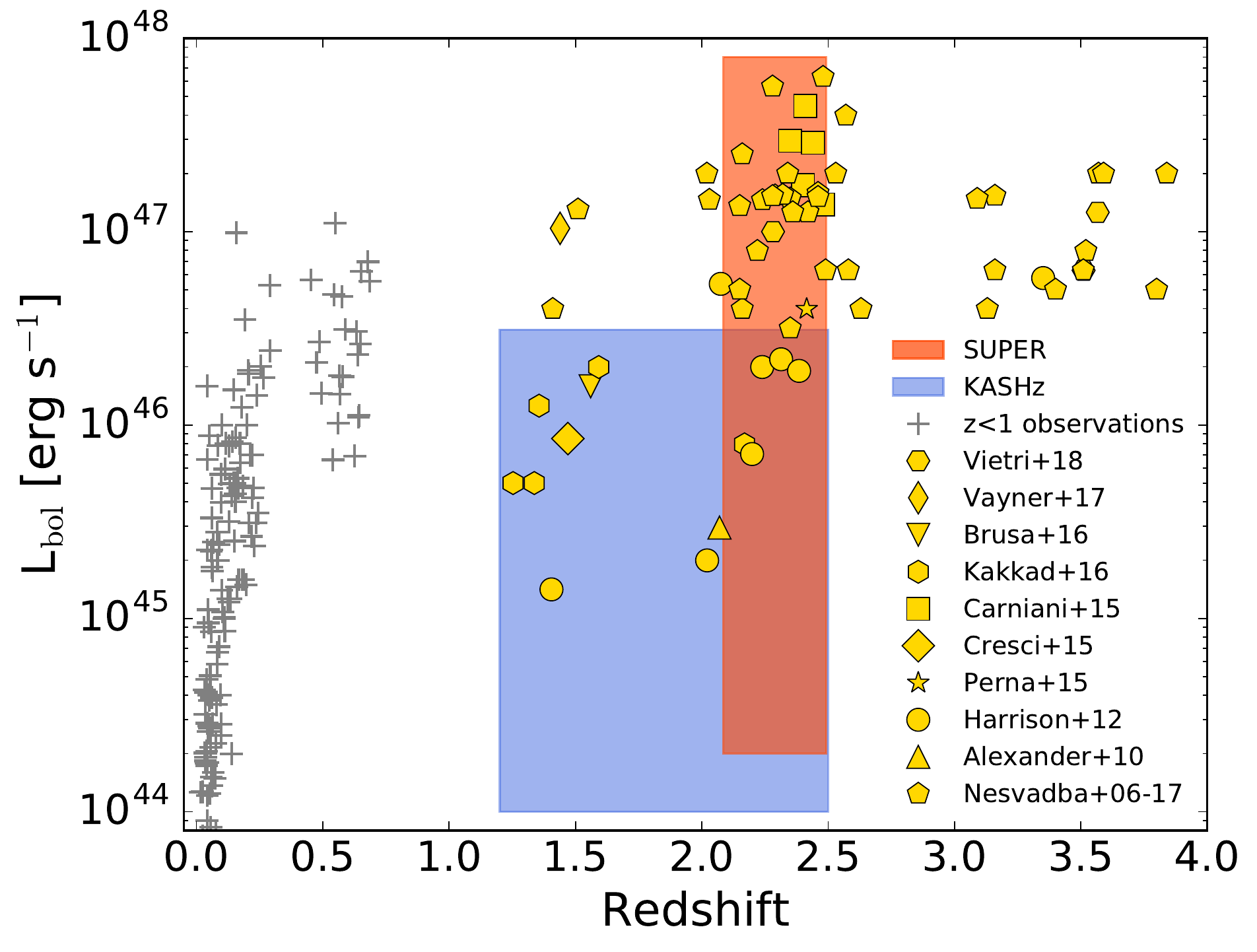}
  \hspace{2mm}
  \includegraphics[width=8.75cm]{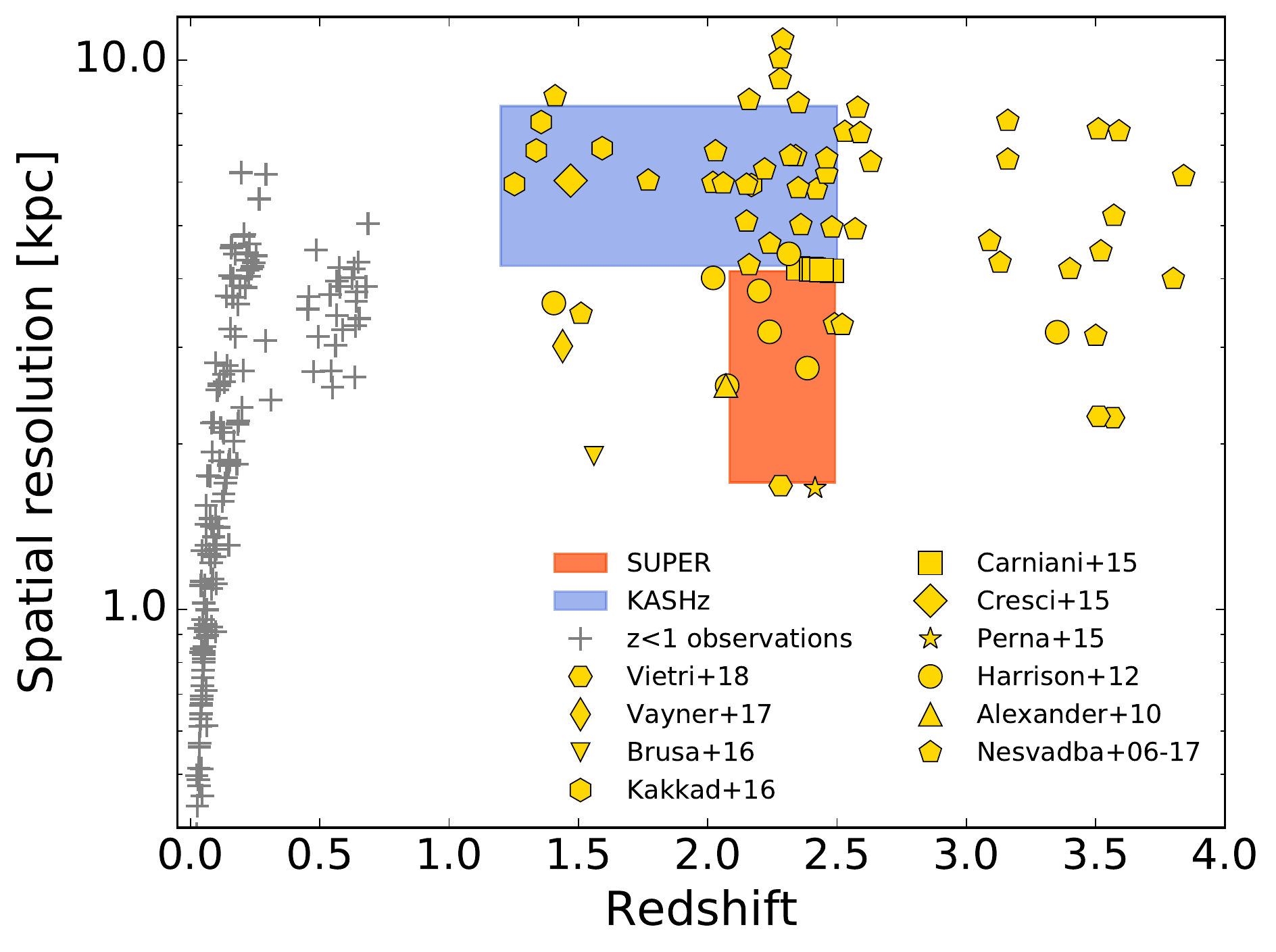}
  \caption{Summary of IFS observations from the literature characterizing ionized outflows through the [\ion{O}{iii}]$\lambda$5007 emission line in AGN host galaxies. \textit{Left}: For each observation we plot the AGN bolometric luminosity of the source, in units of erg s$^{-1}$, as a function of redshift. The red and blue shaded areas show the parameter space probed by SUPER and KASHz \citep[][Harrison et al., in prep.]{harrison16}, respectively. Excluding these two surveys, current observations at $z>1$ are limited to a smaller number of objects, mainly at high luminosity ($L_{\textnormal{bol}} > 10^{46}$ erg s$^{-1}$) and focused on targets mostly selected to increase the chances to detect an outflow. SUPER will be able to explore a wide range in bolometric luminosities ($10^{44} < L_{\textnormal{bol}} < 10^{48}$ erg s$^{-1}$) for an unbiased sample of AGN. The gray crosses represent observations at $z<1$ \citep{bae17,rupke17,karouzos16,harrison14,husemann13,husemann14,husemann17,liu13,liu14} covering the parameter space much more uniformly than high-redshift observations available so far \citep{vietri18,vayner17,brusa16,kakkad16,carniani15,cresci15,perna15,harrison12,alexander10,nesvadba06,nesvadba07,nesvadba08,nesvadba11,nesvadba17a,nesvadba17b}. All AGN bolometric luminosities, when not available in the papers, have been obtained consistently either as indicated in the papers themselves or from the observed [\ion{O}{iii}] luminosity adopting a conversion factor of 3500 \citep{heckman04}. \textit{Right}: Spatial resolution, in kpc, of the observations shown in the left panel as a function of redshift. The angular resolutions from which the values plotted are derived, are taken from the respective papers and given by the seeing of the observations or from the size of the PSF. SUPER observations will allow us to reach an unprecedented spatial resolution (i.e. $\sim 1.7 - 4$ kpc) for a sizeable sample of 39 AGN, obtained just by a few single-object studies so far at similar redshift.}
  \label{fig:summary_IFS}
\end{figure*}

Long-slit optical and near-infrared (NIR) spectroscopy is a useful technique to reveal outflow signatures \citep[e.g.,][]{das05,crenshaw07,brusa15}. However, it is able to provide spatial information along one direction, therefore lacking a detailed mapping of the outflow distribution in the host galaxy together with its velocity. In recent years, integral-field spectroscopy (IFS) studies have offered a more direct way to identify and interpret outflows, allowing astronomers to spatially resolve the kinematics of ionized gas \citep[e.g.,][]{cresci09,alexander10,gnerucci11,foerster_schreiber14,harrison14}. Nevertheless, the observational evidence available so far at $z>1$, the crucial cosmic epoch to study AGN-driven outflows and on which this paper will be focused, is sparse, mainly limited to bright objects or observations performed in seeing limited conditions and therefore not able to resolve scales below $3-4$ kpc, which limits how well the observations can constrain model predictions \citep{harrison18}. 
In Figure \ref{fig:summary_IFS} we collect IFS results from the literature tracing ionized outflows in AGN host galaxies through the [\ion{O}{iii}] emission line. The left panel compares AGN bolometric luminosities and redshift for each target, in order to summarize the state-of-the-art of ionized AGN outflow IFS studies. Contrary to the uniform coverage of the parameter space at $z<1$ (gray crosses in Figure \ref{fig:summary_IFS}, \textit{left panel}), at $z>1$ it is limited to a small number of objects, mainly at high luminosity ($L_{\textnormal{bol}} > 10^{46}$ erg s$^{-1}$, see points in Figure \ref{fig:summary_IFS}, \textit{left panel}). The targets of previous studies are mostly selected to increase the chances to detect an outflow, meaning because they are powerful AGN (e.g., in the IR or radio regime), they have already known outflows or characteristics suitable for being in an outflowing phase \citep[e.g., high mass accretion rate of the SMBH and high column density;][]{brusa15,kakkad16}. 
Because of this observational bias, it is still controversial how common these outflows are especially in sources with low AGN bolometric luminosity. Nevertheless, detailed single object studies have provided evidence that powerful outflows may suppress star formation in the regions where they are detected \citep[e.g.,][]{cano_diaz12,cresci15,carniani16}, although it is still not clear the impact that such outflows may have on the global star-forming activity occurring in the host galaxy (i.e., including regions of the galaxies not affected by the outflow). In addition to negative feedback mechanisms, outflows have been proved to be responsible for positive feedback mechanisms in a few cases by triggering star formation \citep[e.g.,][]{cresci15,molnar17,cresci18}.

In order to draw a coherent picture and definitively address the impact of such outflows on the galaxy population evolution it is necessary to conduct systematic and unbiased searches for outflows in large samples of objects. The KMOS AGN Survey at High redshift \citep[KASHz;][Harrison et al., in prep.; blue rectangle in Figure \ref{fig:summary_IFS}, \textit{left panel}]{harrison16} has first started to provide spatially-resolved information for hundreds of X-ray selected AGN. These observations are seeing limited, which sets a limit on the spatial scales that can be resolved at $z>1$. The range of spatial scales resolved in current observations is shown in the right panel of Figure \ref{fig:summary_IFS}, plotted as a function of redshift for the same collection of data as in the left panel. At $z>1$, the spatial resolution is mainly in the range $3-10$ kpc (i.e. $> 0.5''$).

To provide higher spatial resolutions (down to $\sim 2$ kpc at $z\sim2$), one needs to exploit the possibilities offered by adaptive optics (AO), which corrects for the distortion caused by the turbulence of the Earth's atmosphere. This has been done by, e.g., \citet{perna15}, \citet{brusa16}, \citet{vayner17} and \cite{vietri18}. 
Such observations require a larger amount of observing time, therefore it is necessary to focus on smaller but still representative samples. Our on-going ESO Large Programme called SUPER (the SINFONI Survey for Unveiling the Physics and Effect of Radiative feedback), represented by the red rectangle in Figure \ref{fig:summary_IFS}, is taking advantage of the AO corrections by reaching angular resolutions of $0.2''$. It combines spatially-resolved AO-assisted IFS observations for a fairly representative sample of sources selected in an unbiased way with respect to the chance of detecting outflows, aiming at investigating the physical properties of AGN outflows and their impact on the star formation activity in the host galaxies as well as connecting the physical properties of AGN and host galaxies to those of ionized outflows. As shown in Figure \ref{fig:summary_IFS}, SUPER probes a wide range of AGN bolometric luminosities, up to four orders of magnitude, with spatial resolutions between $\sim 1.7$ and 4 kpc (i.e. $0.2''-0.5''$). 

This paper is the first of a series of publications dedicated to the survey. It focuses on providing an overview of the survey (i.e., characteristics, goals and sample selection criteria), as well as describing the physical properties of the target sample and the way they have been measured through a uniform multi-wavelength analysis from the X-ray to the radio regime. We derive stellar masses, star formation rates (SFRs) and AGN bolometric luminosities from the multi-wavelength spectral energy distributions (SEDs), X-ray luminosities and column densities from the X-ray spectra and BH masses and Eddington ratios from the optical spectra. The paper is organized as follows: in Sec. \ref{s:survey} we present the properties and the main goals of the survey as well as the sample selection criteria and its X-ray properties. In Sec. \ref{s:Characterization} we describe the multi-wavelength dataset and the SED-fitting code used to derive host galaxy and AGN properties of the targets. These properties are then discussed in Sec. \ref{s:overall_results}, with particular emphasis on stellar masses, SFRs and AGN bolometric luminosities as well as the target properties in the radio regime. We finally summarize our results and discuss future follow-up work in Sec. \ref{s:Conclusions}. In this paper we adopt a \textit{WMAP9} cosmology \citep{hinshaw13}, $H_{0} = 69.3$ km s$^{-1}$ Mpc$^{-1}$, $\Omega_{\textnormal{M}} = 0.287$ and $\Omega_{\Lambda} = 0.713$. 

\section{The survey}\label{s:survey}
SUPER\footnote{\url{https://www.super-survey.org}} (PI: Mainieri - 196.A-0377) is a Large Programme at the ESO's Very Large Telescope (VLT). The survey has been allocated 280 hours of observing time in AO-assisted mode with the aim of providing high-resolution, spatially-resolved IFS observations of multiple emission lines for a carefully-selected sample of 39 X-ray AGN at $z \sim 2$. The AO correction is performed in Laser Guide Star-Seeing Enhancer (LGS-SE) mode, which has demonstrated the capability to achieve a point spread function (PSF) full width at half maximum (FWHM) of $\sim 0.3''$ under typical weather conditions in Paranal \citep{nfs18}, i.e., average seeing of $\sim 0.55''$ in \textit{K} band \citep{sarazin08}. We have selected for all our targets the 50 mas/pixel scale of SINFONI which corresponds to a total field of view FOV$=3.2''\times3.2''$. The selected plate scale corresponds to a spectral  resolution of about R$\approx 2730$ in \textit{H} band and R$\approx 5090$ in \textit{K} band.

The redshift range covered by SUPER is crucial to investigate AGN feedback, being at the peak epoch of AGN and galaxy assembly. Key emission lines, such as [\ion{O}{iii}], H$\beta$ and H$\alpha$, are covered with \textit{H}- and \textit{K}-band observations in this redshift range. 
We will use asymmetric and spatially-extended [\ion{O}{iii}] line emission, traced by \textit{H}-band observations, to identify outflowing ionized gas as extensively done in the literature \citep[e.g.,][]{alexander10,cresci15,harrison16}. The \textit{K}-band observations will provide the possibility to map the H$\alpha$ emission, with the aim to construct spatially-resolved maps of the on-going star formation in the host from the narrow component of the line, which could be less affected by AGN emission, and compare it with the outflow geometry derived from the [\ion{O}{iii}] line profile \citep[see, e.g.,][]{cano_diaz12,cresci15,carniani16}. The comparison between these two tracers will give us the opportunity to constrain systematically the role of AGN outflows in regulating star formation.  
Thanks to the extensive set of AGN and host galaxy physical properties (AGN bolometric luminosity, BH mass, Eddington ratio, obscuring column density, radio emission, stellar mass and SFR), derived in a uniform way for each target as explained in the present paper, and outflow parameters which will be extracted from the \textit{H}-band observations (such as mass outflow rate, kinetic power, momentum rate, size), SUPER will explore the potential relations among these quantities \citep{fiore17}.

The science goals of our survey are:
\begin{itemize}
\item Systematic study of the occurrence of outflows in AGN host galaxies and investigation of any possible link between the physical properties of both SMBHs and their hosts, and the outflow properties.
\item Mapping AGN ionized outflow morphology on kpc scale using [\ion{O}{iii}] and constraining their impact on the on-going star formation in the host galaxies using the narrow component of H$\alpha$. If the signal-to-noise of the latter is not good enough to produce spatially-resolved maps of star formation, we should still be able to compare the outflow properties with the integrated SFR (as derived by SED fitting).
\item Investigating the variation of outflow properties as a function of the host galaxy location with respect to the main sequence of star-forming galaxies \citep[MS, e.g.,][]{noeske07}, in order to investigate empirically the relation between galaxy and AGN.
\end{itemize}

An important further goal of this survey will be the comparison of our results to a mass-matched control sample of normal star-forming galaxies at the same redshift and with similar AO-assisted observations \citep[e.g., the SINS/zC-SINF survey,][see Sec. \ref{s:MS_comp}]{nfs18}, to investigate the differences between galaxies hosting active and inactive SMBHs.

In the following we describe the criteria adopted to select our sample.

\subsection{\label{s:selection}Sample selection}

Our Large Programme is designed to conduct a blind search for AGN-driven outflows on a representative sample of AGN. Therefore, we do not preselect AGN with already known outflows or with characteristics suitable for being in an outflowing phase \citep{brusa15,kakkad16}. Instead, aiming at performing a statistical investigation of this phenomenon, the first goal is to cover the widest possible range in AGN properties. 

One of the most efficient tracers of AGN activity is offered by their X-ray emission, since it probes directly the active nucleus with a negligible contamination from the host galaxy, providing the largest AGN surface density \citep[e.g.,][]{padovani17}. We identified our targets by combining X-ray catalogs from several surveys characterized by different depths and areas. While shallow and wide-field surveys provide a better census of the rare high-luminosity AGN, deep and small-area surveys, limited to a few deg$^{2}$, are able to reveal fainter sources \citep[see Fig. 3 in ][]{brandt15}. By adopting this ``wedding cake'' approach we are able to cover a wide range in AGN bolometric luminosity, $10^{44} < L_{\textnormal{bol}} < 10^{48}$ erg s$^{-1}$ (see Fig. \ref{fig:summary_IFS}), spanning both faint and bright AGN. The selection was performed by adopting as a threshold an absorption-corrected X-ray luminosity $L_{X} \geq 10^{42}$ erg s$^{-1}$ from the following surveys:
\begin{itemize}
\item The \textit{Chandra} Deep Field-South \citep[CDF-S;][]{luo17}, the deepest X-ray survey to date which covers a global area of 484.2 arcmin$^{2}$ observed for a total \textit{Chandra} exposure time of $\sim 7$ Ms, reaching a sensitivity of $\sim 1.9 \times 10^{-17}$ erg cm$^{-2}$ s$^{-1}$ in the full $0.5-7.0$ keV band. 
\item The \textit{COSMOS-Legacy} survey \citep{civano16,marchesi16}, a 4.6 Ms \textit{Chandra} observation of the COSMOS field, which offers a unique combination of deep exposure over an area of about 2.2 deg$^{2}$ at a limiting depth of $8.9 \times 10^{-16}$ erg cm$^{-2}$ s$^{-1}$ in the $0.5-10$ keV band. 
\item The wide-area \textit{XMM-Newton} XXL survey \citep{pierre16}, where we focus in particular on the equatorial sub-region of the XMM-XXL North, a $\sim$25 deg$^{2}$ field surveyed for about 3 Ms by \textit{XMM-Newton} with a sensitivity in the full $0.5-10$ keV band of $2 \times 10^{-15}$ erg cm$^{-2}$ s$^{-1}$.
\item The Stripe 82 X-ray survey \citep[Stripe82X;][]{lamassa16,ananna17}, $\sim$980 ks of observing time with \textit{XMM-Newton} covering 31.3 deg$^{2}$ of the Sloan Digital Sky Survey (SDSS) Stripe 82 Legacy Field and a flux limit of $2.1 \times 10^{-15}$ erg cm$^{-2}$ s$^{-1}$ in the full $0.5-10$ keV band.
\item The WISE/SDSS selected Hyper-luminous quasars sample \citep[WISSH;][]{bischetti17,duras17,martocchia17,vietri18}, with both proprietary and archival \textit{Chandra} and \textit{XMM-Newton} observations available, described in \citet{martocchia17}. 
\end{itemize}
The choice of the fields was driven by their visibility from Paranal and the rich multi-wavelength photometric coverage from the UV to the far-infrared (FIR), needed to obtain robust measurements of the target properties by using an SED-fitting technique. 
Our targets are then selected to meet the following criteria:
\begin{enumerate}
\item Spectroscopic redshift in the range $z = 2.0-2.5$, whose quality was flagged as ``Secure'' in the respective catalogs. This redshift range was chosen in order to have H$\beta$ and [\ion{O}{iii}] included in \textit{H}-band and H$\alpha$ in \textit{K}-band together with their potential broad line components, by allowing a margin of 10\,000 km s$^{-1}$ between the peak of the lines and the edges of the filter bands.
\item Observed wavelengths for [\ion{O}{iii}] and H$\alpha$ characterized by a low contamination from the strong telluric OH lines, which affect NIR observations.
\end{enumerate}
The resulting sample consists of 39 AGN (namely 6 from CDF-S, 16 from COSMOS, 10 from XMM-XXL, 4 from Stripe82X and 3 from the WISSH sample), whose IDs, coordinates, redshifts as well as $H-$ and $K-$band magnitudes (AB) are reported in Table\,\ref{tab:sample}. 
This sample results from an optimization between size, the amount of observing time required to carry out the observations, and a wide and uniform coverage in AGN bolometric luminosities, Eddington ratios, and column densities. All our targets have spectroscopic redshifts based on optical spectroscopic campaigns: for example VLT/VIMOS and FORS2 surveys for the CDF-S \citep{balestra10,kurk13}; for the COSMOS field, a master spectroscopic catalog is available within the COSMOS collaboration (Salvato et al., in prep.) and includes results from several spectroscopic surveys of this field \citep[see][]{marchesi16};  
SDSS-BOSS spectra for the XMM-XXL field \citep{menzel16}; SDSS-DR12 for Stripe82X \citep{lamassa16}; SDSS-DR10 and LBT/LUCI1 redshifts for the WISSH subsample \citep{bischetti17}. 
Thanks to the parameter space covered by the survey (Fig. \ref{fig:summary_IFS}, \textit{left panel}), we will be able to probe AGN bolometric luminosities in the range $44 \lesssim \log(L_{\textnormal{bol}}/\textnormal{erg}\,\textnormal{s}^{-1}) \lesssim 48$, not covered so far by a coherent high spatial resolution observing program at this redshift. 

\renewcommand{\arraystretch}{1.1}
\begin{table*}[h!]
\caption{\label{tab:sample} Summary of the target AGN sample.}
\centering
\begin{tabular}{ccccccc}
\hline\hline
Field & ID & RA[J2000] & DEC[J2000] & $z_{\textnormal{spec}}$ & \textit{H}-band mag & \textit{K}-band mag \\
$(1)$ & $(2)$ & $(3)$ & $(4)$ & $(5)$ & $(6)$ & $(7)$ \\
\hline
XMM-XXL & X\_N\_160\_22 & 02:04:53.81 & $-$06:04:07.82 & 2.445 & 19.22 & 18.79 \\
			  & X\_N\_81\_44 & 02:17:30.95 & $-$04:18:23.66 & 2.311 & 18.78 & 18.43 \\
              & X\_N\_53\_3 & 02:20:29.84 & $-$02:56:23.41 & 2.434 & 20.60 & - \\
              & X\_N\_66\_23 & 02:22:33.64 & $-$05:49:02.73 & 2.386 & 20.56 & 20.33 \\
              & X\_N\_35\_20 & 02:24:02.71 & $-$05:11:30.82 & 2.261 & 22.07 & 21.70 \\
              & X\_N\_12\_26 & 02:25:50.09 & $-$03:06:41.16 & 2.471 & 19.83 & 19.53 \\
              & X\_N\_44\_64 & 02:27:01.46 & $-$04:05:06.73 & 2.252 & 21.31 & 20.77 \\
              & X\_N\_4\_48 & 02:27:44.63 & $-$03:42:05.46 & 2.317 & 19.57 & 20.43 \\
              & X\_N\_102\_35 & 02:29:05.94 & $-$04:02:42.99 & 2.190 & 18.76 & 18.19 \\
              & X\_N\_115\_23 & 02:30:05.66 & $-$05:08:14.10 & 2.342 & 19.79 & 19.26 \\              
CDF-S & XID36 & 03:31:50.77 & $-$27:47:03.41 & 2.259 & 21.49 & 20.80 \\
		& XID57\tablefootmark{a} & 03:31:54.40 & $-$27:56:49.70 & 2.298 & 23.49 & 22.19 \\
		& XID419 & 03:32:23.44 & $-$27:42:54.97 & 2.145 & 22.44 & 21.84\\
        & XID427 &  03:32:24.20 & $-$27:42:57.51 & 2.303 & 22.48 & 21.83 \\
        & XID522 & 03:32:28.50 & $-$27:46:57.99 & 2.309 & 22.98 & 22.27 \\
        & XID614 & 03:32:33.02 & $-$27:42:00.33 & 2.448 & 22.59 & 21.82 \\
COSMOS & cid\_166 & 09:58:58.68 & +02:01:39.22 & 2.448 & 18.55 & 18.23 \\
			 & lid\_1289 & 09:59:14.65 & +01:36:34.99 & 2.408 & 22.29 & 21.51 \\
             & cid\_1057 & 09:59:15.00 & +02:06:39.65 & 2.214 & 21.70 & 21.09 \\
             & cid\_1605 & 09:59:19.82 & +02:42:38.73 & 2.121 & 20.63 & 20.14 \\
			 & cid\_337 & 09:59:30.39 & +02:06:56.08 & 2.226 & 22.12 & 21.54 \\
             & cid\_346 &  09:59:43.41 & +02:07:07.44 & 2.219 & 19.24 & 18.95 \\
             & cid\_451 & 10:00:00.61 & +02:15:31.06 & 2.450 & 21.88 & 21.37 \\
             & cid\_1205 & 10:00:02.57 & +02:19:58.68 & 2.255 & 21.64 & 20.72\\
             & cid\_2682 & 10:00:08.81 & +02:06:37.66 & 2.435 & 21.46 & 21.17 \\
             & cid\_1143 & 10:00:08.84 & +02:15:27.99 & 2.492 & 22.90 & 22.27 \\
             & cid\_467 & 10:00:24.48 & +02:06:19.76 & 2.288 & 19.34 & 18.91 \\
             & cid\_852 & 10:00:44.21 & +02:02:06.76 & 2.232 & 21.53 & 21.05 \\
             & cid\_971 & 10:00:59.45 & +02:19:57.44 & 2.473 & 22.58 & 22.10 \\
             & cid\_38 & 10:01:02.83 & +02:03:16.63 & 2.192 & 20.42 & 20.21 \\
             & lid\_206 & 10:01:15.56 & +02:37:43.44 & 2.330 & 22.38 & 21.97 \\
             & cid\_1253 & 10:01:30.57 & +02:18:42.57 & 2.147 & 21.30 & 20.72 \\
WISSH & J1333$+$1649 & 13:33:35.79 & +16:49:03.96 & 2.089 & 15.72 & 15.49 \\
		& J1441$+$0454 & 14:41:05.54 & +04:54:54.96 & 2.059 & 17.15 & 16.53 \\
		& J1549$+$1245 & 15:49:38.73 & +12:45:09.20 & 2.365 & 15.92 & 15.34 \\
Stripe82X & S82X1905 & 23:28:56.35 & $-$00:30:11.74 & 2.263 & 19.72 & 19.15 \\
			& S82X1940 & 23:29:40.28 & $-$00:17:51.68 & 2.351 & 20.80 & 20.15 \\
            & S82X2058 & 23:31:58.62 & $-$00:54:10.44 & 2.308 & 19.79 & 19.29\\
			& S82X2106 & 23:32:53.24 & $-$00:33:35.35 & 2.281 & 20.56 & 20.23 \\
\hline
\end{tabular}
\tablefoot{
(1) Field where the targets are located. (2) Source identification number from the catalogs corresponding to each field, i.e., \citet{menzel16}, \citet{luo17}, \citet{civano16}, \citet{martocchia17} and \citet{lamassa16}, respectively (see also Sec. \ref{s:selection}). (3) RA and (4) DEC, given for the optical counterpart: the XMM-XXL targets have an SDSS counterpart whose coordinates are given in \citet{menzel16}; for the targets in the CDF-S we report the CANDELS coordinates when available (we use the GEMS coordinates from \citet{haussler07} for the targets XID36 and XID57 since they are outside the CANDELS area), as given in \citet{luo17}; for the COSMOS field we list the \textit{i}-band coordinates taken from \citet{marchesi16}; the information for the WISSH subsample are available in \citet{martocchia17}; for the targets in Stripe82X we give the SDSS coordinates from \citet{lamassa16}. (5) Spectroscopic redshift, taken from the papers listed above. (6) \textit{H}-band and (7) \textit{K}-band AB magnitudes. \\
\tablefootmark{a}{We took the redshift available in \citet{xue16}, since \citet{luo17} provide a different redshift flagged as ``Insecure''.}\\

}
\end{table*}

\subsection{X-ray properties of the sample}\label{s:X-ray}

As described in Sec. \ref{s:selection}, our survey sample is selected from available X-ray AGN surveys. Apart from the source detection, these X-ray observations provide us with important information on the AGN properties from the analysis of their X-ray spectra. Since the obscuring column densities $N_{\textnormal{H}}$ and the X-ray luminosities $L_{\textnormal{X}}$ available from the various survey catalogs may be affected by inhomogeneities due to the adoption of different analysis methods and spectral models, we decided to perform a new systematic analysis of all the X-ray spectra, by using XSPEC v.12.9.1\footnote{\url{https://heasarc.gsfc.nasa.gov/xanadu/xspec/}} \citep{arnaud96}. For this purpose, we followed the method described in \citet{lanzuisi13} and \citet{marchesi16X} for \textit{XMM-Newton} and \textit{Chandra} data respectively, which have been extensively tested in the low count regime typical of the current data set. \textit{Chandra} and \textit{XMM-Newton} spectra of sources in the COSMOS field are extracted following \citet{lanzuisi13} and \citet{mainieri11}, respectively. For sources in the CDF-S we followed the approach described in \citet{vito13}, applied to the full 7 Ms data set \citep{luo17}. For sources in XMM-XXL, Stripe82X and WISSH (SDSS targets), we extracted new \textit{XMM-Newton} spectra adopting a standard data reduction procedure\footnote{We used SAS v.16.0.0, \url{https://heasarc.gsfc.nasa.gov/docs/xmm/xmmhp_analysis.html}} (background flare removal, ``single'' and ``double'' event selection, CCD edge and bad pixels removal) and standard source, background and response matrix extraction from circular regions, whose radii were chosen to maximize the
signal-to-noise ratio \citep[e.g.,][for XMM-XXL]{liu16}. Typical background regions are $\sim 10$ times the source extraction regions. We considered the $0.5-7.0$ keV band for \textit{Chandra} and $0.5-10$ keV band for \textit{XMM-Newton}. All the fits were performed by using the Cash statistic \citep{cash79} and the direct background option \citep{wachter79}. The spectra are binned to 1 count per bin to avoid empty channels.

For sources with more than 30 (50) net counts (reported in Table \ref{tab:summary_results}) for \textit{Chandra} (\textit{XMM-Newton}), we performed a simple spectral fit, modeling the emission with an absorbed power law plus Galactic absorption as well as a secondary power law to reproduce any excess in the soft band, due to scattering or partial covering in obscured sources. In 11 cases out of 39 this second component gave a significant contribution to the fit, while in the other cases its normalization was consistent with 0. The photon index was left free to vary during the spectral analysis for spectra with more than $\sim 100$ net counts (typical values within $\Gamma = 1.5-2.5$), otherwise we fixed it to the canonical value of 1.8 \citep[e.g.,][]{piconcelli05}, being mainly interested in deriving reliable $N_{\textnormal{H}}$ and $L_{\textnormal{X}}$ values. For targets with less than 30 (50) counts, we relied on hardness ratios ($HR = \frac{H-S}{H+S}$, where $H$ and $S$ are the number of counts in the hard $2-7$ keV and soft $0.5-2$ keV bands, respectively), converted into $N_{\textnormal{H}}$ values at the source redshift following \citet{lanzuisi09}.

Both in the case of spectral analysis and HR, we propagated the uncertainty on $N_{\textnormal{H}}$ when deriving the errors on the intrinsic luminosity. This is in fact the main source of uncertainty in $L_{\textnormal{X}}$, at least for obscured sources. We compared our results for the targets in the COSMOS field with those presented by \citet{marchesi16X}, who performed X-ray spectral analysis for all the targets with more than 30 counts in the $0.5 - 7$ keV band. Ten out of sixteen of our COSMOS targets were analyzed in \citet{marchesi16X}, for which the comparison results in an average $\langle \log(L_{\textnormal{X, literature}}/L_{\textnormal{X, this work}})\rangle\,=-0.08$ dex and $\langle\log(N_{\textnormal{H, literature}}/N_{\textnormal{H, this work}})\rangle\,=0.07$ dex, as well as a standard deviation of 0.2 and 0.3 dex, respectively. 

The results derived for column densities and $2-10$ keV absorption-corrected luminosities are listed in Table \ref{tab:summary_results}. X-ray luminosities range between $L_{\textnormal{X}} = 1.6 \times 10^{43}$ erg s$^{-1}$ and $6.5 \times 10^{45}$ erg s$^{-1}$, therefore including AGN with Seyfert-like X-ray luminosities ($L_{\textnormal{X}} \sim 10^{42}-10^{44}$ erg s$^{-1}$) and quasar-like ones ($L_{\textnormal{X}} > 10^{44}$ erg s$^{-1}$). In terms of column densities, the target sample covers uniformly a range from unobscured ($N_{\textnormal{H}} \le 10^{20}$ cm$^{-2}$, given by the Galactic value) to obscured and Compton-thick AGN ($N_{\textnormal{H}} > 10^{24}$ cm$^{-2}$), with values up to $2 \times 10^{24}$ cm$^{-2}$. For the objects whose column density derived from the X-ray spectral analysis is $\sim 10^{20}$ cm$^{-2}$ we provide 90\% confidence level upper limits. 
From an X-ray point of view, AGN are classified as unobscured when $N_{H} < 10^{22}$ cm$^{-2}$ and obscured vice versa. Overall the sample is split in almost an equal number of unobscured and obscured objects based on the X-ray classification. Further discussion about these results, in relation to other physical properties of our targets, is presented in Sec. \ref{s:X-ray_SED}.

\renewcommand{\arraystretch}{1.3}
\onecolumn
\begin{landscape}
\begin{longtable}{ccccccccccc}
\caption{\label{tab:summary_results} Target sample properties summary, for both AGN and host galaxies.}\\
\hline\hline
ID & AGN type & $\log \frac{M_{\ast}}{M_{\odot}}$ & $\log \frac{L_{\textnormal{FIR}}}{\textnormal{erg\,s$^{-1}$}}$ & SFR [$M_{\odot}$ yr$^{-1}$] & $\log \frac{L_{\textnormal{bol}}}{\textnormal{erg\,s$^{-1}$}}$ & X-ray net counts & $\log \frac{N_{\textnormal{H}}}{\textnormal{cm$^{-2}$}}$ & $\log \frac{L_{[2-10\,\textnormal{keV}]}}{\textnormal{erg\,s$^{-1}$}}$ & $\log \frac{M_{\textnormal{BH}}}{M_{\odot}}$ & $\log \frac{P_{\textnormal{1.4 GHz}}}{\textnormal{W Hz$^{-1}$}}$ \\
(1) & (2) & (3) & (4) & (5) & (6) & (7) & (8) & (9) & (10) & (11) \\
\hline
X\_N\_160\_22 & BL & - & - & - & $46.74 \pm 0.02$ & $24.5 \pm 4.9$ & $<22.32$ & $44.77^{+0.14}_{-0.19}$ & $8.9 \pm 0.2$ & $<24.87$ \\
X\_N\_81\_44 & BL & $11.04 \pm 0.37$ & $45.93 \pm 0.20$ & $229 \pm 103$ & $46.80 \pm 0.03$ & $94.8 \pm 9.7$ & $<21.86$ & $44.77^{+0.07}_{-0.09}$ & $9.2 \pm 0.1$ & $<24.81$ \\
X\_N\_53\_3 &  BL & - & $46.41 \pm 0.11$ & $686 \pm 178$ & $46.21 \pm 0.03$ & $25.8 \pm 5.1$ & $22.77^{+0.37}_{-0.67}$ & $44.80^{+0.10}_{-0.13}$ & $8.7 \pm 0.1$ & $<24.86$ \\
X\_N\_66\_23 & BL & $10.96 \pm 0.29$ & $< 46.00$ & $< 268$ & $46.04 \pm 0.02$ & $118.7 \pm 10.9$ & $<21.51$ & $44.71^{+0.06}_{-0.08}$ & $8.3 \pm 0.2$ & $<24.84$ \\
X\_N\_35\_20 & BL & - & - & - & $45.44 \pm 0.02$ & $38.1 \pm 6.2$ & $<22.27$ & $44.00^{+0.07}_{-0.40}$ & $8.7 \pm 0.1$ & $<24.79$ \\
X\_N\_12\_26 & BL & - & - & - & $46.52 \pm 0.02$ & $61.3 \pm 7.8$ & $<20.90$ & $44.56^{+0.13}_{-0.12}$ & $8.9 \pm 0.1$ & $<24.88$ \\
X\_N\_44\_64 & BL & $11.09 \pm 0.25$ & $45.93 \pm 0.15$ & $229 \pm 80$ & $45.51 \pm 0.07$ & $51.8 \pm 7.2$ & $<21.97$ & $44.21^{+0.11}_{-0.17}$ & $9.1 \pm 0.2$ & $<24.78$ \\
X\_N\_4\_48 & BL & - & - & - & $46.16 \pm 0.02$ & $58.2 \pm 7.6$ & $<21.85$ & $44.52^{+0.09}_{-0.16}$ & $9.1 \pm 0.2$ & $<24.81$ \\
X\_N\_102\_35\tablefootmark{a} & BL & - & - & - & $46.82 \pm 0.02$ & $79.0 \pm 8.9$ & $<22.17$ & $45.37^{+0.05}_{-0.11}$ & $8.9 \pm 0.1$ & $27.05 \pm 0.01$\\
X\_N\_115\_23 & BL & - & - & - & $46.49 \pm 0.02$ & $131.8 \pm 11.5$ & $<22.26$ & $44.93^{+0.08}_{-0.10}$ & $8.4 \pm 0.1$ & $<24.82$ \\
XID36\tablefootmark{a} & NL & $10.68 \pm 0.07$ & $45.84 \pm 0.02$ & $184 \pm 9$ & $45.70 \pm 0.06$ & $47.2 \pm 6.9$ & \tablefootmark{b}$>24.1$ & $43.84^{+0.31}_{-0.63}$ & - & $25.40 \pm 0.01$ \\
XID57 & NL & $10.49 \pm 0.11$ & $< 45.10$ & $< 34$ & $44.26 \pm 0.18$ & $58.9 \pm 7.7$ & \tablefootmark{b}$23.30^{+0.32}_{-0.39}$ & $44.04^{+0.17}_{-0.24}$ & - & $<24.20$ \\
XID419 & NL & $10.89 \pm 0.02$ & $45.20 \pm 0.04$ & $42 \pm 4$ & $45.54 \pm 0.05$ & $70.2 \pm 8.4$ & \tablefootmark{b}$24.28^{+0.19}_{-0.31}$ & $43.84^{+0.29}_{-0.44}$ & - & $<24.12$ \\
XID427 & NL & $10.87 \pm 0.08$ & $< 45.43$ & $< 72$ & $44.60 \pm 0.13$ & $324.2 \pm 18.0$ & $22.43^{+0.24}_{-0.34}$ & $43.20^{+0.06}_{-0.06}$ & - & $<24.20$ \\
XID522 & NL & $10.42 \pm 0.02$ & $46.26 \pm 0.02$ & $492 \pm 25$ & $45.02 \pm 0.02$ & $35.1 \pm 5.9$ & \tablefootmark{b}$>22.5$ & $43.51^{+0.76}_{-0.87}$ & - & $24.37 \pm 0.05$ \\
XID614 & NL & $10.78 \pm 0.08$ & $45.97 \pm 0.02$ & $247 \pm 12$ & $44.97 \pm 0.13$  & $78.3 \pm 8.8$ & $24.25^{+0.19}_{-0.18}$ & $43.61^{+0.18}_{-0.18}$ & - & $<24.26$ \\
cid\_166 & BL & $10.38 \pm 0.22$ & $< 45.92$ & $< 224$  & $46.93 \pm 0.02$ & $717.8 \pm 26.8$ & $<21.25$ & $45.15^{+0.03}_{-0.02}$ & $9.3 \pm 0.1$\tablefootmark{c} & $<24.00$ \\
lid\_1289 & NL & $9.59 \pm 0.14$ & $< 44.98$ & $< 25$ & $45.09 \pm 0.08$ & $123.4 \pm 11.1$ & $22.50^{+0.29}_{-0.22}$ & $44.69^{+0.26}_{-0.13}$ & - & $<23.98$ \\
cid\_1057 & NL & $10.84 \pm 0.07$ & $45.50 \pm 0.02$ & $85 \pm 4$ & $45.91 \pm 0.06$ & $36.1 \pm 6.0$ & $23.98^{+0.24}_{-0.28}$ & $44.53^{+0.26}_{-0.30}$ & - & $<23.89$ \\
cid\_1605 & BL & - & $< 45.54$ & $< 94$ & $46.03 \pm 0.02$ & $327.9 \pm 18.1$ & $21.77^{+0.51}_{-0.75}$ & $44.69^{+0.06}_{-0.04}$ & $8.4 \pm 0.2$ & $<23.84$ \\
cid\_337 & NL & $11.13 \pm 0.04$ & $45.63 \pm 0.03$ & $115 \pm 9$ & $45.34 \pm 0.09$ & $83.1 \pm 9.1$ & $<22.76$ & $44.22^{+0.11}_{-0.12}$ & - & $23.99 \pm 0.07$ \\
cid\_346\tablefootmark{a} & BL & $11.01 \pm 0.22$ & $46.13 \pm 0.06$ & $362 \pm 49$ & $46.66 \pm 0.02$ & $124.1 \pm 11.1$ & $23.05^{+0.17}_{-0.19}$ & $44.47^{+0.08}_{-0.09}$ & $8.9 \pm 0.1$\tablefootmark{c} & $24.86 \pm 0.07$\tablefootmark{d} \\
cid\_451\tablefootmark{a} & NL & $11.21 \pm 0.05$ & $< 45.67$ & $< 125$ & $46.44 \pm 0.07$ & $136.9 \pm 11.7$ & $23.87^{+0.19}_{-0.15}$ & $45.18^{+0.23}_{-0.19}$ & - & $26.43 \pm 0.01$\tablefootmark{d}\\
cid\_1205 & NL & $11.20 \pm 0.10$ & $46.16 \pm 0.04$ & $384 \pm 33$ & $45.75 \pm 0.17$ & $33.9 \pm 5.8$ & $23.50^{+0.27}_{-0.27}$ & $44.25^{+0.21}_{-0.23}$ & - & $24.10 \pm 0.06$ \\
cid\_2682 & NL & $11.03 \pm 0.04$ & $< 45.54$ & $< 93$ & $45.48 \pm 0.10$ & $35.5 \pm 6.0$ & $23.92^{+1.01}_{-0.20}$ & $44.30^{+0.96}_{-0.27}$ & - & $<23.99$ \\
cid\_1143\tablefootmark{a} & NL & $10.40 \pm 0.17$ & $45.61 \pm 0.07$ & $108 \pm 18$ & $44.85 \pm 0.12$ & $51.3 \pm 7.2$ & $24.01^{+0.77}_{-0.29}$ & $44.83^{+0.43}_{-0.36}$ & - & $24.39 \pm 0.05$ \\
cid\_467 & BL & $10.10 \pm 0.29$ & $< 45.74$ & $< 147$ & $46.53 \pm 0.04$ & $446.8 \pm 21.1$ & $22.31^{+0.23}_{-0.32}$ & $44.87^{+0.04}_{-0.05}$ & $8.9 \pm 0.6$ & $<23.92$ \\
cid\_852 & NL & $11.17 \pm 0.02$ & $< 45.57$ & $< 100$ & $45.50 \pm 0.11$ & $25.0 \pm 5.0$ & $24.30^{+0.38}_{-0.37}$ & $45.20^{+1.14}_{-0.76}$ & - & $<23.90$ \\
cid\_971 & NL & $10.60 \pm 0.12$ & - & $<96$\tablefootmark{e} & $44.71 \pm 0.24$ & $33.1 \pm 5.8$ & $<23.68$ & $43.87^{+0.36}_{-0.38}$ & - & $<24.01$ \\
cid\_38 & NL & $11.01 \pm 0.12$ & $< 45.98$ & $< 258$ & $45.78 \pm 0.04$ & $159.4 \pm 12.6$ & $<22.95$ & $44.41^{+0.16}_{-0.13}$ & - & $24.10 \pm 0.05$ \\
lid\_206 & BL & $10.30 \pm 0.25$ & - & $63 \pm 27$\tablefootmark{e} & $44.77 \pm 0.12$ & $40.2 \pm 6.3$ & $<22.55$ & $43.91^{+0.39}_{-0.29}$ & $7.9 \pm 0.1 $ & $<23.94$ \\
cid\_1253 & NL & $10.99 \pm 0.25$ & $46.02 \pm 0.30$ & $280 \pm 194$ & $45.08 \pm 0.18$ & $36.1 \pm 6.0$ & $23.22^{+0.47}_{-0.39}$ & $43.92^{+0.29}_{-0.31}$ & - & $24.30 \pm 0.08$\tablefootmark{d}\\
J1333$+$1649\tablefootmark{a} & BL & - & - & - & $47.91 \pm 0.02$ & $174.5 \pm 13.2$ & $21.81^{+0.22}_{-0.34}$ & $45.81^{+0.07}_{-0.06}$ & $9.79 \pm 0.3$ & $28.15 \pm 0.01$ \\
J1441$+$0454 & BL & - & - & - & $47.55 \pm 0.02$ & $74.5 \pm 8.6$ & $22.77^{+0.18}_{-0.21}$ & $44.77^{+0.10}_{-0.11}$ & $10.2 \pm 0.3$ & $25.78 \pm 0.03$\\
J1549$+$1245 & BL & - & - & - & $47.73 \pm 0.04$ & $1023.1 \pm 32.0$ & $22.69^{+0.09}_{-0.11}$ & $45.38^{+0.02}_{-0.02}$ & $10.1 \pm 0.3$\tablefootmark{f} & $25.91 \pm 0.03$ \\
S82X1905 & BL & - & - & - & $46.50 \pm 0.02$ & $31.3 \pm 5.6$ & $22.95^{+0.35}_{-0.17}$ & $44.91^{+0.50}_{-0.50}$ & $9.3 \pm 0.1$ & $<24.79$ \\
S82X1940 & BL & - & - & - & $46.03 \pm 0.02$ & $33.7 \pm 5.8$ & $<20.50$ & $44.72^{+0.30}_{-0.30}$ & $8.7 \pm 0.2$ & $<24.83$ \\
S82X2058 & BL & - & - & - & $46.39 \pm 0.02$ & $29.5 \pm 4.3$ & $<20.50$ & $44.67^{+0.30}_{-0.30}$ & $8.9 \pm 0.3$ & $<24.81$ \\
S82X2106 & BL & - & - & - & $46.08 \pm 0.03$ & $94.5 \pm 9.7$ & $<22.08$ & $45.08^{+0.08}_{-0.11}$ & $9.2 \pm 0.1$ & $<24.80$ \\
\hline
\end{longtable}
\tablefoot{
(1) Target ID, see also Table \ref{tab:sample}; (2) AGN classification into broad line (BL) and narrow line (NL) according to the optical spectra; (3) Galaxy stellar mass and 1$\sigma$ error; (4) FIR luminosity in the $8-1000$ $\mu$m range and 1$\sigma$ error; (5) SFR from the FIR luminosity and 1$\sigma$ error; (6) AGN bolometric luminosity and 1$\sigma$ error, derived from SED fitting; (7) X-ray net counts (i.e., background subtracted) in the full band and respective error, computed assuming a Poisson statistic; (8) Absorbing hydrogen column density and 90\% confidence level error; (9) Absorption-corrected X-ray luminosity in the hard band ($2-10$ keV) and 90\% confidence level error; (10) Black hole mass and 1$\sigma$ error. For the XMM-XXL and Stripe82X targets the values are from \citet{shen15}, for the WISSH targets are from \citet{weedman12}, and for COSMOS we have re-analyzed zCOSMOS and FMOS spectra (Schulze et al., in prep.); (11) Radio power at 1.4 GHz and 1$\sigma$ error. \\
\tablefootmark{a}{Targets classified as ``jetted'' \citep{padovani17_comment}, according to the comparison between their IR and radio properties (see Fig. \ref{fig:radio_FIR}).}\\
\tablefootmark{b}{Targets fit with the physical model MYtorus \citep{my09} that self-consistently takes into account photoelectric absorption, Compton scattering, cold reflection and fluorescent emission in a fixed toroidal geometry. The analysis was performed as described in \citet{lanzuisi18}.}\\ 
\tablefootmark{c}{BH mass derived using the H$\beta$ emission line from Subaru/FMOS spectroscopy (Schulze et al., in prep.).}\\
\tablefootmark{d}{Targets detected at 1.4 GHz as part of the VLA-COSMOS survey \citep{schinnerer07}. The luminosities reported here for these targets are derived from the 1.4 GHz fluxes. The predicted 3 GHz fluxes (assuming $\alpha = 0.7$) are consistent with the observed 1.4 GHz ones for cid\_1253 and cid\_346, while for cid\_451 the predicted flux is a factor of 2 higher than the measured 1.4 GHz one. The measurements reported for the other COSMOS targets are derived from 3 GHz fluxes (see Sec. \ref{s:radio}).} \\
\tablefootmark{e}{Average SFR over the last 100 Myr of the galaxy history as obtained from the modeling of the stellar component with SED fitting.}\\
\tablefootmark{f}{BH mass derived using the H$\beta$ emission line from \citet{bischetti17}.} \\
}
\end{landscape}
\twocolumn

\section{Target sample characterization}\label{s:Characterization}

To draw a wide and complete picture of the physical properties of our AGN and host galaxies, a full multi-wavelength support is needed. We make use of the rich suite of multi-wavelength ancillary data available for these targets, which are unique in terms of amount and depth. They range from the X-rays (Sec. \ref{s:selection}) to the optical, NIR and FIR regimes, and up to the radio (see Sec. \ref{s:radio}). This allows us to gather information about AGN quantities, such as obscuring column density, X-ray and bolometric luminosity, BH mass, as well as galaxy ones, e.g., stellar mass and SFR.

In this Section we describe the ancillary data collected for this work from the UV to the FIR when available, as well as the code used to perform the SED-fitting analysis of the target sample.

\subsection{Multi-wavelength dataset}\label{s:dataset}

The counterparts to the X-ray sources in the CDF-S and COSMOS are provided along with the optical-to-MIR multi-wavelength photometry by the original catalogs \citep[][]{hsu14,laigle16}, where in both cases images were previously registered at the same reference and the photometry was PSF-homogenized. The counterparts to the SUPER targets in XMM-XXL and Stripe82X are known in the SDSS optical images and the corresponding associations to the X-ray sources are likewise given in the original catalogs \citep{fotopoulou16,lamassa16,ananna17}. The remaining SDSS targets are WISE selected with follow-up in the X-ray band \citep{martocchia17}. 

We complemented the UV-to-MIR photometry with further FIR data from \textit{Herschel}/PACS and SPIRE, when available, using a positional matching radius of $2''$, taking into account that we used 24 $\mu$m-priored catalogs which in turn are IRAC-3.6 $\mu$m priored. Here we briefly describe the multi-wavelength data set used for this study but further information can be found in the specific papers mentioned for each field in the following. In Table\,\ref{tab:dataset} we summarize the wavelength bands used to build the SEDs of our AGN. The column description of the photometric catalog is available in Appendix \ref{sec:photo_cat}.

\subsubsection{CDF-S} 
The multi-wavelength catalog used for this field, presented in \citet{hsu14}, provides UV-to-MIR photometric data for all the sources detected in the Extended \textit{Chandra} Deep Field-South \citep[E-CDF-S;][]{xue16,lehmer05}, combining data from CANDELS \citep{guo13}, MUSYC \citep{cardamone10} and TENIS \citep{hsieh12}. 
MIR and FIR photometry at 24 $\mu$m with \textit{Spitzer}/MIPS and at 70, 100 and 160 $\mu$m with \textit{Herschel}/PACS is presented by \citet{magnelli13}, combining observations from the PACS Evolutionary Probe \citep[PEP;][]{lutz11} and the GOODS-\textit{Herschel} programs \citep[GOODS-H;][]{elbaz11}. 
\textit{Herschel}/SPIRE fluxes at 250, 350 and 500 $\mu$m are taken from the \textit{Herschel} Multi-tiered Extragalactic Survey \citep[HerMES;][]{roseboom10,roseboom12,oliver12} DR3. We point out that the HerMES team provides also \textit{Herschel}/PACS photometry for the same field. However, we decided to take advantage of the deeper data released by the PEP team, after verifying the consistency of the fluxes obtained by both teams. Two out of the six targets in the CDF-S are outside the area covered by CANDELS (namely XID36 and XID57). Therefore, we adopted FIR observations at 100 and 160 $\mu$m from the PEP DR1 \citep{lutz11}, since the data products released by \citet{magnelli13} cover the GOODS-S field only. The prior information for these FIR catalogs is given by IRAC-3.6 $\mu$m source positions. To this data set, we added ALMA data in Band 7 and 3 available from the ALMA Archive and 
\citet{scholtz18}.

\subsubsection{COSMOS} 
The UV-to-MIR photometry is taken from the COSMOS2015 catalog presented in \citet{laigle16}, combining existing data from previous releases \citep[e.g.,][]{capak07,ilbert09,ilbert13} and new NIR photometry from the UltraVISTA-DR2 survey, Y-band observations from Subaru and infrared data from \textit{Spitzer}. The source detection is based on deep NIR images and all the photometry is obtained from images registered at the same reference. 
\textit{Spitzer}/MIPS photometry at 24 $\mu$m and \textit{Herschel}/PACS at 100 and 160 $\mu$m is taken from the PEP DR1 \citep{lutz11} extracted using IRAC-3.6 $\mu$m source position priors, as mentioned above. The 24 $\mu$m data for the targets cid\_971 and lid\_206 were provided by Le Floc'h (priv. comm.), since they are particularly faint in this band and therefore not reported in the original catalog. \textit{Herschel}/SPIRE photometry at 250, 350 and 500 $\mu$m is retrieved from the data products presented in \citet{hurley17}, who describe the 24 $\mu$m prior-based source extraction tool XID+, developed using a probabilistic Bayesian method. The resulting flux probability distributions for each source in the catalog are described by the 50th, 84th and 16th percentiles. We assumed Gaussian uncertainties by taking the maximum between the 84th-50th percentile and the 50th-16th percentile. As done for the CDF-S, we added ALMA data in Band 7 and 3 available from the ALMA Archive and 
\citet{scholtz18}.

\subsubsection{XMM-XXL North} 
The multi-wavelength photometry from UV-to-MIR for this field is obtained by merging the photometric SDSS and CFHTLenS \citep{erben13} optical catalogs \citep[see][]{fotopoulou16,georgakakis17}. These data were complemented with GALEX/NUV photometry, 
\textit{YZJHK} band photometry from VISTA  
as well as \textit{u} and \textit{i} bands from CFHT \citep[see][]{fotopoulou16}. 
We considered total magnitudes for CFHTLenS data and model mag for SDSS. As for IRAC, we considered aperture 2 (1.9$''$) photometry corrected to total. 
The WISE data are taken from \citet{lang16}, who provide forced photometry of the WISE All-sky imaging at SDSS positions. 
\textit{Herschel}/PACS and SPIRE data are those released by the HerMES collaboration in the Data Release 4 and 3 respectively \citep{oliver12}. Both sets of data are extracted using the same \textit{Spitzer}/MIPS 24 $\mu$m prior catalog, whose fluxes are available along with the SPIRE data. We use aperture fluxes in smaller apertures, that is 4$''$ diameter.

\subsubsection{Stripe82X} 
We used the photometry that was made public in \citet{ananna17} and was used for the computation of the photometric redshifts in the field. The data are homogeneously deep in optical \citep{fliri16}, but in the NIR and MIR a patchwork of surveys was used \citep[see][and its Fig. 1]{ananna17}. Similarly to the XMM-XXL photometry, we took WISE data from \citet{lang16}. 

\subsubsection{WISSH} 
For these targets we collected UV-to-MIR photometry from the WISSH photometric catalog (Duras et al., in prep.), which includes SDSS photometry, NIR data from the 2MASS as well as WISE photometry from 3 to 22 $\mu$m \citep[see][for further details]{duras17}. 

All the data used in this work are corrected for Galactic extinction \citep{schlegel98}. The resulting photometry, from NUV to FIR, spans a maximum of 31, 36, 27, 12 and 17 wavebands overall for XMM-XXL, CDF-S, COSMOS, WISSH and Stripe82X, respectively. However there is some overlap among bands from different surveys, which reduces the number of unique wavebands. As far as the mid and far-IR photometry from 24 to 500 $\mu$m is concerned, we considered as detections only photometric points with $S/N>3$, where the total noise is given by the sum in quadrature of both the instrumental and the confusion ones \citep{lutz11,oliver12,magnelli13}. The detections below this threshold were converted to $3\sigma$ upper limits. The number of targets with ($\geq 3\sigma$) \textit{Herschel} detections in at least one PACS band is 7 out of 39, while there are 12 out of 39 targets with at least one SPIRE band detection. Five sources present detections in both PACS and SPIRE filters. All the targets have photometric data available from the UV to the MIR. 

Although these data enable a detailed SED modeling, they are collected and/or stacked over many years, so that issues related to variability (intrinsic properties of AGN) can potentially arise \citep[e.g.,][]{simm16}. While we cannot correct for variability in case of stacked images, we were able to correct this issue for the AGN whose photometry was taken in the same wavebands from different surveys. Clear variability was shown by the XMM-XXL targets X\_N\_4\_48, X\_N\_35\_20 and X\_N\_44\_64, for which the SDSS photometry was brighter than the CFHT one by up to two magnitudes. We have taken the latter since it is closer in time to the X-ray observations. 

\renewcommand{\arraystretch}{1.1}
\begin{table*}[h!]
\footnotesize
\caption{\label{tab:dataset} Summary of the photometric data used for the SED-fitting modeling.} 
\centering
\begin{tabular}{ccccc}
\hline\hline
Field & $\lambda$ range & Reference & Telescope/Instrument & Bands \\ 
\hline
XMM-XXL & UV to MIR & \citet{georgakakis17} and & GALEX & NUV \\
              &                 & \citet{fotopoulou16}  & CFHT   & \textit{u, g, r, i, z} \\
              &                 &            & SDSS   & \textit{u, g, r, i, z} \\
              &                 &            & VISTA & \textit{z, Y, J, H, K} \\
              &                &            & \textit{Spitzer}/IRAC & 3.6, 4.5, 5.8, 8.0 $\mu$m \\
              &            &  \citet{lang16}  & WISE & \textit{W1, W2, W3, W4} \\
              & $24-500$ $\mu$m & \citet{oliver12} & \textit{Spitzer}/MIPS & 24 $\mu$m \\
              &                               &                      & \textit{Herschel}/PACS & 70, 100, 160 $\mu$m \\
              &                               &                      & \textit{Herschel}/SPIRE & 250, 350, 500 $\mu$m \\
\hline
CDF-S & UV to MIR & \citet{hsu14} & CTIO-Blanco/Mosaic-II & \textit{U} \\ 
         &               &                       & VLT/VIMOS & \textit{U}\\ 
         &               &                       & HST/ACS & F435W, F606W, F775W, F814W, F850LP \\
         &               &                       & HST/WFC3 & F098M, F105W, F125W, F160W  \\
         &               &                       & ESO-MPG/WFI & \textit{UU$_{38}$BVRI} \\
         &               &                       & CTIO-Blanco/Mosaic-II & \textit{z}-band \\
         &               &                       & NTT/SofI & \textit{H}-band \\
         &               &                       & CTIO-Blanco/ISPI & \textit{J, K} \\
         &				 &						 & VLT/ISAAC & \textit{K$_{S}$} \\
         &               &                       & VLT/HAWK-I & \textit{K$_{S}$} \\
         &               &                       & \textit{Spitzer}/IRAC & 3.6, 4.5, 5.8, 8.0 $\mu$m \\
         & $24-160$ $\mu$m & \citet{magnelli13} or & \textit{Spitzer}/MIPS & 24 $\mu$m \\
         &                               & \citet{lutz11}     & \textit{Herschel}/PACS & 70, 100, 160 $\mu$m \\
         & $250-500$ $\mu$m & \citet{oliver12} & \textit{Herschel}/SPIRE & 250, 350, 500 $\mu$m \\
         & $>1000$ $\mu$m & \citet{scholtz18} and & ALMA & Band 7 (800$-$1100 $\mu$m) \\
         &				 & ALMA Archive & & Band 3 (2600$-$3600 $\mu$m) \\
\hline
COSMOS & UV to MIR & \citet{laigle16} & GALEX & NUV \\
             &                &                         & CFHT/MegaCam & \textit{u$^{\ast}$} \\
             &                &                         & Subaru/Suprime-Cam & \textit{B, V, r, i$^{+}$, z$^{++}$,} \\
             &                &                         & Subaru/HSC & \textit{Y} \\
             &                &                         & VISTA/VIRCAM & \textit{Y, J, H, K$_{s}$} \\
             &                &                         & CFHT/WIRCam & \textit{H, K$_{s}$} \\
             &                &                         & \textit{Spitzer}/IRAC & 3.6, 4.5, 5.8, 8.0 $\mu$m \\
             & $24-160$ $\mu$m & \citet{lutz11} & \textit{Spitzer}/MIPS & 24 $\mu$m \\
             &                               &                             & \textit{Herschel}/PACS & 70, 100, 160 $\mu$m \\
             & $250-500$ $\mu$m & \citet{hurley17} & \textit{Herschel}/SPIRE & 250, 350, 500 $\mu$m \\
             & $>1000$ $\mu$m & \citet{scholtz18} and & ALMA & Band 7 (800$-$1100 $\mu$m) \\
         &				 & ALMA Archive & & Band 3 (2600$-$3600 $\mu$m) \\
\hline
WISSH & UV to MIR & Duras et al. (in prep.) & SDSS & \textit{u, g, r, i, z} \\ 
     &	           &  & 2MASS & \textit{J, H, K} \\
     &				&	& WISE & \textit{W1, W2, W3, W4} \\
\hline
Stripe 82X  &  UV to MIR & \citet{ananna17} & SDSS & \textit{u, g, r, i, z} \\
                &                &          & UKIDSS & \textit{J, H, K} \\
                &                &          & VISTA & \textit{J, H, K} \\
                &                &          & \textit{Spitzer}/IRAC & 3.6, 4.5 $\mu$m \\
                &                & \citet{lang16} & WISE & \textit{W1, W2, W3, W4} \\
\hline
\end{tabular}
\end{table*}

\subsection{Data modeling}\label{s:code}

The analysis presented in this work is performed by using the Code Investigating GALaxy Emission \citep[CIGALE\footnote{\url{https://cigale.lam.fr}};][]{noll09}, a publicly available state-of-the-art galaxy SED-fitting technique. CIGALE adopts a multi-component fitting approach in order to disentangle the AGN contribution from the emission of its host galaxy and estimate in a self-consistent way AGN and host galaxy properties from the integrated SEDs. Moreover, it takes into account the energy balance between the UV-optical absorption by dust and the corresponding re-emission in the FIR. 
Here we provide a brief description of the code and we refer the reader to \citet{noll09}, \citet{buat15} and \citet{ciesla15} for more details. In this work we used the version 0.11.0.

CIGALE accounts for three main distinct emission components: (i) stellar emission, dominating the wavelength range $0.3-5$ $\mu$m; 
(ii) emission by cold dust heated by star formation which dominates the FIR; (iii) AGN emission, appearing as direct energy coming from the accretion disk at UV-optical wavelengths and reprocessed emission by the dusty torus peaking in the MIR.  
The code assembles the models, according to a range of input parameters, which are then 
compared to the observed photometry by computing model fluxes in the observed filter bands and performing an evaluation of the $\chi^{2}$. The output parameters as well as the corresponding uncertainties are determined through a Bayesian statistical analysis: the probability distribution function (PDF) for each parameter of interest is built by summing the exponential term $\exp({-\chi^{2}/2})$ related to each model in given bins of the parameter space. The output value of a parameter is the mean value of the PDF and the associated error is the standard deviation derived from the PDF \citep{noll09}. 
The values of the input parameters used for the fitting procedure are listed in Table\,\ref{tab:input_parameters}. In the following we describe the assumptions and the models adopted.

(i) To create the stellar models we assumed a star formation history (SFH) represented by a delayed $\tau$-model (exponentially declining) with varying e-folding time and stellar population ages (see Table\,\ref{tab:input_parameters}), defined as:

\begin{equation}
\textnormal{SFR}(t) \propto t \times \exp{(-t/\tau)}
\end{equation}

where $\tau$ is the e-folding time of the star formation burst. The stellar population ages are constrained to be younger than the age of the Universe at the redshift of the source sample. 
The SFH is then convolved with the stellar population models of \citet{bc03} and a \citet{chabrier03} initial mass function (IMF). The metallicity is fixed to solar (0.02)\footnote{The impact of lower metallicity on the SED-fitting output was tested by fixing the metallicity to a value 0.3 dex lower than the solar one. The results of the fitting procedure are well within the uncertainties.}. 
To account for the role played by dust in absorbing the stellar emission in the UV/optical regime we applied an attenuation law to the stellar component. One of the most used ones, also at high redshift, is the \citet{calzetti00} law. 
However, in the literature there is evidence for shapes of the attenuation law different from the standard Calzetti one \citep[e.g.,][but see also \citealt{cullen18}]{salvato09,buat11,buat12,reddy15,lofaro17}. We used the modified version of 
the \citet{calzetti00} curve, which is multiplied in the UV range by a power law with a variable slope $\delta$, where the attenuation is given by $A(\lambda) = A(\lambda)_{Calz.} \times (\lambda/550 \,\textnormal{nm})^{\delta}$. In this recipe, negative slopes of the additional power law produce steeper attenuation curves and vice versa positive values give a flatter curve, while a slope equal to 0 reproduces the \citet{calzetti00} curve. We did not include the bump feature at 2175 \AA. The same law is applied to both old ($> 10$ Myr) and young ($< 10$ Myr; \citealt{charlot00}) stars. Moreover we took into account that stars of different ages can suffer from differential reddening 
by applying a reduction factor of the visual attenuation to the old stellar population \citep{calzetti00}. The reduction factor, $E(B-V)_{\textnormal{old}}/E(B-V)_{\textnormal{young}}$, is fixed to 0.93 as derived by \citet{puglisi16}.

(ii) The reprocessed emission from dust heated by star formation is modeled using the library presented by \citet{dale14}, which 
includes the contributions from dust heated by both star formation and AGN activity. In order to treat the AGN emission separately by adopting different models, and therefore estimate the contribution from star formation only with this library, we assumed an AGN contribution equal to 0. This family of models is made of a suite of templates constructed with synthetic and empirical spectra which represent emission from dust exposed to a wide range of intensities of the radiation field. These templates are combined in order to model the total emission and their relative contribution is given by a power law, whose slope is the parameter $\alpha_{\textnormal{SF}}$. 
For higher values of the slope the contribution of weaker radiation fields is more important and the dust emission peaks at longer wavelengths. The dust templates are linked to the stellar emission by a normalization factor which takes into account the energy absorbed by dust and re-emitted in the IR regime.

(iii) Accounting for the AGN contribution is essential for the determination of the host galaxy properties. 
To reproduce the AGN emission component we chose the physical models presented by \citet{fritz06}, who solved the radiative transfer equation for a flared disk geometry with a smooth dust distribution composed by silicate and graphite grains. Although a clumpy or filamentary structure has been observed for nearby AGN \citep[e.g.,][]{jaffe04} and is more physical, in this work we focus on the global characterization of the SED, for which 
both clumpy and smooth models provide good results and are widely used in the literature. As claimed by \citet{feltre12}, the major differences in the SEDs produced by the two dust distributions are due to different model assumptions and not to their intrinsic properties. The main AGN parameter we want to reliably constrain from the SED is the AGN bolometric luminosity, therefore the details of the dust distribution are not fundamental in this work. The law describing the dust density within the torus is variable along the radial and the polar coordinates and is given by:

\begin{equation}
\rho(r, \theta) = \alpha r^{\beta} e^{-\gamma \mid \cos(\theta)\mid}
\end{equation}

where $\alpha$ is proportional to the equatorial optical depth at 9.7 $\mu$m ($\tau_{9.7}$), $\beta$ and $\gamma$ are related to the radial and angular coordinates respectively. Other parameters describing the geometry are the ratio between the outer and the inner radii of the torus, $R_{max}/R_{min}$, and the opening angle of the torus, $\Theta$. The inclination angle of the observer's line of sight with respect to the torus equatorial plane, the parameter $\psi$ with values in the range between 0$^{\circ}$ and 90$^{\circ}$, allows one to distinguish between type 1 AGN (unobscured) for high inclinations and type 2 AGN (obscured) for low inclinations. Intermediate types are usually associated to $\psi \simeq 40^{\circ}-60^{\circ}$ depending on the dust distribution. The central engine is assumed to be a point-like source emitting isotropically with an SED described by a composition 
of power laws parameterizing the disk emission. This emission is partially obscured when the line of sight passes through the dusty torus. 
Another important input parameter that handles the normalization of the AGN component to the host galaxy emission is the AGN fraction, which is the contribution of the AGN emission to the total ($8-1000$ $\mu$m) IR luminosity and is given by $f_{\textnormal{AGN}}=L^{\textnormal{AGN}}_{\textnormal{IR}}/L^{\textnormal{TOT}}_{\textnormal{IR}}$, with $L^{\textnormal{TOT}}_{\textnormal{IR}}=L^{\textnormal{AGN}}_{\textnormal{IR}}+L^{\textnormal{starburst}}_{\textnormal{IR}}$ \citep{ciesla15}. The input values available in the code are based on the results presented by \citet{fritz06}. However, as described by \citet{hatziminaoglou08}, using all the possible values would produce degeneracies in the model templates. 
Therefore we cannot determine the torus geometry in an unequivocal way 
and the parameter proving to be best constrained is the bolometric luminosity. The values of the above-mentioned physical parameters related to the torus geometry should be taken as indicative. For this reason we decided to narrow down the grid of input values and to fix some of them. Our selected values (see Table\,\ref{tab:input_parameters}) are partly based on the analysis performed by \citet{hatziminaoglou08}, who presented a restricted grid of input parameters. 
Differently from their setup, we fixed $R_{\textnormal{max}}/R_{\textnormal{min}}$ and the opening angle to a single value, as well as using a less dense grid for the optical depth. 

To the main emission components described above we also added templates reproducing nebular emission, ranging from the UV to the FIR. 
These templates are based on the models presented by \citet{inoue11} and represent the emission from \ion{H}{ii} regions. They include recombination lines, mainly from hydrogen and helium, and continuum emission due to free-free, free-bound and 2-photon processes of hydrogen. This SED component is proportional to the rate of Lyman continuum photons ionizing the gas and takes into account the Lyman continuum escape fraction and the absorption of the ionizing photons by dust. The templates do not include lines from photo-dissociation regions and nebular lines due to AGN emission. Therefore they do not reproduce the AGN contribution to the emission lines which may contaminate the photometric data. 
We fixed the parameters of the nebular emission model (see Table\,\ref{tab:input_parameters}) as in \citet{boquien16}. 

\renewcommand{\arraystretch}{1.1}
\begin{table*}
\caption{\label{tab:input_parameters} Input parameter values used in the SED-fitting procedure.}
\centering
\begin{tabular}{cccc}
\hline\hline
Template & Parameter & Value and range& Description \\
\hline
{\it Stellar emission} & IMF & \citet{chabrier03}& \\
      & Z & 0.02 & Metallicity \\
      & Separation age & 10 Myr & Separation age between the young \\
      & & & and the old stellar populations \\
Delayed SFH & Age &  0.10, 0.25, 0.5, 1.0, 1.5, 2.0, 2.5 Gyr & Age of the oldest SSP \\
                   & $\tau$ & 0.10, 0.25, 0.5, 1.0, 3.0, 5.0, 10.0 Gyr & e-folding time of the SFH \\
Modified Calzetti & $E(B-V)$ & 0.05, 0.1, 0.3, 0.5, 0.7, 0.9, 1.1, 1.3 & Attenuation of the \\
                  &          &   & young stellar population \\
attenuation law & Reduction factor & 0.93 & Differential reddening applied to \\
				&					&		& the old stellar population \\
                & $\delta$ & -0.6, -0.4, -0.2, 0.0 & Slope of the power law multiplying \\
                &			&		& the Calzetti attenuation law \\
\hline
{\it Dust emission} & $\alpha_{\textnormal{SF}}$ & 0.5, 1.0, 1.5, 2.0, 2.5, 3.0 & Slope of the power law combining \\
					&		&		& the contribution of different dust templates \\
\hline
{\it AGN emission} & $R_{\textnormal{max}}/R_{\textnormal{min}}$ & 60 & Ratio of the outer and inner radii \\
			& $\tau_{9.7}$ & 0.6, 3.0, 6.0 & Optical depth at 9.7 $\mu$m \\
            & $\beta$ & 0.00, -0.5, -1.0 & Slope of the radial coordinate \\
            & $\gamma$ & 0.0, 6.0 & Exponent of the angular coordinate \\
            & $\Theta$ & 100 degrees & Opening angle of the torus \\
            & $\psi$ & 0, 10, 20, 30, 40, 50, 60, 70, 80, 90 degrees & Inclination of the observer's line of sight\\
            & $f_{\textnormal{AGN}}$ & 0.05, 0.1, 0.15, 0.2, 0.25, 0.3, 0.35, 0.4, 0.45, & AGN fraction \\
            &	& 0.5, 0.55, 0.6, 0.65, 0.7, 0.75, 0.8, 0.85, 0.9 & \\
\hline
{\it Nebular emission} & $U$ & $10^{-2}$ & Ionization parameter \\
				& $f_{\textnormal{esc}}$ & 0\% & Fraction of Lyman continuum \\
                &           &     & photons escaping the galaxy \\
                & $f_{\textnormal{dust}}$ & 10\% & Fraction of Lyman continuum \\
                &			&  & photons absorbed by dust \\
\hline
\end{tabular}
\end{table*}

\section{Overall properties of the target sample}\label{s:overall_results}

In this section we provide a detailed picture of the multi-wavelength properties of our target sample obtained using SED fitting and spectral analysis. In particular, we focus on the main AGN and host galaxy physical parameters that we aim to connect to the outflow properties, as traced by our on-going SINFONI observations.

The AGN sample is characterized by a wide range of column densities, up to $2 \times 10^{24}$ cm$^{-2}$, derived from the X-ray spectra (see Section \ref{s:X-ray}). This translates into different levels of contamination of the AGN to the galaxy emission at UV-to-NIR wavelengths. Therefore, host galaxy properties for the targets where this contamination is low, that is obscured AGN, can be robustly determined. At the same time, AGN properties are better constrained for unobscured targets, whose emission prevails in the UV-to-IR portion of the SED. In general, the classification of AGN into obscured and unobscured sources can be performed based on different criteria, such as X-ray spectral analysis, optical spectral properties, and shape of the UV-to-NIR SED \citep[see][]{merloni14}. We adopt the following nomenclature: from an optical point of view, the classification depends on the presence of broad (FWHM $>1\,000$ km s$^{-1}$) or narrow (FWHM $<1\,000$ km s$^{-1}$) permitted lines in their spectra, defining broad-line (BL) or narrow-line (NL) AGN respectively; according to the shape of the UV-to-NIR SED, we can constrain the AGN type based on the inclination of the observer's line of sight with respect to the obscuring torus; 
finally, AGN are classified as unobscured or obscured when the column density is smaller or larger than $10^{22}$ cm$^{-2}$ (see Sec. \ref{s:X-ray}). As explained later in Sec. \ref{s:X-ray_SED}, the three classification methods broadly agree with each other. However, as final classification, we decided to adopt the optical spectroscopic classification (BL/NL, Table \ref{tab:summary_results}). In the following we will refer to type 1 and type 2 AGN as based on the optical spectroscopic classification. 

\subsection{SED-fitting results}\label{s:results_sec}

\begin{figure*}[h!]
\centering
  \includegraphics[width=8cm]{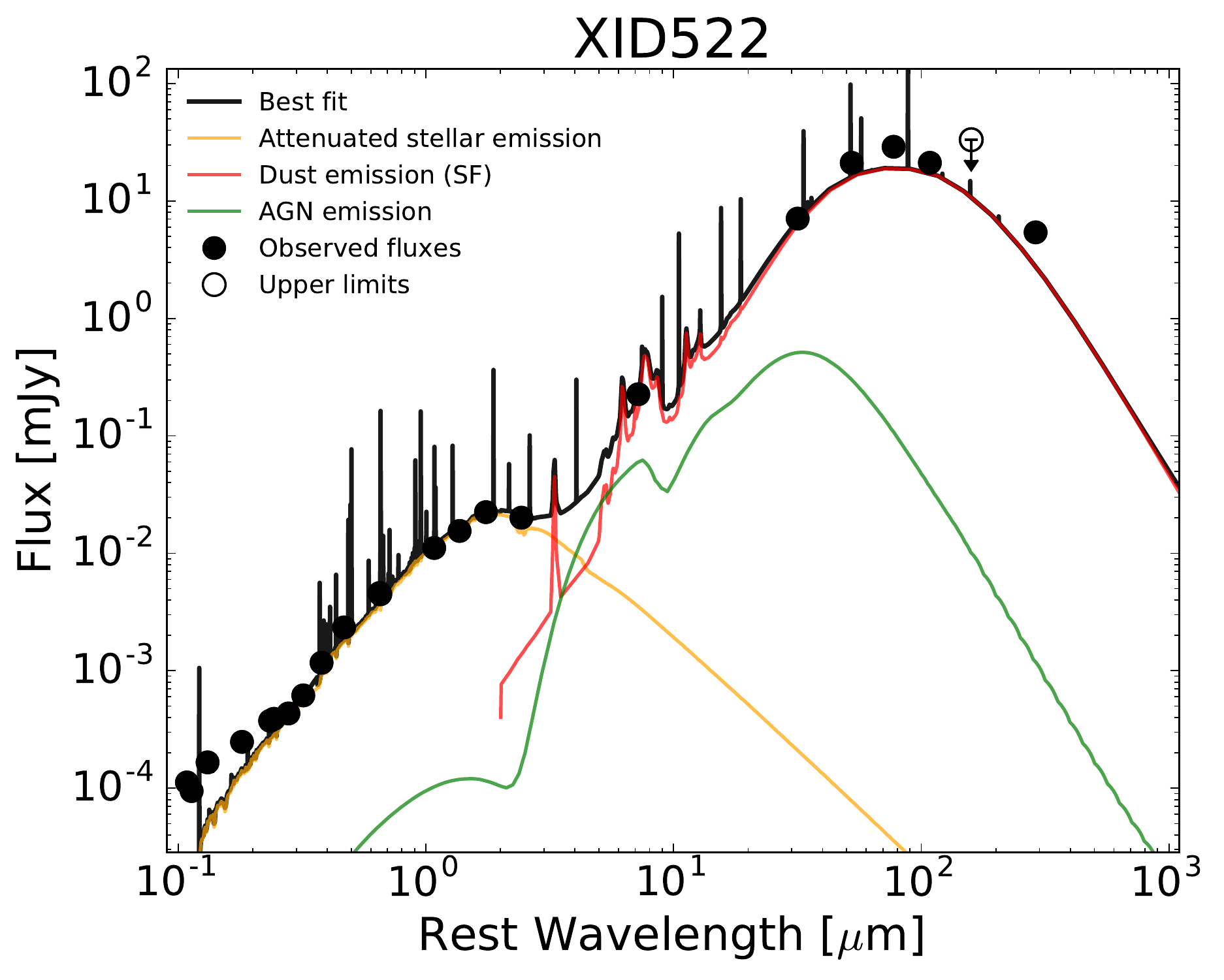}
  \hspace{2mm}
  \includegraphics[width=8cm]{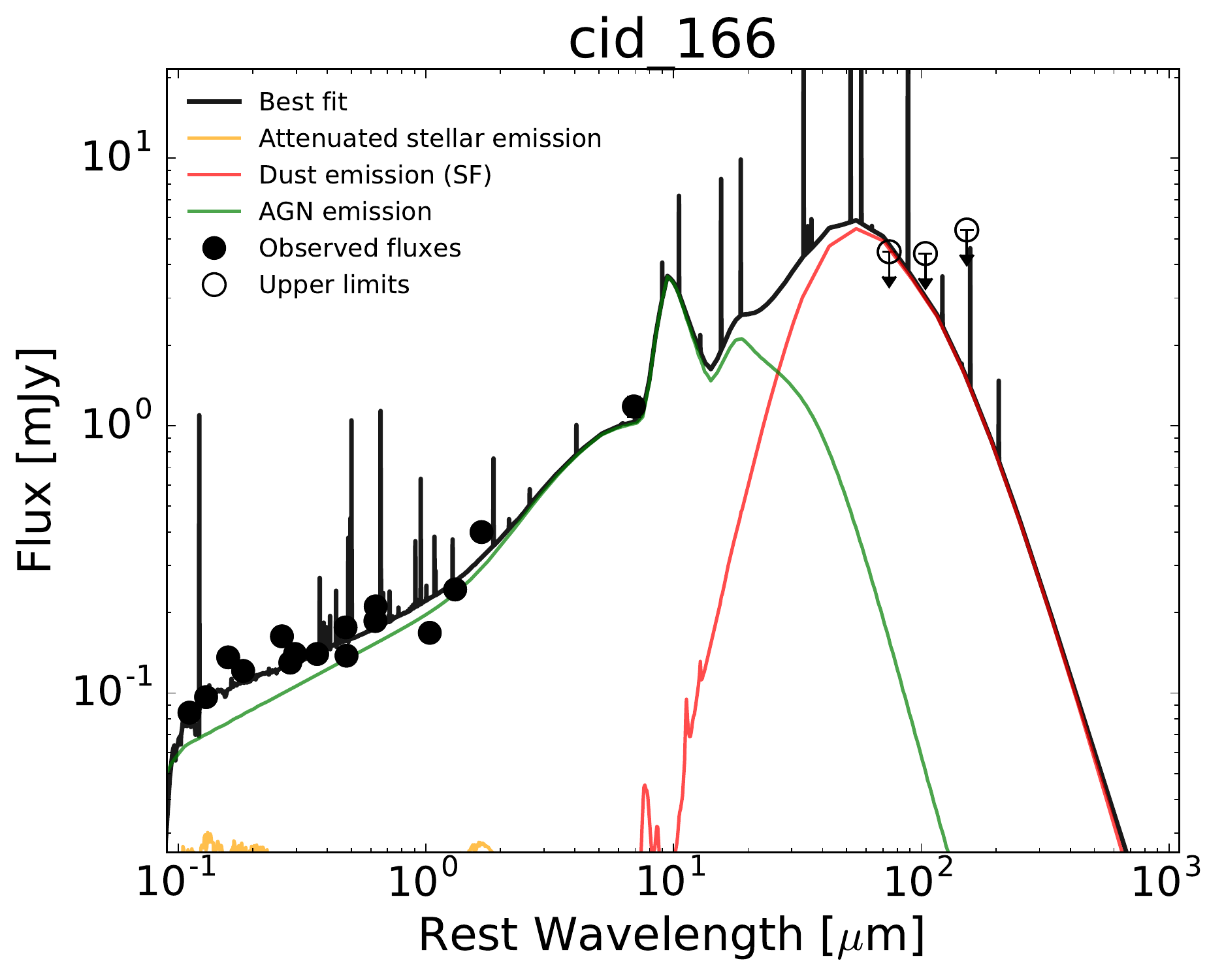}
  \caption{Two examples of rest-frame SEDs obtained for a type 2 (XID522, \textit{left}) and a type 1 (cid\_166, \textit{right}) AGN. The black dots represent the observed multi-wavelength photometry, while the empty dots indicate 3$\sigma$ upper limits. The black solid line is the total best-fit model, the orange curve represents the stellar emission attenuated by dust, the green template reproduces the AGN emission, the red curve accounts for dust emission heated by star formation. Emission lines in the black curves are part of the nebular emission component, included in the overall SED.}\label{fig:sed}
\end{figure*}

The main output parameters obtained with CIGALE are reported in Table\,\ref{tab:summary_results}, that is stellar mass, SFR, and AGN bolometric luminosity together with their 1$\sigma$ uncertainties. Two representative examples of SEDs are shown in Figure\,\ref{fig:sed} for a type 2 (\textit{left panel}) and a type 1 (\textit{right panel}) AGN from CDF-S and COSMOS, respectively. The SEDs of the whole sample are presented in Appendix \ref{sec:SEDs}.

\subsubsection{Stellar masses}

Stellar masses ($M_{\ast}$) are probed by rest-frame NIR flux densities shifted to the MIR at this redshift, which are dominated by old stellar populations. The uncertainty associated to stellar masses increases with the level of AGN contamination.  
As shown by the green template in the left panel of Figure\,\ref{fig:sed}, in type 2s there is a negligible AGN contribution in the UV-to-NIR regime. Conversely, for type 1s the green template in the right panel outshines the galaxy emission (orange curve) preventing a derivation of the stellar mass as robust as for type 2s. However, estimates of the stellar mass for type 1 AGN can still be recovered albeit with larger uncertainties \citep[e.g.,][]{bongiorno12}, apart from very bright type 1s (e.g., Stripe82X, WISSH and some XMM-XXL targets in our sample) for which the uncertainties on this parameter are much larger than the parameter value itself and therefore an estimate of the stellar mass is meaningless. For these targets, we do not report a value of $M_{\ast}$ in Table\,\ref{tab:summary_results}. 
Our results range between $\sim 4 \times 10^{9}$ M$_{\odot}$ and $\sim 1.6 \times 10^{11}$ M$_{\odot}$, with an average 1$\sigma$ uncertainty of 0.1 dex for type 2s and 0.3 dex for type 1s\footnote{\label{foot}In general, the statistical uncertainties in the determination of $M_{\ast}$ and SFR through SED modeling are typically around 0.3 dex for stellar masses and larger for SFRs \citep[e.g.,][]{mancini11,santini15}, usually underestimated by the SED-fitting tools. Moreover, systematic differences in the results are due to the models used, degeneracies and a priori assumptions as well as the discrete coverage of the parameter space.}.

\subsubsection{Star Formation Rates}

SFRs are derived from the IR luminosity integrated in the rest-frame wavelength range $8-1000$ $\mu$m, when possible,  assuming the \citet{kennicutt98} SFR calibration converted to a \citet{chabrier03} IMF (i.e., by subtracting 0.23 dex). This value is an indication of the SFR averaged over the last 100 Myr of the galaxy history and is produced by emission from dust heated by young stars as well as from evolved stellar populations. The AGN also contributes to the IR luminosity (whose percentage is given by the AGN fraction, see Sec. \ref{s:code}), although it usually dominates the emission only up to 30 $\mu$m rest-frame as described in \citet{mullaney11} \citep[see also][]{symeonidis16}. 
Since our SED fitting allows us to disentangle the contribution of the two components (AGN and SF), we estimate the IR luminosity from SF removing the AGN contamination\footnote{AGN fractions (derived for the targets with FIR detections) range between 0.05 and 0.90, with a median value of 0.36.}. 
However this is affected by intrinsic degeneracies that cannot be solved with the current data sampling at MIR and FIR wavelengths. Therefore an over-estimation of the AGN fraction will result in an under-estimation of the IR emission from the galaxy and thus of the SFR and vice versa \citep[e.g.,][]{ciesla15}. We provide a 3$\sigma$ upper limit on the SFR, derived as the 99.7th percentile of the FIR luminosity PDF, for the targets with only upper limits at $\lambda > 24$ $\mu$m. For the subset of targets without data at observed $\lambda > 24$ $\mu$m we did not include the dust templates in the fitting procedure. Therefore we report the average SFR over the last 100 Myr of the galaxy history as obtained from the modeling of the stellar component in the UV-to-NIR regime with SED fitting. This has been done for cid\_971 and lid\_206, since their 24 $\mu$m flux was not available in the catalog used as a prior for the extraction of the FIR photometry (see Sec. \ref{s:dataset}). The targets without an estimate of the SFR are instead bright type 1s, therefore no information about SFR, and stellar mass, can be retrieved from the UV-optical regime. SFRs determined for our targets are in the range between $\sim 25$ M$_{\odot}$ yr$^{-1}$ and $\sim 680$ M$_{\odot}$ yr$^{-1}$ with an average 1$\sigma$ uncertainty of 0.15 dex for type 1 and 0.06 for type 2 AGN (see footnote \ref{foot}). The SFRs derived from the FIR luminosity and through the modeling of the stellar emission in the UV-to-NIR regime are in very good agreement (when the comparison is possible), with the low scatter due to the energy-balance approach used \citep[see also][]{bongiorno12}.

\subsubsection{Comparison of $M_{\ast}$ and SFRs to literature results}

We compared our results with those presented by \citet{santini15} for the targets in the CDF-S and \citet{chang17}, \citet{delvecchio17} as well as \citet{suh17} for the COSMOS targets. \citet{santini15} collected $M_{\ast}$ measurements of the targets in the CANDELS field from several teams which used different SED-fitting codes and assumptions, in order to study the influence of systematic effects on the final output. The resulting estimates turned out to be clustered around the median value with a scatter of $25\%-35\%$. Their results are available for all of our CDF-S targets covered by CANDELS. \citet{chang17} derived physical parameters for galaxies over the whole COSMOS field, \citet{delvecchio17} dealt with a sub-sample of AGN as part of the VLA-COSMOS 3 GHz Large Project, while \citet{suh17} provided physical properties for a sample of X-ray selected type 2s. 16, 6 and 7 out of 16 of our COSMOS targets have a match in these catalogs, respectively. However, the values from \citet{delvecchio17} have been recomputed by adopting the same photometry used in this work (Delvecchio, priv. comm.). The overall comparison for stellar masses is quite satisfactory, with the average $\langle \log(M_{\ast,\,\textnormal{literature}}/M_{\ast,\,\textnormal{this work}}) \rangle$ equal to 0.30 dex (this result includes both type 2 and type 1 AGN), 0.03 dex and 0.18 dex for \citet{chang17}, \citet{delvecchio17} and \citet{suh17}, and 0.20 dex for \citet{santini15}. The standard deviation is 0.38, 0.3 and 0.19 dex for the COSMOS targets and 0.19 dex for the CDF-S ones. The fits performed in \citet{santini15} do not take into account the AGN contribution. 
As for the SFRs, the results are similar, with an average $\langle \log(\textnormal{SFR}_{\textnormal{literature}}/\textnormal{SFR}_{\textnormal{this work}}) \rangle\,=0.39$, -0.30 and 0.03 dex and standard deviation 0.44, 0.34 and 0.47 dex for \citet{chang17}, \citet{delvecchio17} and \citet{suh17}, respectively. The AGN contribution was subtracted in all the estimates. For this comparison we did not consider the SFRs reported in \citet{santini15}, because their SED fitting did not include the FIR fluxes which are crucial to properly constrain the total SFR. 
The larger discrepancies for SFRs are mainly attributed to different and looser constraints in the FIR regime. In general, other sources of uncertainties are the diverse models used and the sparser data with large error bars (often just upper limits) compared to the UV-to-NIR regime. 

Although the SFR is a key quantity to be compared with AGN activity in order to understand the feedback processes, measuring the current SFR in AGN hosts is a well-known challenge, since the tracers are usually contaminated by AGN emission. Thanks to the SINFONI data that will be available for our targets, we will be
able to compare various SF tracers (e.g., narrow H$\alpha$ vs. $L_{\textnormal{FIR}}$) in order to explore the systematic effects in this kind of measurements.

\subsubsection{AGN bolometric luminosities}

As described in Section \ref{s:code}, we used the \citet{fritz06} models to reproduce the overall AGN emission. According to a comparison discussed in \citet{ciesla15}, type 2 AGN templates from the \citet{fritz06} library are cooler than the SEDs obtained empirically by \citet{mullaney11}, which may indicate that those models do not reproduce all the physical properties of the AGN obscuring structure. Moreover, there are several models (both theoretical and empirical) in the literature reproducing the dusty torus emission \citep[e.g.,][]{nenkova08,mor12,stalevski12,lani17} and approximations of the intrinsic AGN continuum \citep[e.g.,][]{telfer02,richards06,stevans14}. Nevertheless, it is important to stress that our main goal is not the detailed determination of the torus or accretion disk specific characteristics but just recovering the AGN bolometric luminosity. To test the reliability of the derived quantity we explored the input parameter space described in Section \ref{s:code} (see also Table\,\ref{tab:input_parameters}) by fixing the input parameters to different values and comparing $L_{\textnormal{bol}}$ to those obtained using the whole grid of models used in this work. Even though the best-fit geometry varied through the different runs, the bolometric luminosity proved to be constrained within a variation of 0.2 dex. The same trend emerged for the dust luminosity due to star formation, which is related to the AGN luminosity by the AGN fraction. Moreover, we compared our results with available literature values for the targets in the COSMOS field \citep[from][]{chang17,delvecchio17} and those from the WISSH catalog \citep[Duras, priv. comm.,][]{duras17}. \citet{duras17} modeled the AGN emission combining models from \citet{feltre12} and \citet{stalevski16}; \citet{chang17} used empirical templates by \citet{richards06}, \citet{polletta07}, \citet{prieto10} and \citet{mullaney11}; \citet{delvecchio17} adopted the \citet{feltre12} library. In spite of the variety of models used by the different authors, the comparison is satisfactory and all the results are within 0.3 dex scatter: the average $\langle \log(L_{\textnormal{bol, literature}}/L_{\textnormal{bol, this work}}) \rangle$ is equal to 0.03, 0.18 and -0.09 dex for \citet{chang17}, \citet{delvecchio17} and Duras (priv. comm.) respectively, with standard deviation 0.26, 0.30 and 0.14 dex. From our SED fitting, average 1$\sigma$ uncertainties of the bolometric luminosity are on the order of 0.03 and 0.1 dex for type 1s and type 2s respectively, with best-fit values in the range $2 \times 10^{44} - 8 \times 10^{47}$ erg s$^{-1}$. As pointed out for $M_{\ast}$ and SFRs, the uncertainties estimated through SED fitting can be underestimated. According to the comparison mentioned above, a more realistic typical uncertainty can be fixed to 0.3 dex.

\begin{figure}
\centering
  \includegraphics[width=9cm]{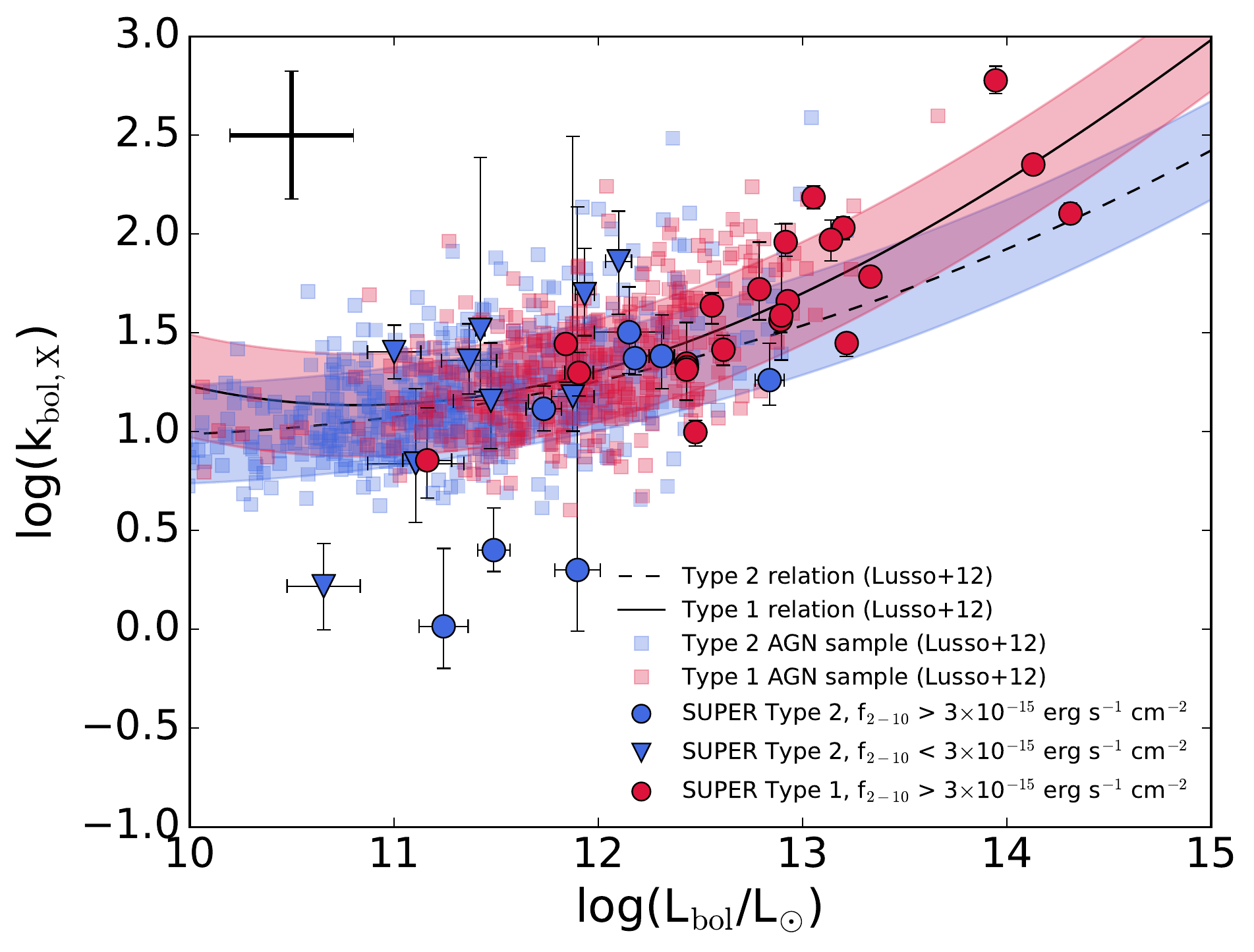}
  \caption{Bolometric corrections in the hard $2-10$ keV band versus bolometric luminosities. Circles and triangles mark the SUPER targets with $2-10$ keV fluxes higher and lower than $3 \times 10^{-15}$ erg s$^{-1}$ cm$^{-2}$ respectively, while type 1 and type 2 AGN are plotted in red and blue. The solid and dashed lines show the relations obtained by \citet{lusso12} for type 1s and type 2s with fluxes higher than $3 \times 10^{-15}$ erg s$^{-1}$ cm$^{-2}$, respectively. The shaded areas depict the scatter of these relations. We plot as red and blue squares the sample of type 1 and type 2 AGN, respectively, analyzed by \citet{lusso12} to show the dispersion of the data around the best-fit relations. The error bar in the upper-left corner takes into account a systematic error of 0.3 dex on $L_{\textnormal{bol}}$. The SUPER data points are well consistent with the trends found for the bolometric correction.
}
  \label{fig:kbol}
\end{figure}

The physical quantities available for the SUPER sample give us the opportunity to study the distribution of the X-ray bolometric correction in the hard $2-10$ keV band (defined as $k_{\textnormal{bol, X}} = L_{\textnormal{bol}}/L_{[2-10\,\textnormal{keV}]}$) versus $L_{\textnormal{bol}}$. This can be done over a wide range of bolometric luminosities with a set of values determined in a uniform way. In Figure \ref{fig:kbol} we compare our results to the relation derived by \citet{lusso12} for a sample of more than 900 AGN (both type 1 and type 2) selected from the COSMOS field \citep[see also][]{lusso16}. At variance with their AGN selection, which includes sources with hard X-ray fluxes larger than $3 \times 10^{-15}$ erg s$^{-1}$ cm$^{-2}$, 23\% of our targets reach fainter values (down to $\approx 5 \times 10^{-17}$ erg s$^{-1}$ cm$^{-2}$, triangles in Fig. \ref{fig:kbol}). Moreover, we can probe the $k_{\textnormal{bol, X}} - L_{\textnormal{bol}}$ relation for targets with bolometric luminosities an order of magnitude higher \citep[see also][]{martocchia17}. We plot the relations obtained by \citet{lusso12} for type 1 and type 2 AGN (solid and dashed lines respectively), although they do not differ too much. The shaded areas depict the dispersion of these relations, while the red and blue squares represent the sample of type 1 and type 2 AGN, respectively, from which the relations were obtained. Our results for both type 1 (shown in red) and type 2 (shown in blue) are well consistent with the trends found by \cite{lusso12} with only the presence of three targets outside the $\pm 1\sigma$ scatter, according to the errorbars. 
The rest of the SUPER targets are within the scatter shown by their sample and also the most luminous AGN are well represented by those curves. One of the outliers has faint hard-band flux (marked by different symbols to distinguish the targets below the threshold adopted by \citealt{lusso12}) and a total number of X-ray counts $< 60$. We note that the error bars in the plot are given by the error on $L_{\textnormal{bol}}$ provided by the SED-fitting code. In the upper-left corner of the panel we plot a median error bar taking into account a systematic error of 0.3 dex on $L_{\textnormal{bol}}$ which is more representative. Accounting for the scatter of the data presented by \citet{lusso12} around the best fits and the underestimated error bars for our $L_{\textnormal{bol}}$, our estimates result to be in agreement with the literature trends and therefore we can consider $L_{\textnormal{bol}}$ and $L_{\textnormal{X}}$ obtained with our analysis reliable parameters.

\begin{figure}
\centering
  \includegraphics[width=9cm]{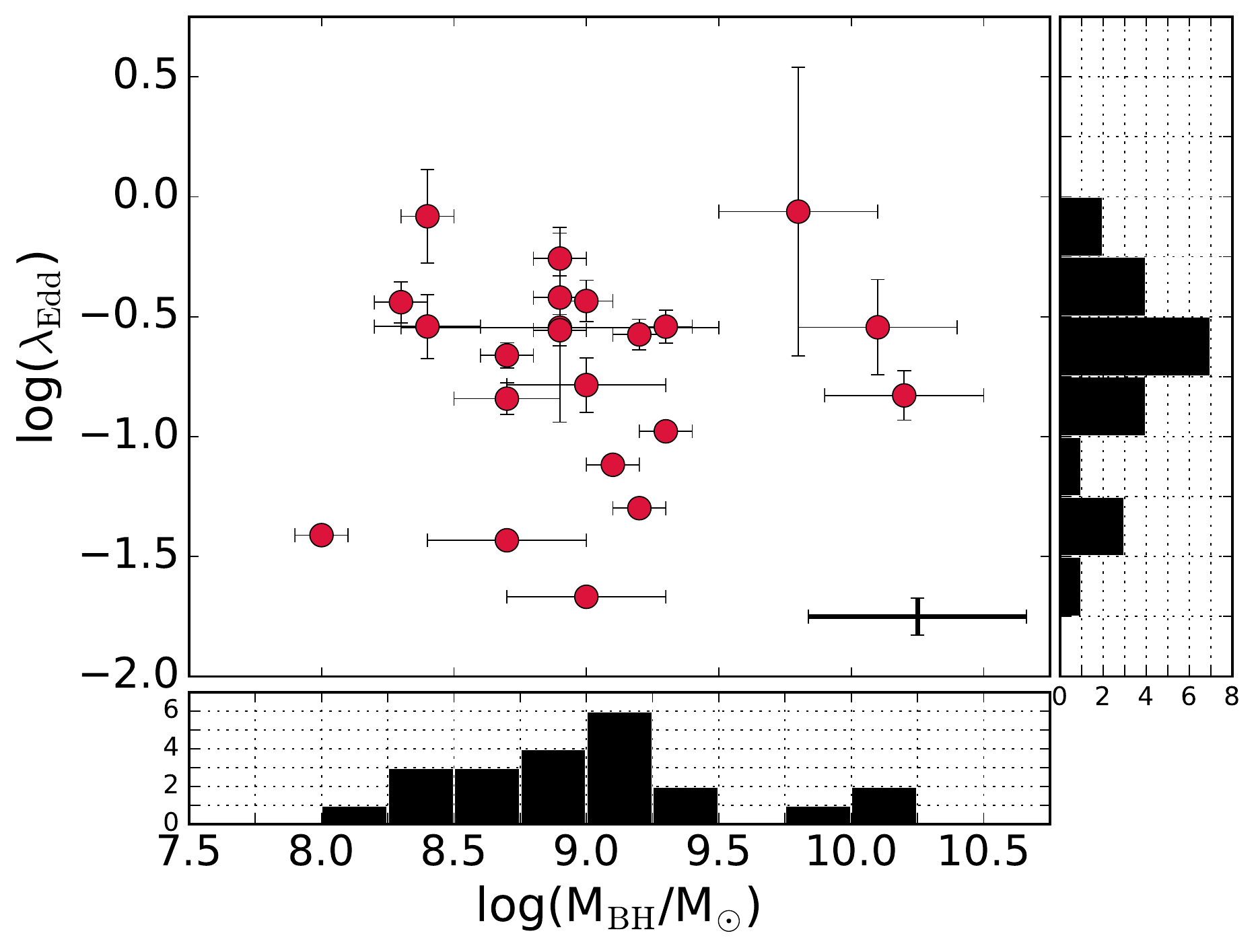}
  \caption{Eddington ratios versus BH masses of the 22 type 1 AGN in the target sample. The Eddington ratio is given by $\lambda_{\textnormal{Edd}} = L_{\textnormal{bol}}/L_{\textnormal{Edd}}$, with bolometric luminosities estimated through SED-fitting analysis. BH masses (given in Table \ref{tab:summary_results}) are derived via the ``virial method'' mainly using the broad \ion{C}{iv} emission line and the calibration of \citet{vest_pet06}. We plot a representative error bar at the bottom-right corner of the plot which takes into account a systematic error of 0.4 dex on $M_{\textnormal{BH}}$. The black histograms show the projected distribution of the two quantities along each axis. SUPER will allow us to sample both accretion rates close to the Eddington limit and more moderate ones and to connect these quantities to the potential outflows detected by SINFONI.}
  \label{fig:Mbh_eddRatio}
\end{figure}

Bolometric luminosities can be also combined with black hole masses, when available, in order to obtain the Eddington ratio $\lambda_{\textnormal{Edd}} = L_{\textnormal{bol}}/L_{\textnormal{Edd}}$, where $L_{\textnormal{Edd}} = 1.5 \times 10^{38} (M_{\textnormal{BH}}/M_{\odot})$ erg s$^{-1}$. BH masses for type 1 AGN, ranging between $8 \times 10^{7}$ and $1.6 \times 10^{10}$ $M_{\odot}$, are reported in Table \ref{tab:input_parameters} together with the respective references. These values are derived via the ``virial method'' mainly using the broad \ion{C}{iv} $\lambda 1549$ emission line and the calibration of \citet{vest_pet06}. Such method is affected by well-known limitations since the \ion{C}{iv} emitting gas could be affected by non-virial motion \citep{trak_netz12}. However, in the present paper we only want to give a broad idea of the coverage in the $\lambda_{\textnormal{Edd}}$-$M_{\textnormal{BH}}$ plane that will be provided by our survey. As shown in Fig. \ref{fig:Mbh_eddRatio}, where we plot the distribution of BH masses and Eddington ratios, we will be able to sample both accretion rates close to the Eddington limit and more moderate ones ($\sim 10^{-2}$ the Eddington limit) and to connect these quantities to the potential outflows that will be detected by SINFONI. To take into account the heavy uncertainties \ion{C}{iv}-based BH mass estimates are affected by, we assume in Fig. \ref{fig:Mbh_eddRatio} a systematic error on $M_{\textnormal{BH}}$ equal to 0.4 dex and plot a median error bar as a reference. Importantly, SINFONI observations will allow us to derive accurate estimates of $M_{\textnormal{BH}}$ combining broad H$\beta$ and H$\alpha$ line profiles with continuum luminosities verifying, and improving upon, the \ion{C}{iv}-based measurements.

\subsubsection{Comparison to the Main Sequence of star-forming galaxies}\label{s:MS_comp}

\begin{figure*}
\centering
  \includegraphics[width=12cm]{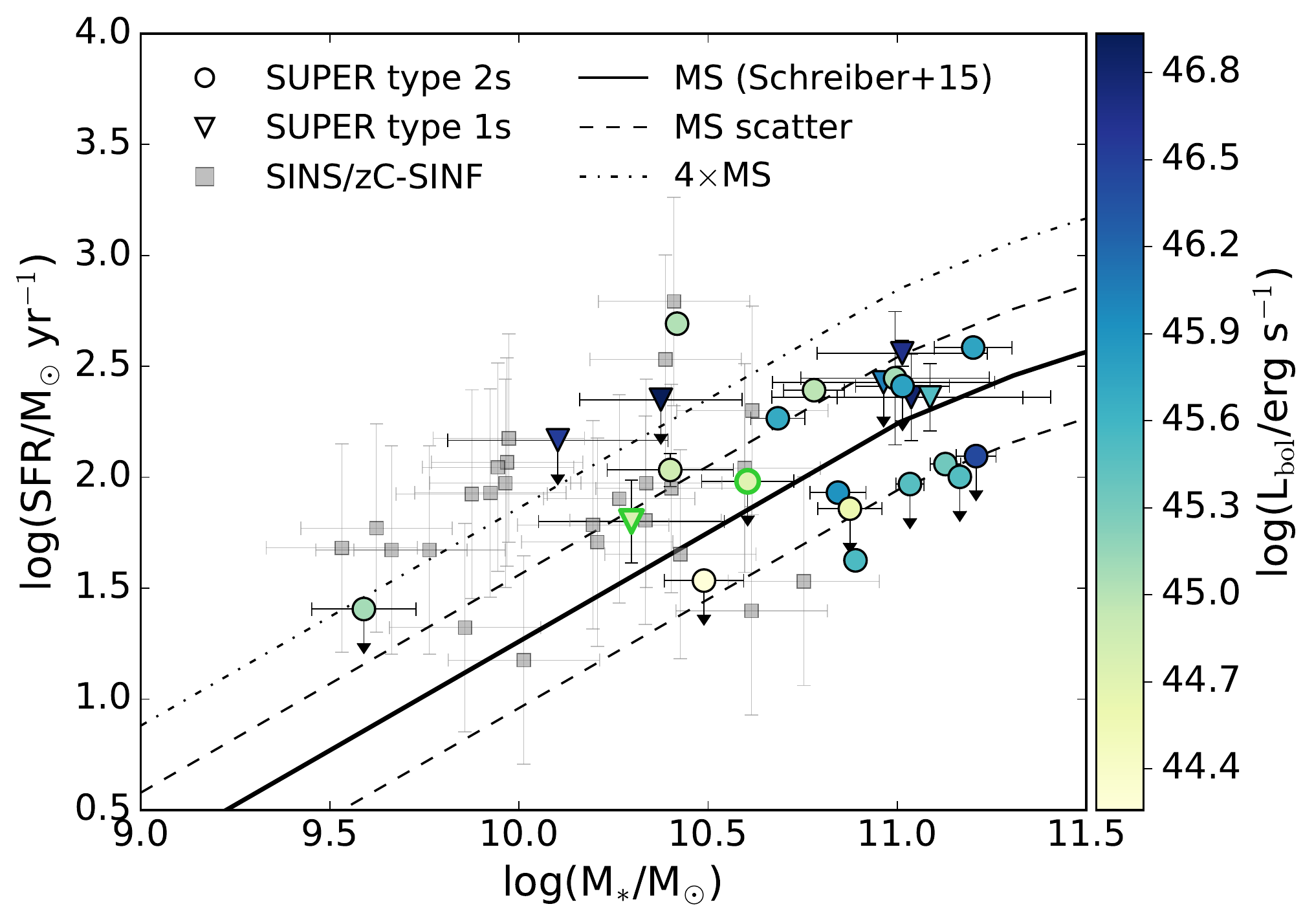}
  \caption{Distribution of host galaxy properties in the SFR-$M_{\ast}$ plane for the 24 AGN (type 1s marked by triangles and type 2s marked by circles) with star formation constraints in our sample as given in Table \ref{tab:summary_results}. The two data points with green edges represent the targets with SFR derived through modeling of the stellar emission with SED fitting. The color coding indicates the AGN bolometric luminosity for each object of this subsample. The black solid line reproduces the main sequence (MS) of star-forming galaxies from \citet{schreiber15} at the average redshift of our target sample (i.e. $\sim 2.3$). The dashed lines mark the scatter of the main sequence (equal to 0.3 dex) while the dot-dashed line represents the locus 4 times above the main sequence along the SFR axis \citep[as defined by][]{rodighiero11}. The gray squares trace the properties of the 25 star-forming galaxies targeted by the SINS/zC-SINF survey \citep{nfs18} without AGN signatures. We note that their selection based on a minimum SFR or H$\alpha$ flux results in a preferentially higher sSFRs than the overall population of normal galaxies at those redshifts \citep[see discussion in][]{nfs09,mancini11}. These galaxies, with IFU data comparable to the SUPER ones, will be our non-AGN comparison sample in future analyses (see text for more details).}
  \label{fig:mainSeq}
\end{figure*}

In Figure\,\ref{fig:mainSeq} we show the location of our targets in the SFR-$M_{\ast}$ plane for the objects with an estimate of both parameters, that is obscured AGN and a subsample of unobscured ones (24 targets, those for which we provide $M_{\ast}$ and SFR in Table \ref{tab:summary_results}). SFRs are already corrected for the AGN contribution. 
The distribution of our targets is compared to the so-called main sequence of star-forming galaxies \citep[e.g.,][]{noeske07}. We adopted the parametrization derived by \citet{schreiber15}, who performed a stacking analysis of deep \textit{Herschel} data in several extragalactic fields (GOODS, UDS, COSMOS), finding a flattening of the MS at high stellar masses ($\log(M_{\ast}/M_{\odot})>10.5$) and a SFR dispersion of 0.3 dex. Our sample covers in a quite uniform way the SFR-$M_{\ast}$ plane, probing a wide range in terms of SFRs. About 46\% of the targets are within the $\pm 1 \sigma$ scatter of the main sequence at the average redshift of the sample $z \sim 2.3$, while the rest are subdivided above (33\%) and below (20\%) it. 
As far as the stellar mass range is concerned, our AGN reside in massive hosts (median $M_{\ast}$ of $10^{10.88}$ $M_{\odot}$). This can be ascribed to a selection effect, as already pointed out by, e.g., \citet{bongiorno12} and \citet{aird12}. In particular, they found that AGN with a low Eddington ratio are more numerous than AGN with a high one. 
At a fixed X-ray flux limit there is a bias toward galaxies hosting an AGN with higher stellar masses, given the relation between $L_{\textnormal{Edd}}$, $M_{\textnormal{BH}}$ and $M_{\ast}$. Over a sample of 1700 AGN in the COSMOS field analyzed by \citet{bongiorno12}, the host galaxy masses range from $10^{10}$ to $10^{11.5}$ M$_{\odot}$, with a peak at $\sim 10^{10.9}$ M$_{\odot}$. 
The color coding in Figure\,\ref{fig:mainSeq} refers to the AGN bolometric luminosity of each target. 
The detailed analysis of potential outflows in our AGN, as a function of their position in the SFR$-M_{\ast}$ plane and their bolometric luminosity, will expand the physical understanding of the impact of AGN outflows on host galaxies by investigating the variation of outflow properties (such as mass outflow rates and energetics) moving from above to below the MS.

Currently the largest AO-assisted NIR IFU observations of galaxies in the same redshift range covered by SUPER is represented by the SINS/zC-SINF survey \citep{nfs18}. These observations focus on the H$\alpha$ and [\ion{N}{ii}] emission lines, probing their distribution and kinematics in the galaxy, with a spatial resolution of $\sim 1.5$ kpc. Excluding objects classified as AGN in \citet{nfs18}, this SINFONI survey includes 25 objects, shown in Figure\,\ref{fig:mainSeq}, in the redshift range $2<z<2.5$ 
of the SUPER sample. The total stellar mass and SFR intervals, used to match the SINS/zC-SINF sample to the SUPER one, span a range which takes into account also the uncertainties on these quantities. As can be seen from Figure\,\ref{fig:mainSeq}, the SUPER and SINS/zC-SINF (excluding AGN) samples have an overlap in this plane, in the stellar mass range $\log (M_{\ast}/M_{\odot}) = [9.5-10.8]$, which will enable an interesting comparison of the properties of galaxies hosting active and inactive SMBHs. 

\subsection{X-ray vs. optical spectroscopic and SED-fitting classification}\label{s:X-ray_SED}

\begin{figure*}
\centering
  \includegraphics[width=13cm]{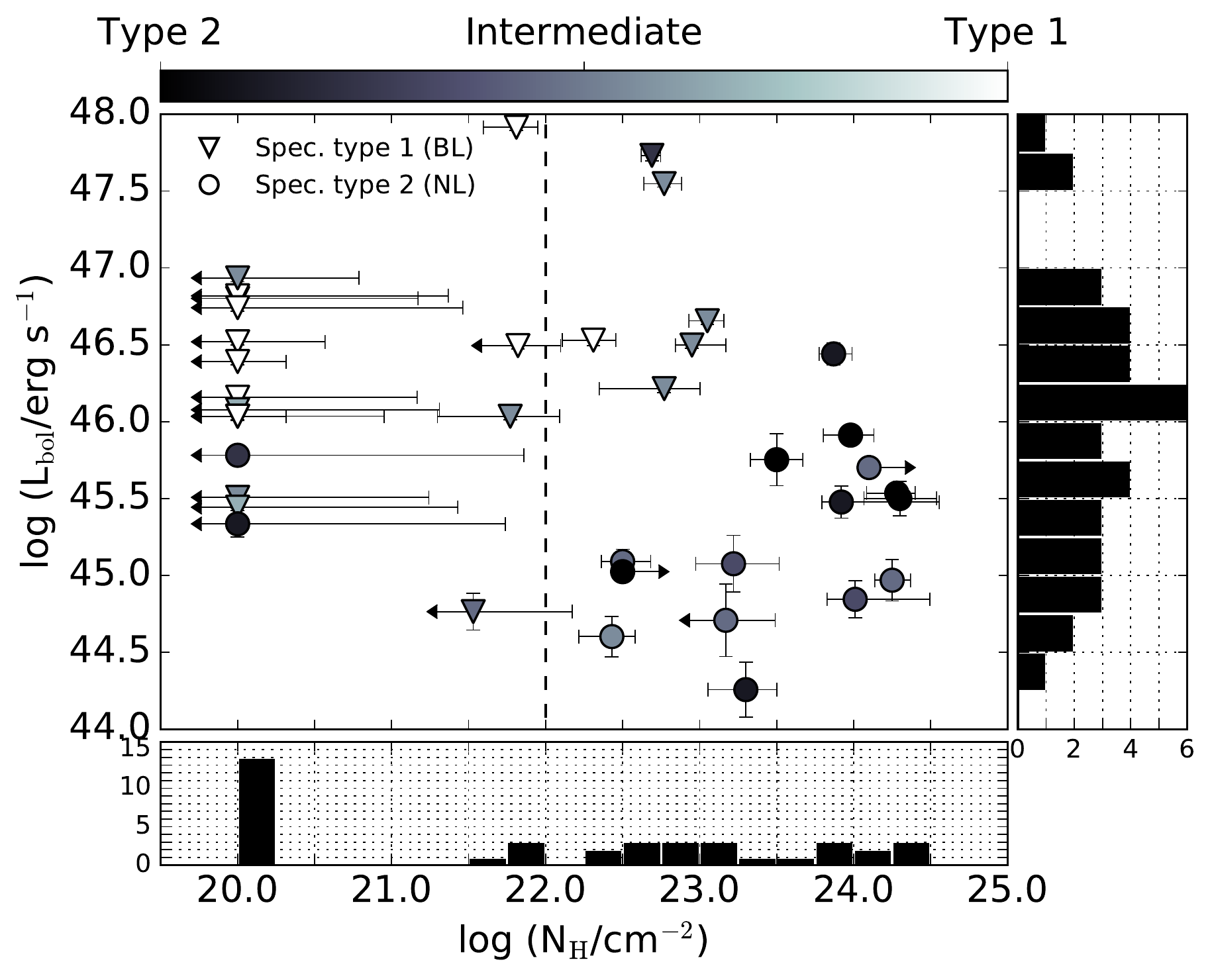}
  \caption{AGN bolometric luminosities versus column densities of the whole target sample. These quantities are derived through SED-fitting modeling and X-ray spectral analysis, respectively. The dashed line at $\log (N_{\textnormal{H}}/ \textnormal{cm}^{-2}) = 22$ marks the assumed separation between X-ray unobscured and obscured AGN. The black histograms show the projected distribution of the two quantities along each axis. The gray color scale depicts the inclination of the observer's line of sight with respect to the dusty torus equatorial plane derived from the SED-fitting analysis, which corresponds to type 2 for dark colors, type 1 for light colors and intermediate (i.e., the transition between the two classes of AGN) in between. The AGN type as derived from the optical spectra is depicted by the different symbols, triangles for type 1s and circles for type 2s. The comparison of the color coding and the different symbols to the location of the targets in the $L_{\textnormal{bol}}-N_{\textnormal{H}}$ plane suggests an agreement between the three classification methods and provides extra confidence in the SED-fitting results. The sample results to be almost equally divided in type 1 and type 2 AGN.}
  \label{fig:lum_columnD}
\end{figure*}

In Fig. \ref{fig:lum_columnD} we plot the distribution of our targets in the AGN bolometric luminosity and column density plane. 
The coverage of this parameter space is quite uniform. The bolometric luminosity probed by our survey ranges from $\sim$10$^{44}$ erg s$^{-1}$ up to $\sim$10$^{48}$ erg s$^{-1}$, spanning almost 4 orders of magnitude. In terms of column density, the sample covers uniformly a range from unobscured ($N_{\textnormal{H}} \le N_{\textnormal{H}}^{\textnormal{gal}}$) to heavily obscured objects, with values up to $2 \times 10^{24}$ cm$^{-2}$. We adopt a separation value of $10^{22}$ cm$^{-2}$ between obscured and unobscured AGN \citep{mainieri02,szokoly04}. In Fig. \ref{fig:lum_columnD} we also compare the X-ray and optical (spectroscopy and SED fitting) diagnostics to distinguish between obscured and unobscured AGN as introduced at the beginning of Sec. \ref{s:overall_results}. The diagnostic recovered from the SED fitting is the 
inclination of the observer's line of sight with respect to the torus equatorial plane, shown by the color coding in Fig. \ref{fig:lum_columnD}. The optical spectroscopic diagnostic (i.e., the presence of broad or narrow lines in the spectra) is depicted with different markers for broad- and narrow-line AGN. 
As can be clearly seen in Fig. \ref{fig:lum_columnD}, the three diagnostics agree rather well. Upper limits refer mainly to objects for which the column density derived from the X-ray spectral analysis is consistent with $\sim 10^{20}$ cm$^{-2}$ (given by Galactic absorption) and are therefore classified as unobscured from an X-ray point of view, even if the formal upper limit for $N_{\textnormal{H}}$ is larger than $10^{22}$ cm$^{-2}$. In some cases, the SED-fitting procedure is affected by significant degeneracies, since the same SED can sometimes be fit by either a type 1 AGN template and a negligible contribution from the host galaxy or an absorbed AGN template together with a very young and UV-bright set of stellar populations. In most cases the results of the SED fitting were in agreement with the overall classification of the target, returning a robust estimate of the AGN type. For some of the bright (type 1) AGN we restricted the range of inclinations based on the obscuration type suggested by the spectroscopic diagnostic in order to overcome the degeneracy. There are also some ambiguous cases. We find 2 targets classified as unobscured in the X-rays but showing obscured characteristics in the optical regime (both in the spectra and in the SEDs), although the upper error for $N_{\textnormal{H}}$ is very large. Six targets show an obscured X-ray spectrum ($10^{22} <N_{\textnormal{H}}< 10^{23}$ cm$^{-2}$) but with broad lines in the optical spectrum and Intermediate/Type 1 characteristics in the SED \citep[see][]{merloni14}. The final classification is performed according to the optical spectroscopic diagnostic, which divides the sample in almost an equal number of type 1 (22) and type 2 AGN (17).

\subsection{Radio regime}\label{s:radio}

All our AGN are located in fields targeted by radio surveys. In particular, the E-CDF-S  has been observed with the Very Large Array (VLA) at 1.4 GHz \citep{miller13}, with a typical rms of 7.4 $\mu$Jy beam$^{-1}$ ($2''.8 \times 1''.6$ beam size). A catalog of optical and IR counterparts for this survey is provided by \citet{bonzini12}. As for the COSMOS field, we took advantage of the deep 3 GHz VLA-COSMOS project \citep{smolcic17}, characterized by an average rms sensitivity of 2.3 $\mu$Jy beam$^{-1}$ and an angular resolution of $0''.75$. The other targets (from XMM-XXL, Stripe 82X and WISSH) are part of the VLA's FIRST survey at 1.4 GHz \citep{becker95}, with a typical 5$\sigma$ sensitivity of 0.15 mJy beam$^{-1}$ and a resolution of $5''$.

\begin{figure*}
\centering
  \includegraphics[width=8.9cm]{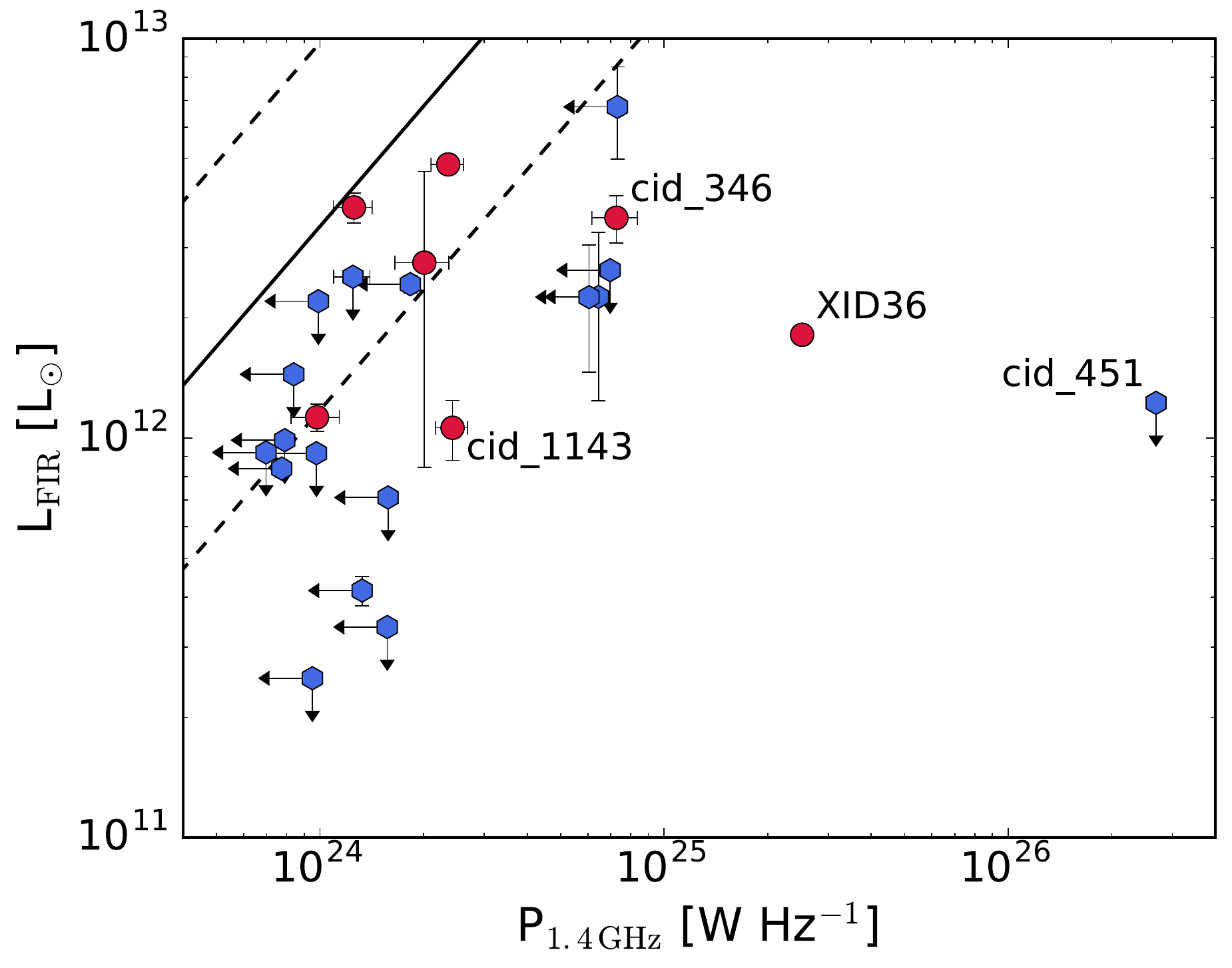}
  \hspace{2mm}
  \includegraphics[width=9.1cm]{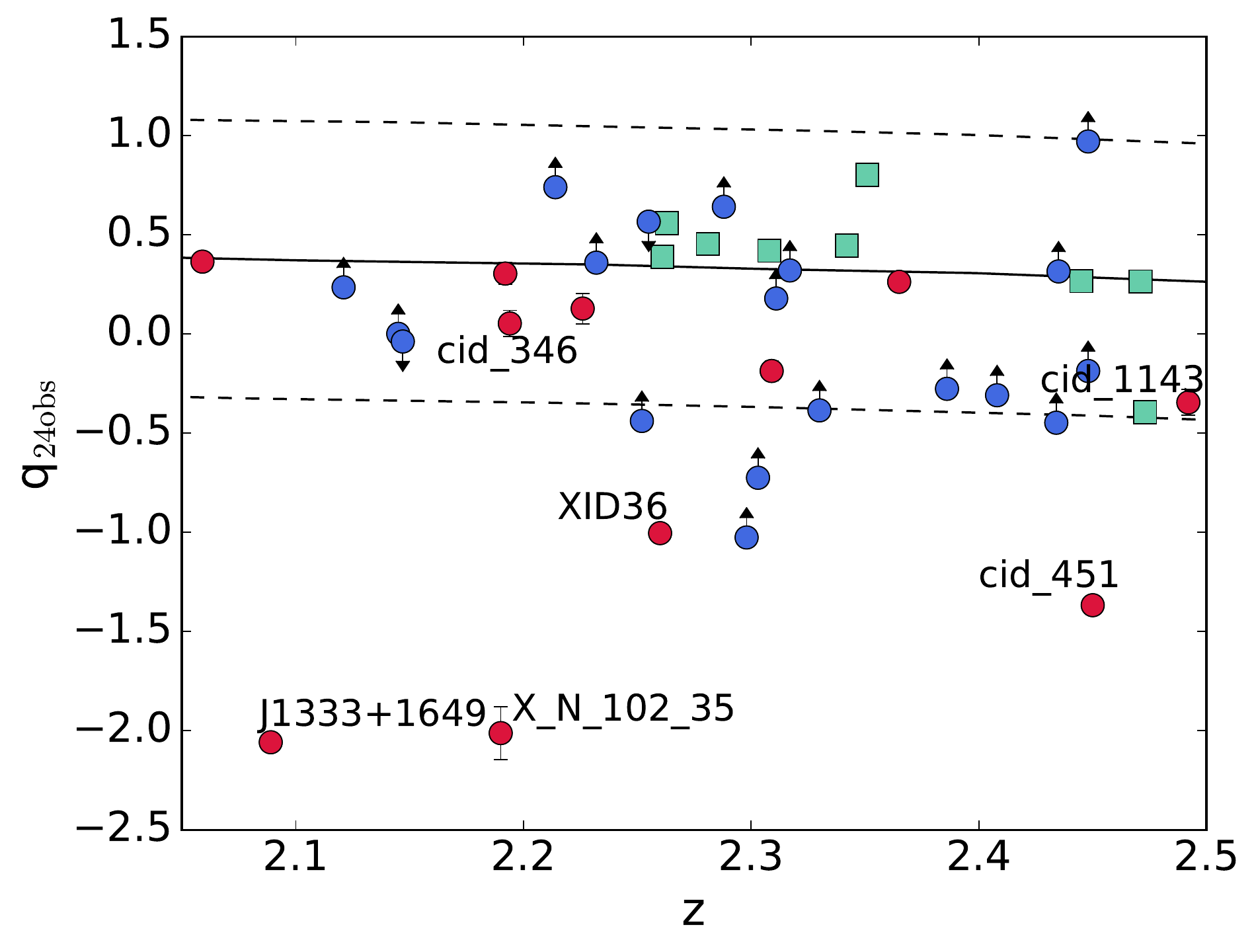}
  \caption{Radio properties of the target sample. \textit{Left}: FIR luminosities due to star formation versus radio power at 1.4 GHz. The solid line shows the \citet{kennicutt98_radio} relation, given by $\log P_{1.4\,GHz} = \log L_{\textnormal{FIR}} + 11.47$, while the dashed lines represent its 2$\sigma$ dispersion. Red circles depict targets with radio detections, while blue hexagons mark targets with upper limits in the radio regime and/or in the FIR. The four AGN outside the 2$\sigma$ dispersion and classified as jetted are marked by their ID. \textit{Right}: $q_{24\,\textnormal{obs}} = \log (S_{24\, \mu \textnormal{m}}/S_{\textnormal{r}})$ plotted as a function of redshift for the whole AGN sample. Red dots mark targets with detections both in the MIR and radio regime; blue-dot upper and lower limits represent sources with detections only in the radio or in the MIR, respectively; green squares depict AGN with upper limits both in the MIR and in the radio. The solid line displays $q_{24\,\textnormal{obs}}$ versus redshift for M82 \citep[from][]{bonzini13}, while dashed lines mark the $\pm 2 \sigma$ dispersion. The six AGN classified as jetted, either from this plot of from the left panel of the figure, are marked by their ID.}
  \label{fig:radio_FIR}
\end{figure*}

We want to study the radio properties of our targets to see, in particular, which ones are
jetted and non-jetted\footnote{We follow \citet{padovani17_comment} and use this new nomenclature, which
supersedes the old ``radio-loud/radio-quiet'' distinction.}. We do this by comparing their FIR and radio luminosities. Namely, when an object lies along the FIR-radio correlation both its radio and FIR emission are supposed to be driven by recent star-formation \citep{yun01}. Instead, if an object is off the correlation its ``radio excess'' is interpreted as evidence for radio emission from strong jets \citep{padovani17_comment}. In Figure \ref{fig:radio_FIR} (\textit{left panel}) we plot these quantities for the 24 targets with detections or upper limits in the FIR regime for which we could derive FIR luminosities through SED-fitting modeling (Section \ref{s:code}). The values are reported in Table\,\ref{tab:summary_results}. We computed the radio power at 1.4 GHz for all sources, converting the 3 GHz flux for the COSMOS targets assuming a radio spectral index $\alpha_{\textnormal{r}} = 0.7$. For the objects without radio detections (blue hexagons in the left panel of Figure \ref{fig:radio_FIR}) we used the 5$\sigma$ sensitivity flux values (0.02, 0.037 and 0.15 mJy beam$^{-1}$ for the COSMOS, CDF-S and XMM-XXL/Stripe82X/WISSH targets respectively) to estimate upper limits for the radio power. The plot includes $\sim 62\%$ of the sample although most of the datapoints are actually radio and/or FIR upper limits. The comparison with the FIR-radio correlation and its 2$\sigma$ dispersion shows the presence of 4 outliers: cid\_451, cid\_346, cid\_1143 and XID36.

Since FIR luminosities are not available for the whole AGN sample, we further explored its radio properties by deriving the so-called \textit{q} parameter, defined as the logarithm of the ratio between IR monochromatic and radio flux densities. The photometric band at the longest wavelength which allows us to use actual detections for most of the sample by keeping the number of upper limits as low as possible, is 24 $\mu$m. We therefore use $q_{24\,\textnormal{obs}} = \log (S_{24\, \mu \textnormal{m}}/S_{\textnormal{r}})$, where $S_{24\, \mu \textnormal{m}}$ is the observed flux density at 24 $\mu$m and $S_{\textnormal{r}}$ is that at 1.4 GHz \citep[see, e.g.,][]{bonzini13}. For the only target which is undetected at 24 $\mu$m and without an upper limit, S82X1940, we used an upper limit of 6 mJy 
given by the WISE All-sky survey 5$\sigma$ sensitivity in the 22 $\mu$m \textit{W4} filter. The distribution of $q_{24\,\textnormal{obs}}$ as a function of redshift is plotted in Figure \ref{fig:radio_FIR} (\textit{right panel}). Red dots mark targets with detections both in the MIR and radio regime; blue-dot upper and lower limits represent sources with detections only in the radio or in the MIR, respectively; green squares depict AGN with upper limits both in the MIR and in the radio, for which the two limits go in opposite directions. As done by \citet{bonzini13}, we compare our results to the $q_{24\,\textnormal{obs}}$ of M82 (as representative of SFGs) and compute $q_{24\,\textnormal{obs}}$ from its SED as a function of redshift \citep[for more details see Section 3.1.1 and Figure 2 in][]{bonzini13}.  
The SFG locus is defined as the region of $\pm 2 \sigma$ around the M82 template (dashed lines in the plot) and sources below this region show a radio excess. We are fully aware of the fact that, using the 24 $\mu$m flux density (which corresponds to $\lambda \sim 7.3$ $\mu$m rest-frame at the average $z$ of the sample), we are actually probing a wavelength regime where the AGN can dominate the total energy budget. 
To have an estimate of the increase in $q_{24\,\textnormal{obs}}$ the AGN emission may produce, we evaluate the average AGN contribution to the total 24 $\mu$m flux from the SEDs where we can model all the emission components (i.e., 24 targets, 62\% of the sample). The median value of this fraction is $\sim 86\%$ which, when subtracted from $q_{24\,\textnormal{obs}}$, would produce a down shift of the data points by $\sim 0.8$ dex. In the right panel of Figure \ref{fig:radio_FIR} the datapoints are already downshifted by such value. After accounting for this correction we find that four of our targets display a clear radio excess: the COSMOS target cid\_451, whose jetted nature is confirmed also from the FIR-radio comparison; the CDF-S target XID36, which was inside the 2$\sigma$ area before the correction but an outlier in the FIR-radio plane; the targets J1333$+$1649 and X\_N\_102\_35, not plotted in the left panel of Figure \ref{fig:radio_FIR} because they lack an FIR detection. The targets cid\_346 and cid\_1143 are still within the dispersion after the correction, but classified as jetted according to the position in the FIR-radio plane. Combining the results of the two panels in Fig. \ref{fig:radio_FIR}, we estimate a jetted AGN fraction of $10-15\%$, which is consistent with the typical values observed in X-ray selected samples. 

\section{Summary and future work}\label{s:Conclusions}

We have presented the sample targeted by SUPER, an on-going ESO's VLT/SINFONI Large Programme assisted by AO facilities, designed to map the ionized gas kinematics down to $\sim 2$ kpc spatial resolution in a representative sample of 39 AGN at $2<z<2.5$. It will provide a systematic investigation of AGN ionized outflows and their effects on star formation in the host galaxies, by exploring a wide range in AGN and host galaxies properties. The sample was selected in an unbiased way with respect to the chance of detecting outflows, with the aim to cover the widest possible range in AGN properties. 
In this first paper we fully characterized the physical properties of the AGN sample, drawn from X-ray surveys (i.e., CDF-S, COSMOS, XMM-XXL, Stripe82X, WISSH) which benefit from a wealth of multi-wavelength data, from the radio to the X-rays as follows:
\begin{itemize}
\item By collecting UV to FIR photometric data we built up the AGN SEDs and performed a detailed SED-fitting modeling which allowed us to derive stellar masses, $\log (M_{\ast}/M_{\odot}) = [9.59-11.21]$, $\textnormal{SFR} = [25 - 680]$ $M_{\odot}$ yr$^{-1}$, and AGN bolometric luminosities, $\log (L_{\textnormal{bol}}/\textnormal{erg}\,\textnormal{s}^{-1}) = [44.3-47.9]$. \\
\item A detailed X-ray spectral fitting was performed to determine column densities $N_{\textnormal{H}}$ up to $2\times10^{24}$ cm$^{-2}$ and X-ray $2-10$ keV luminosities, $\log (L_{\textnormal{X}}/\textnormal{erg}\,\textnormal{s}^{-1})=[43.2-45.8]$. \\
\item For AGN characterized by broad lines in their optical spectra we reported BH masses obtained using the ``virial method'' on the \ion{C}{iv} and H$\beta$ lines, with results in the range $\log (M_{\textnormal{BH}}/M_{\odot}) = [7.9-10.2]$. These values were combined with the bolometric luminosity to compute Eddington ratios for this subsample of AGN which includes BHs accreting at the Eddington limit and down to $10^{-2}$ times $\lambda_{\textnormal{Edd}}$. \\
\item Finally, we retrieved the radio fluxes (or upper limits) for each target and, by comparing their FIR luminosities (when available) or their 24 $\mu$m fluxes, we inferred the presence of at least 6 jetted AGN in our sample. \\
\end{itemize}
As clear from the wide parameter ranges given above, our survey probes a representative sample of AGN, in terms of both host galaxy properties, such as stellar mass and SFR, and AGN ones, like column density, AGN bolometric luminosity, BH mass and Eddington ratio. This will give us the context to place our IFS studies and the opportunity to investigate possible links among all these quantities and connect them to the outflow properties.

To achieve one of the main goals of the survey, namely inferring the impact that outflows may have on the ability of the host galaxy to form stars, we need to quantify their gas content. Molecular gas represents indeed the principal fuel for star formation in galaxies and the fundamental link between SF and AGN activity. In AGN host galaxies, molecular gas fractions ($f_{\textnormal{gas}} = M_{\textnormal{gas}}/M_{\ast}$) and depletion timescales ($t_{\textnormal{dep}} = M_{\textnormal{gas}}/\textnormal{SFR}$) appear to be smaller than the values measured for the parent population of SF galaxies \citep[e.g.,][but see also \citealt{husemann17CO,rosario18}]{brusa17,kakkad17}. The interpretation of these quantities is non-trivial and requires a joint characterization of the cold molecular and ionized gas phase. SUPER will achieve this goal by combining the SINFONI observations with two on-going programs with ALMA and APEX:
\begin{itemize}
\item SUPER-ALMA (PI: Mainieri; Circosta et al., in prep.), which has been allocated 12.6 and 19.5 hours of ALMA Band-3 observing time in Cycle 4 and 5 respectively, to target the CO(3-2) emission line with $1''$ angular resolution over a sample constructed to include the SUPER sources. This project will perform a systematic study of the gas content of AGN hosts, in order to derive gas fractions and depletion timescales, but will also complement the goals of our SINFONI survey. In fact, information about the possible presence of outflows in these targets will be available and will allow us to infer whether there is a causal connection between a lower gas fraction and the presence of an AGN-driven outflow. \\
\item SUPER-APEX (PI: Cicone), a pilot project with the APEX PI230 Rx receiver that was allocated 28.2 hours to observe, in two of our targets, the [\ion{C}{i}](2-1) transition as a tracer of the total amount of cold $H_{2}$ and the CO(7-6) transition, which will trace the warmer and denser phase of $H_{2}$.\\
\end{itemize}

Recent studies showed that a significant fraction of the mass and momentum of AGN-driven outflows can be contained in the molecular gas phase \citep[e.g.,][]{cicone14,carniani15,fiore17}. To obtain a comprehensive picture of the feedback processes, it is crucial to investigate the molecular gas properties. The next step will be to map the molecular gas, tracing the fuel for star formation and feedback, with the same $\sim$kpc spatial resolution of the ionized gas \citep[e.g.,][]{cicone18}. A comprehensive and dynamic view of the evolution of the star formation process and the impact of AGN feedback across the host galaxy, at the peak epoch of galaxy assembly, will be finally possible. This will have far reaching implications on theoretical models and simulations of galaxy-AGN evolution. 
The final goal of the SUPER project is to be a reference legacy survey for future work and to establish a unique statistical sample at high redshift characterized by a wide set of ancillary data. The systematic approach adopted will reveal key clues about outflow physics and feedback in AGN host galaxies.

\begin{acknowledgements}
We thank the anonymous referee for carefully reading the paper and providing comments. C. Circosta also thanks: D. Burgarella and L. Ciesla for helpful advice in using Cigale; I. Baronchelli, G. Calistro Rivera, A. Feltre, E. Hatziminaoglou and D. Rosario for useful discussions about SED fitting; E. Le Floc'h for providing MIPS fluxes for the targets cid\_971 and lid\_206;  Y.-Y. Chang and I. Delvecchio for providing their SED-fitting results, and F. Duras for the WISSH photometric catalog; E. Lusso for providing the data of her sample plotted in Fig. 3 and P. Lang, D. Liu and J. Scholtz for the ALMA data of COSMOS and CDF-S; A. Zanella for helpful discussions and support while the paper was written. C. Circosta acknowledges support from the IMPRS on Astrophysics at the LMU (Munich). CF and CC acknowledge support from the European Union Horizon 2020 research and innovation programme under the Marie Sklodowska-Curie grant agreement No 664931. BM acknowledges support by the Collaborative Research Centre 956, sub-project A1, funded by the Deutsche Forschungsgemeinschaft (DFG). GC acknowledges the support by INAF/Frontiera through the ``Progetti Premiali'' funding scheme of the Italian Ministry of Education, University, and Research. GC has been supported by the INAF PRIN-SKA 2017 programme 1.05.01.88.04. This research project was supported by the DFG Cluster of Excellence ``Origin and Structure of the Universe'' (www.universe-cluster.de). 

This research has made use of the following data: data based on data products from observations made with ESO Telescopes at the La Silla Paranal Observatory under ESO programme ID 179.A-2005 and on data products produced by TERAPIX and the Cambridge Astronomy Survey Unit on behalf of the UltraVISTA consortium. Data from HerMES project (http://hermes.sussex.ac.uk/). HerMES is a Herschel Key Programme utilising Guaranteed Time from the SPIRE instrument team, ESAC scientists and a mission scientist. The HerMES data was accessed through the Herschel Database in Marseille (HeDaM - http://hedam.lam.fr) operated by CeSAM and hosted by the Laboratoire d'Astrophysique de Marseille. HerMES DR3 was made possible through support of the Herschel Extragalactic Legacy Project, HELP (http://herschel.sussex.ac.uk), HELP is a European Commission Research Executive Agency funded project under the SP1-Cooperation, Collaborative project, Small or medium-scale focused research project, FP7-SPACE-2013-1 scheme.
\end{acknowledgements}


\bibliographystyle{aa}
\bibliography{biblio}

\onecolumn
\begin{appendix}

\section{Description of the photometric catalog}\label{sec:photo_cat}

\begin{table*}[h!]
\footnotesize
\caption{\label{tab:photo_summary} Column description of the photometric catalog. Fluxes and errors are given in units of mJy. All fluxes are corrected for Galactic extinction. Refer to Table \ref{tab:dataset} for sets of filters used in specific fields compared to labels and descriptions given here.} 
\centering
\begin{tabular}{ccc}
\hline\hline
Column number & Label & Description \\ 
\hline
1 & ID & ID of each target as given in Table \ref{tab:sample} \\
2 & z & Redshift \\
3, 4 & NUV\_galex, NUV\_galex\_err & GALEX NUV flux and error \\
5, 6 & U\_ctio, U\_ctio\_err & CTIO-Blanco/Mosaic-II \textit{U}-band flux and error \\
7, 8 & U\_vimos, U\_vimos\_err & VLT/VIMOS \textit{U}-band flux and error \\
9, 10 & u\_megacam, u\_megacam\_err & CFHT/MegaCam \textit{u}-band flux and error \\
11-20 & u\_sloan, ..., z\_sloan\_err & SDSS fluxes and errors \\
21-30 & B\_subaru, ..., z\_subaru\_err & Subaru/Suprime-Cam fluxes and errors \\
31-42 & WFI\_U, ..., WFI\_I\_err & ESO-MPG/WFI fluxes and errors \\
43-52 & acs\_f435w, ..., acs\_f850lp\_err & HST/ACS fluxes and errors \\
53-60 & wfc3\_098M, ..., wfc3\_H160W\_err & HST/WFC3 fluxes and errors \\
61-72 & u\_cfhtl, ..., z\_cfhtl\_err & CFHT fluxes and errors \\ 
73-76 & J\_ctio, ..., Ks\_ctio\_err & CTIO-Blanco/ISPI fluxes and errors\\
77, 78 & z\_ctio, z\_ctio\_err & CTIO-Blanco/Mosaic-II \textit{z}-band flux and error \\
79-88 & VISTA\_Z, ..., VISTA\_Ks\_err & VISTA fluxes and errors \\
89-94 & J\_ukidss, ..., K\_ukidss\_err & UKIDSS fluxes and errors \\
95, 96 & H\_sofi, H\_sofi\_err & NTT/SofI \textit{H}-band flux and error \\
97, 98 & isaac\_Ks, isaac\_Ks\_err & VLT/ISAAC \textit{Ks}-band flux and error \\
99, 100 & HAWKI\_Ks, HAWKI\_Ks\_err & VLT/HAWK-I \textit{Ks}-band flux and error \\
101-108 & Y\_uv, ..., K\_uv\_err & VISTA/VIRCAM fluxes and errors \\
109, 110 & Y\_hsc, Y\_hsc\_err & Subaru/HSC \textit{Y}-band flux and error \\
111-114 & H\_w, ..., K\_w\_err & CFHT/WIRCam fluxes and errors \\
115-120 & J\_2mass, ..., Ks\_2mass\_err & 2MASS fluxes and errors \\
121-128 & irac\_ch1, ..., irac\_ch4\_err & \textit{Spitzer}/IRAC fluxes and errors \\
129-136 & W1, ..., W4\_err & WISE fluxes and errors \\
137-138 & mips24, mips24\_err & \textit{Spitzer}/MIPS 24 $\mu$m flux and error \\
139-144 & pacs70, ..., pacs160\_err & \textit{Herschel}/PACS fluxes and errors \\
145-150 & spire250, ..., spire500\_err & \textit{Herschel}/SPIRE fluxes and errors \\
151-154 & ALMA\_band7, ..., ALMA\_band3\_err & ALMA Band 7 and 3 fluxes and errors \\
\hline
\end{tabular}
\end{table*}
\newpage

\section{Spectral energy distributions of the SUPER sample}\label{sec:SEDs}

\begin{figure*}[h!]
\centering
  	\includegraphics[width=7.8cm]{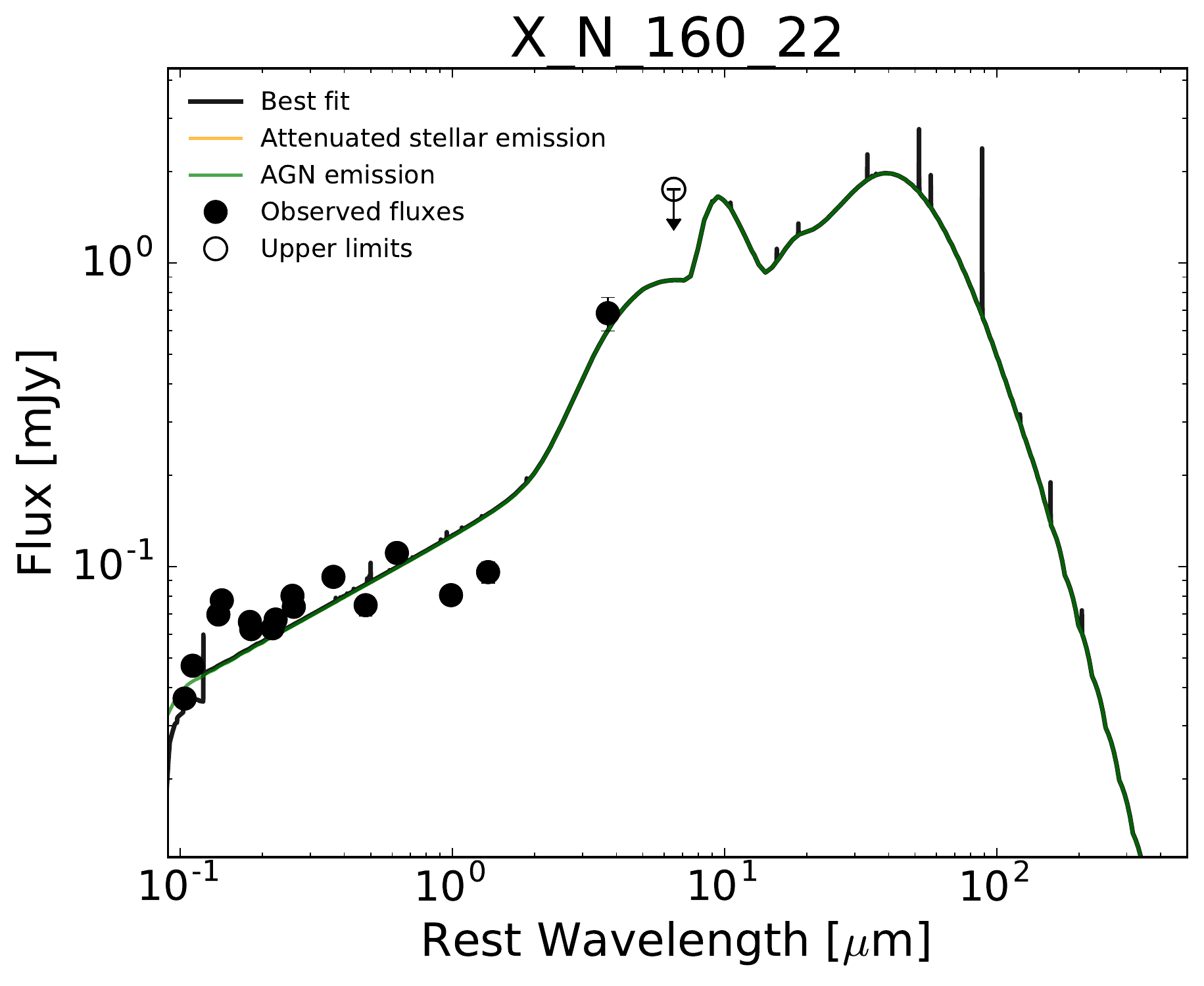} 
  	\hspace{2mm}
  	\includegraphics[width=7.8cm]{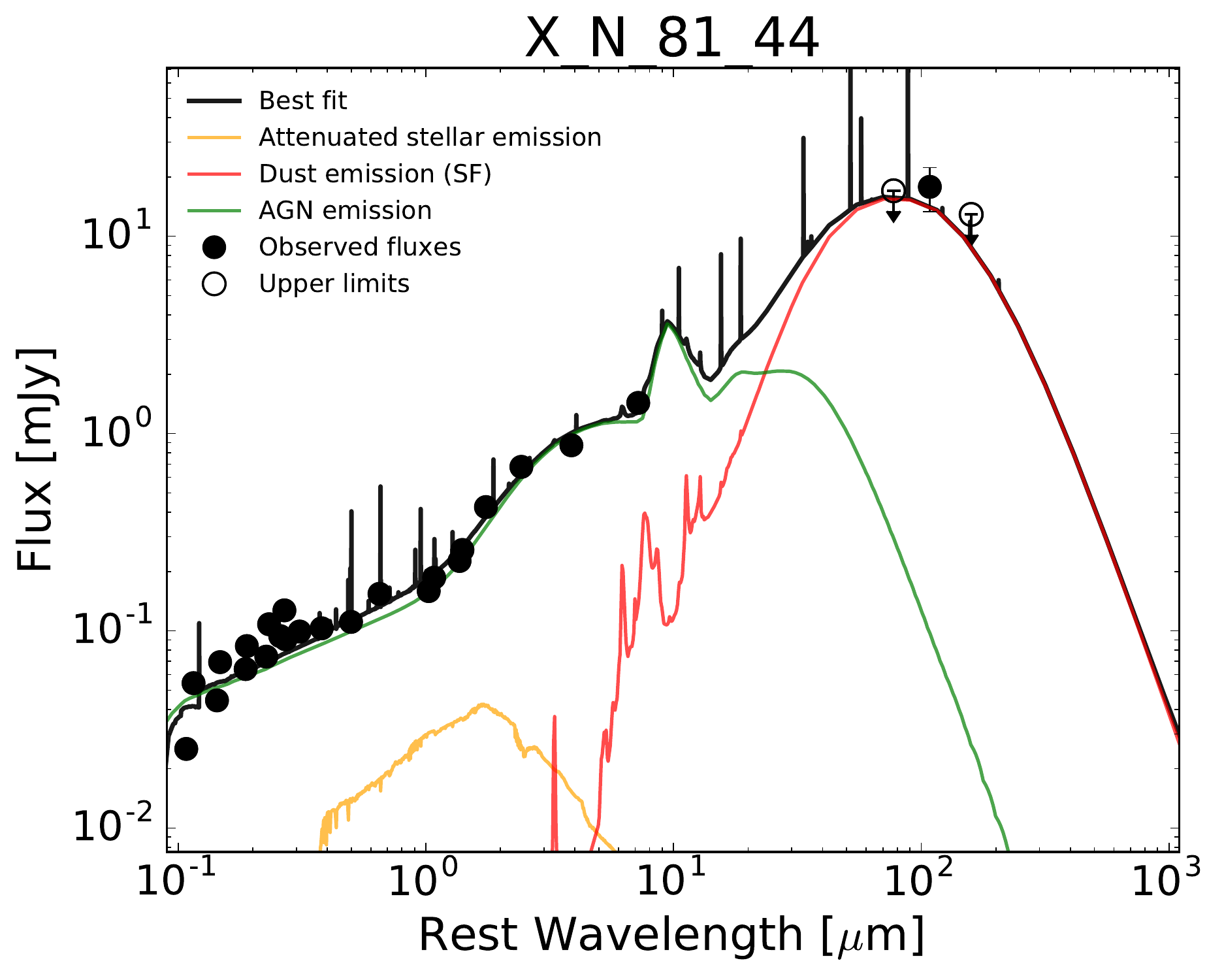} 
  	\hspace{2mm}
  	\includegraphics[width=7.8cm]{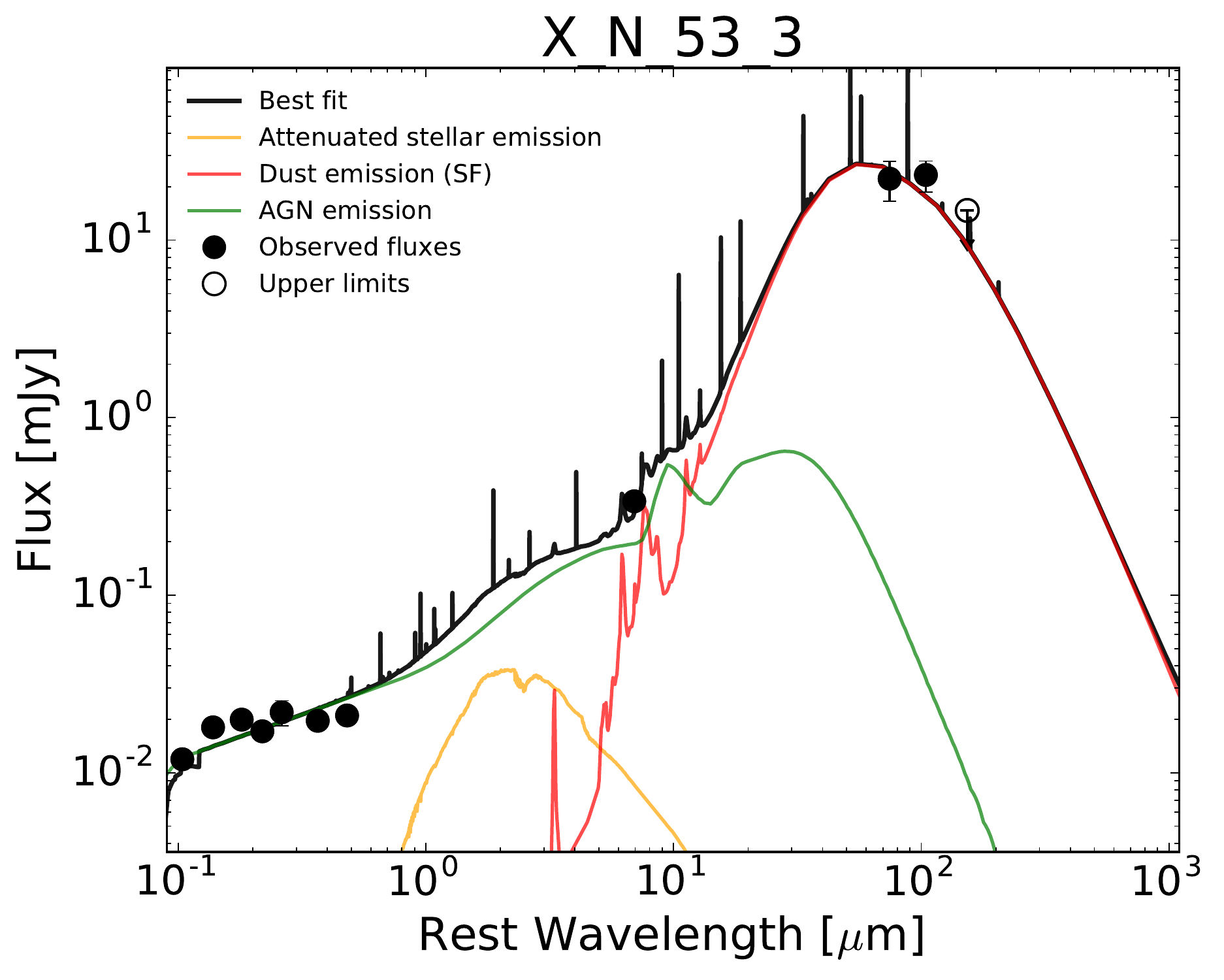} 
 	\hspace{2mm}
 	\includegraphics[width=7.8cm]{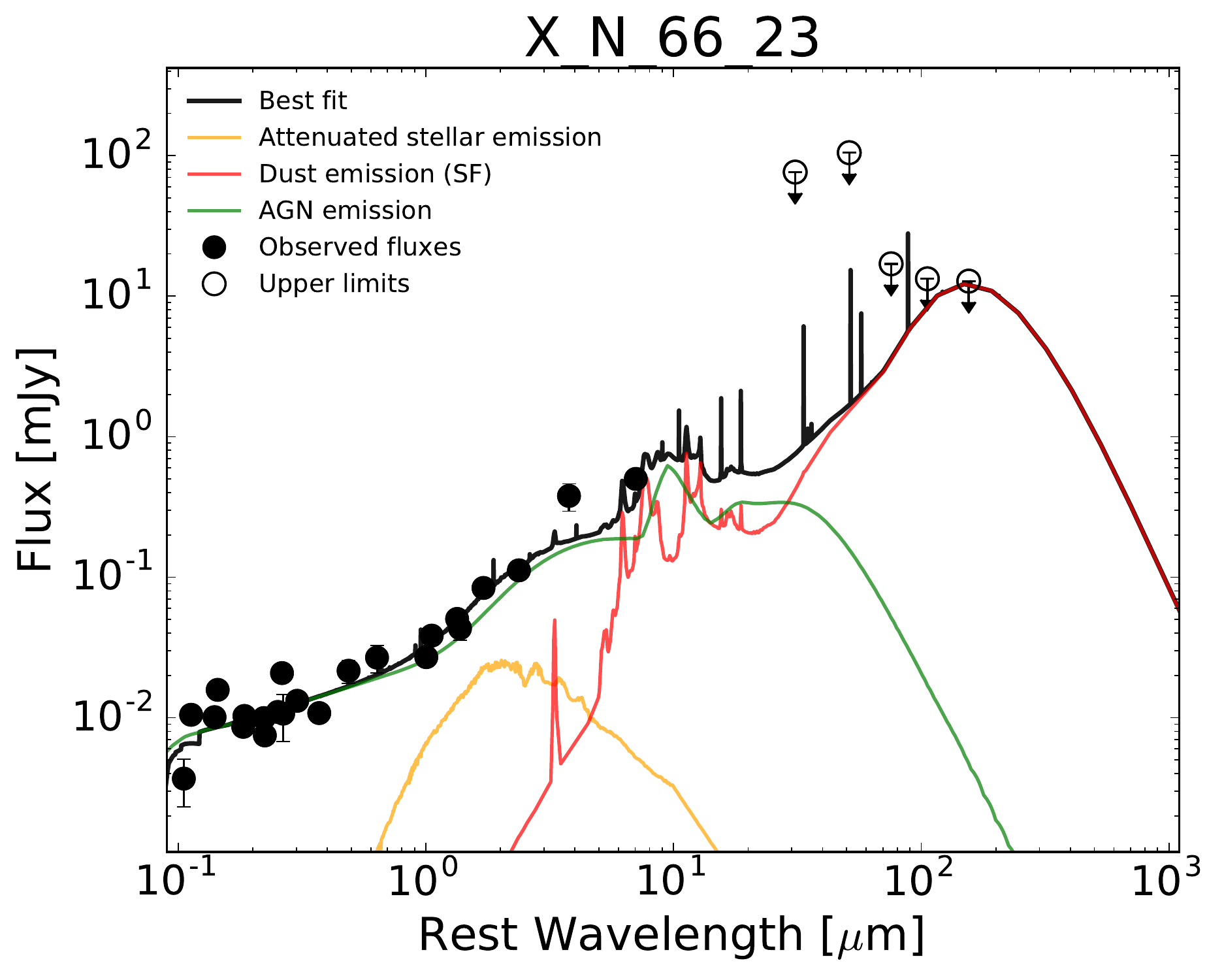} 
  	\hspace{2mm}
 	\includegraphics[width=7.8cm]{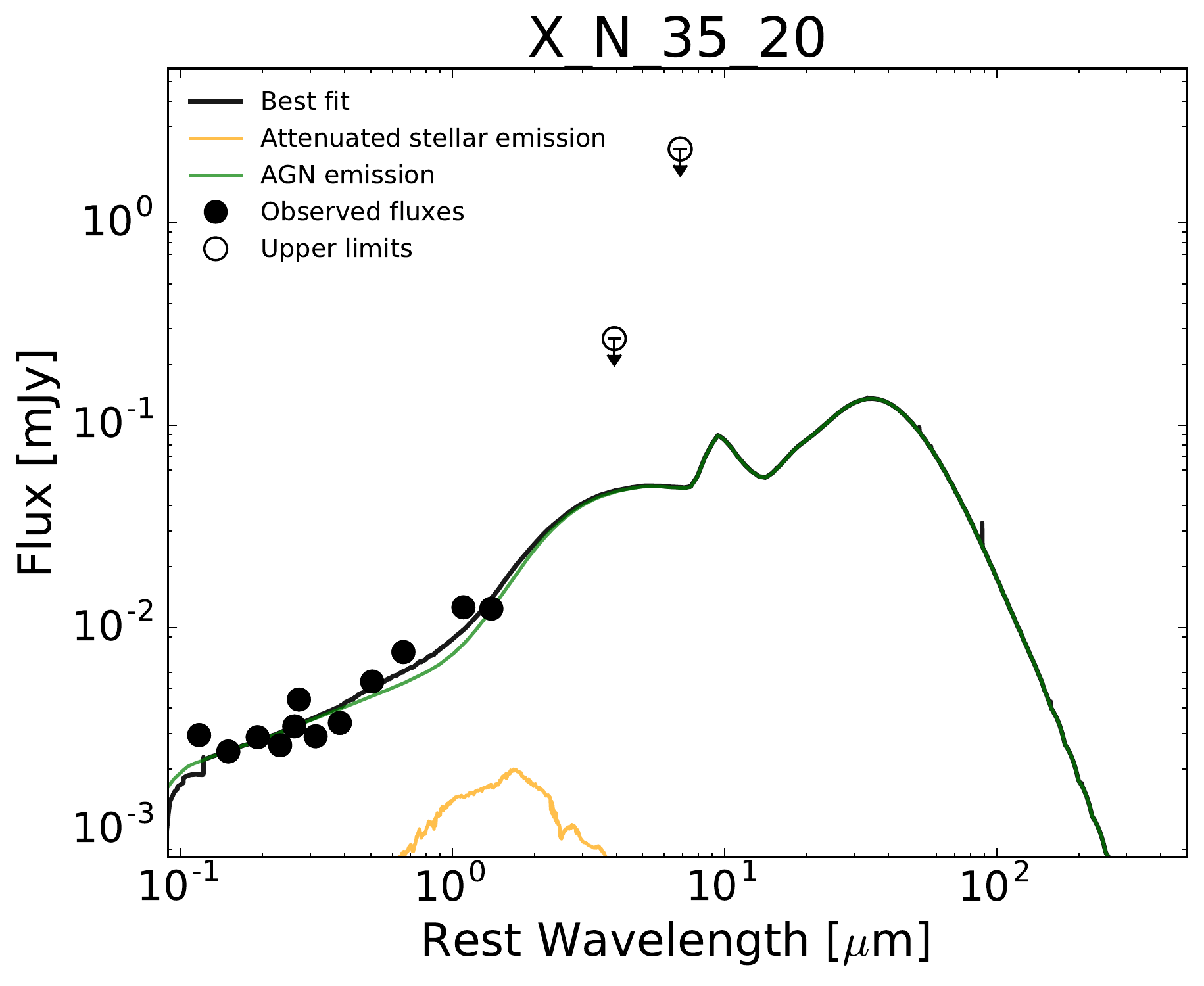} 
 	\hspace{2mm}
  	\includegraphics[width=7.8cm]{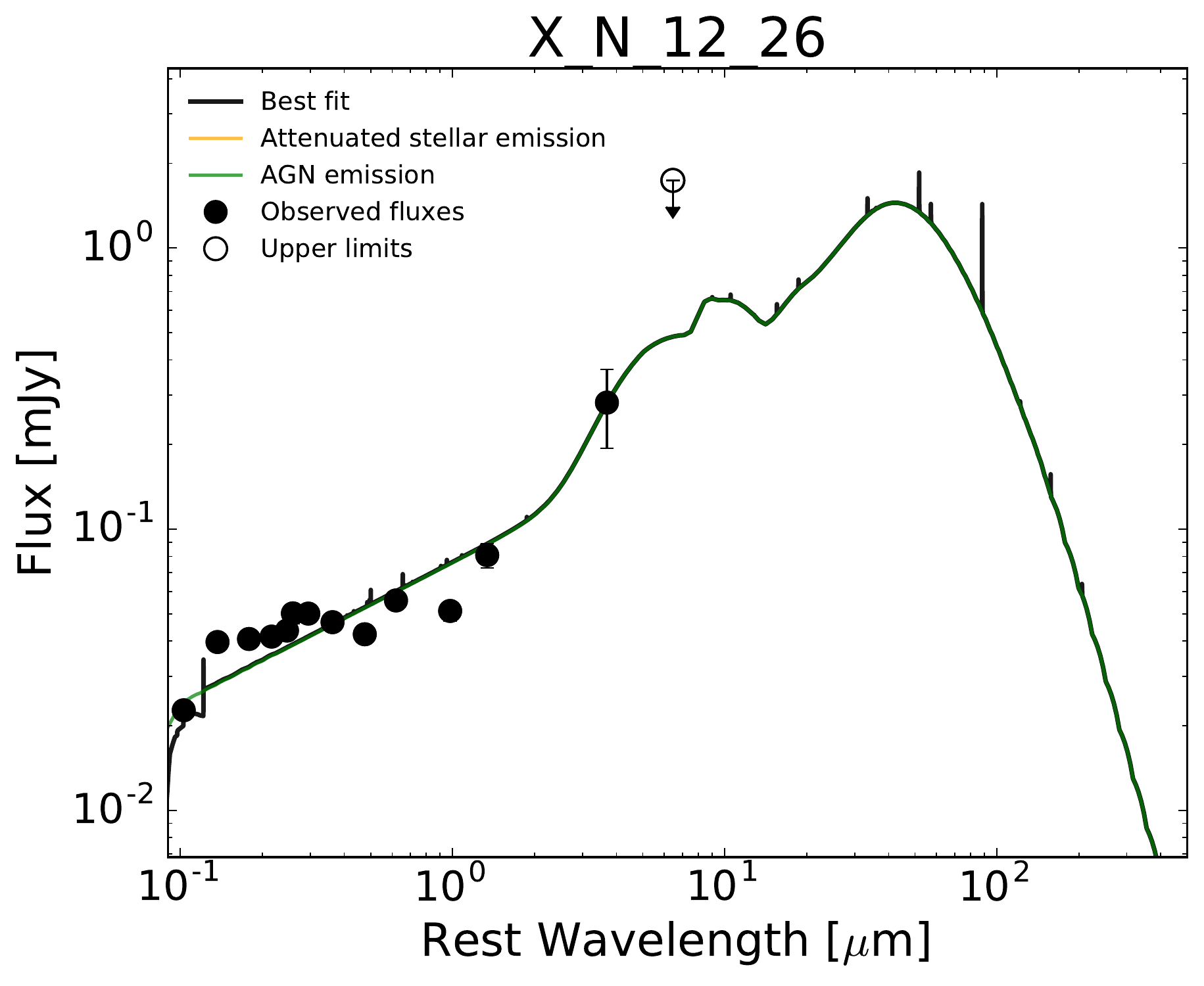}
    \caption{Rest-frame SEDs of the whole SUPER sample. The black dots represent the observed multi-wavelength photometry, while the empty dots indicate 3$\sigma$ upper limits. The black solid line is the total best-fit model, the orange curve represents the stellar emission attenuated by dust, the green template reproduces the AGN emission, the red curve accounts for dust emission heated by star formation. Emission lines in the black curves are part of the nebular emission component, included in the overall SED.}
\end{figure*}

\begin{figure*}
	\centering
  	\includegraphics[width=7.9cm]{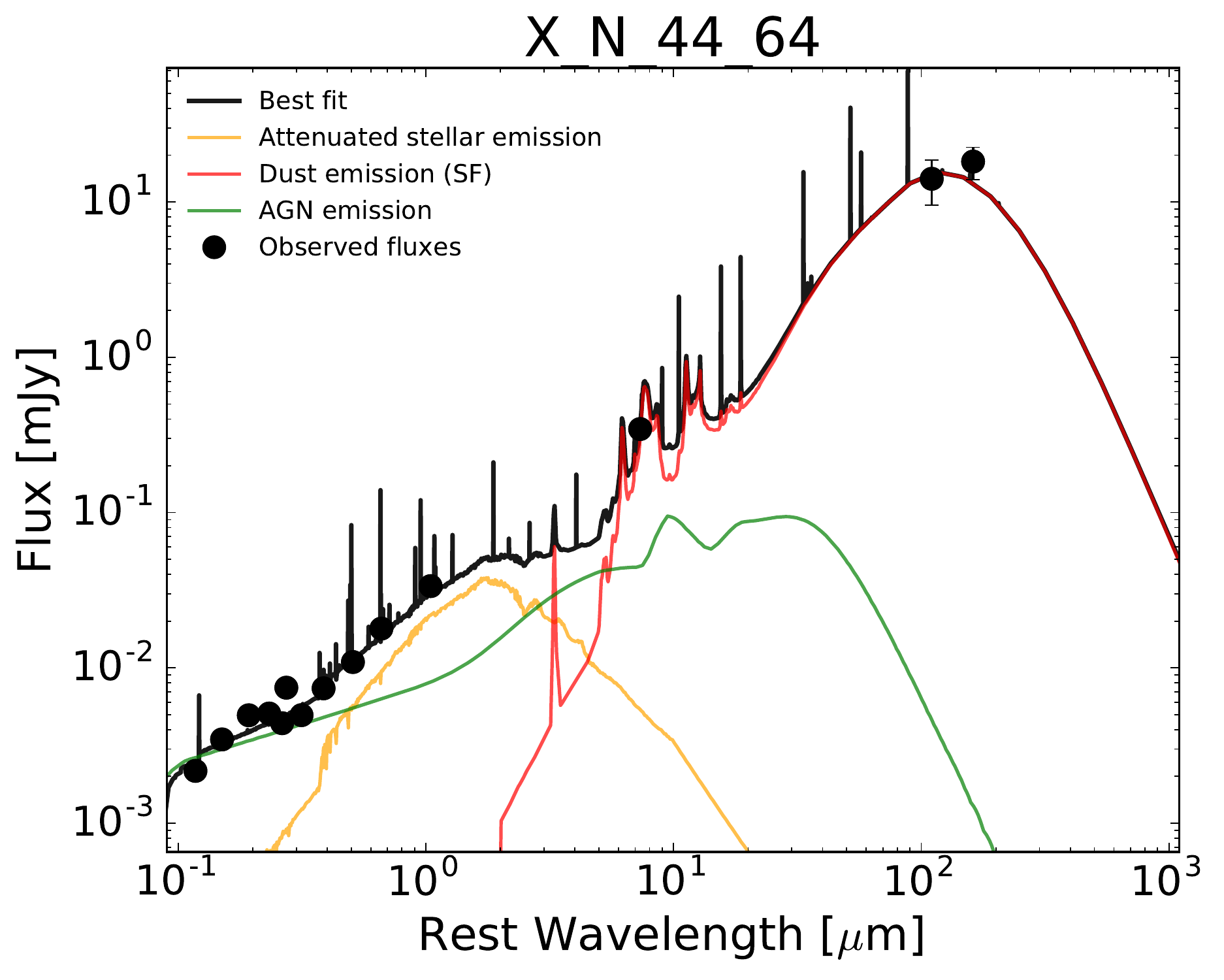} 
  	\hspace{2mm}
  	\includegraphics[width=7.8cm]{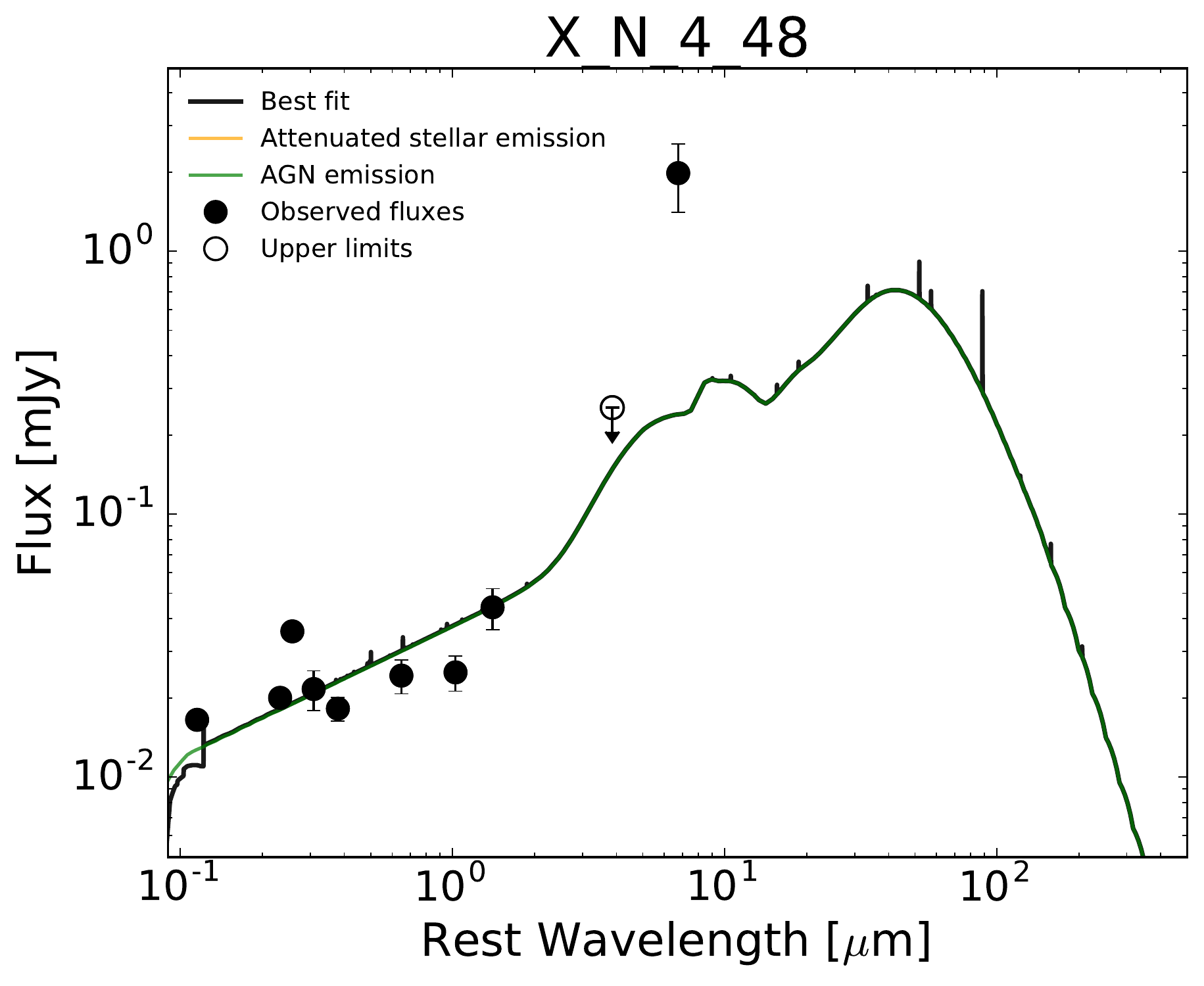} 
  	\hspace{2mm}
  	\includegraphics[width=7.8cm]{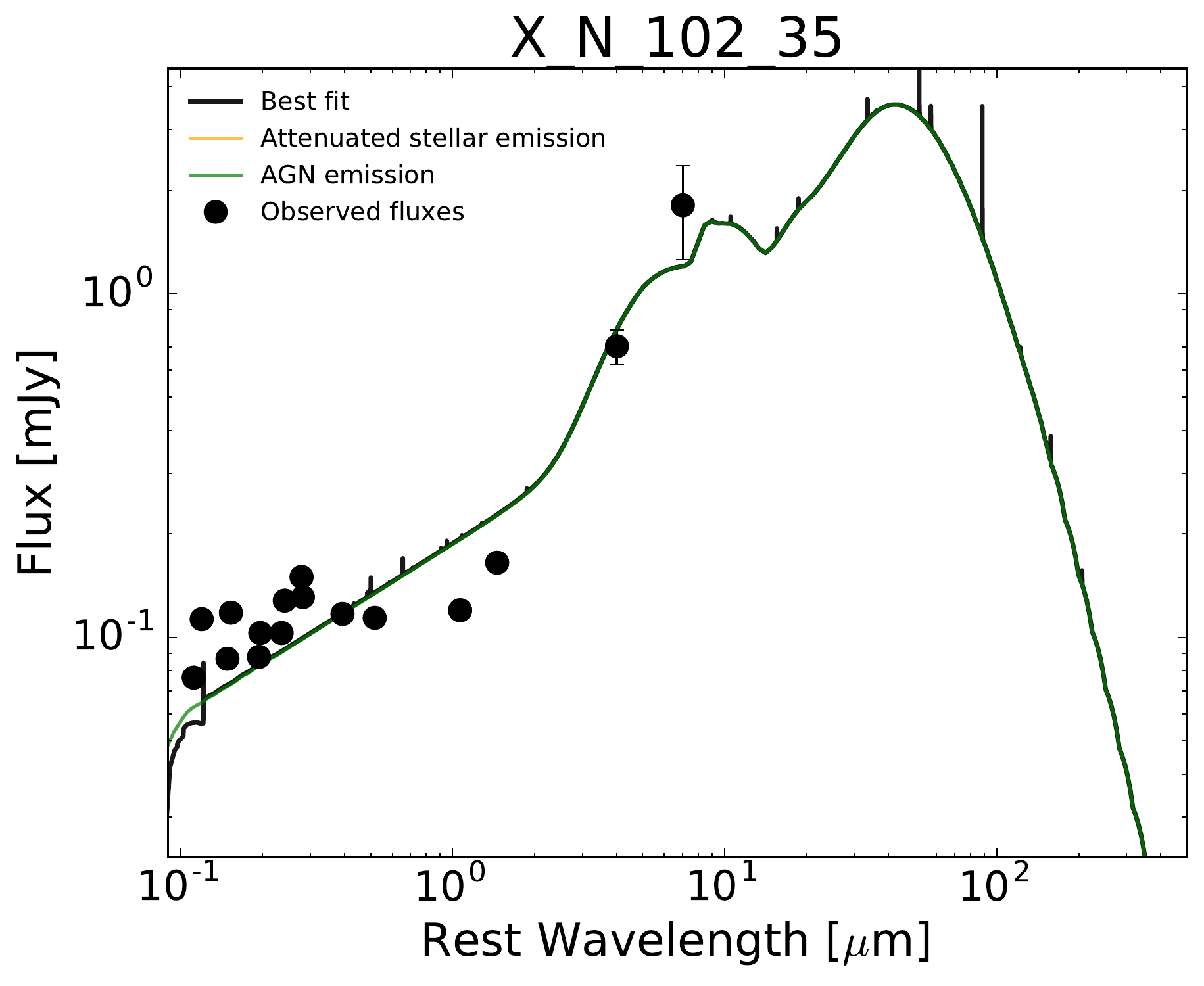}
  	\hspace{2mm}
  	\includegraphics[width=7.8cm]{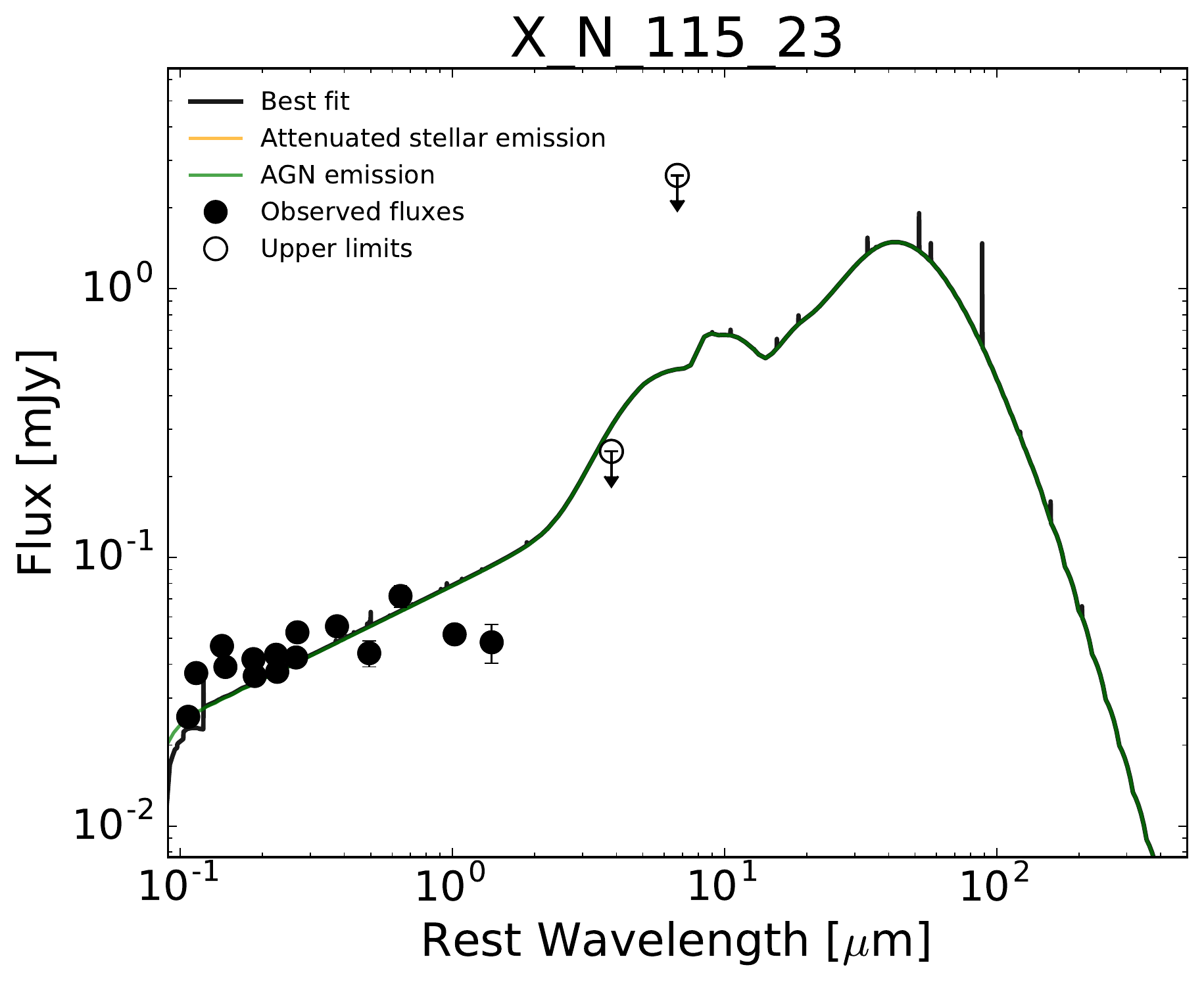}
    \hspace{2mm}
 	\includegraphics[width=7.8cm]{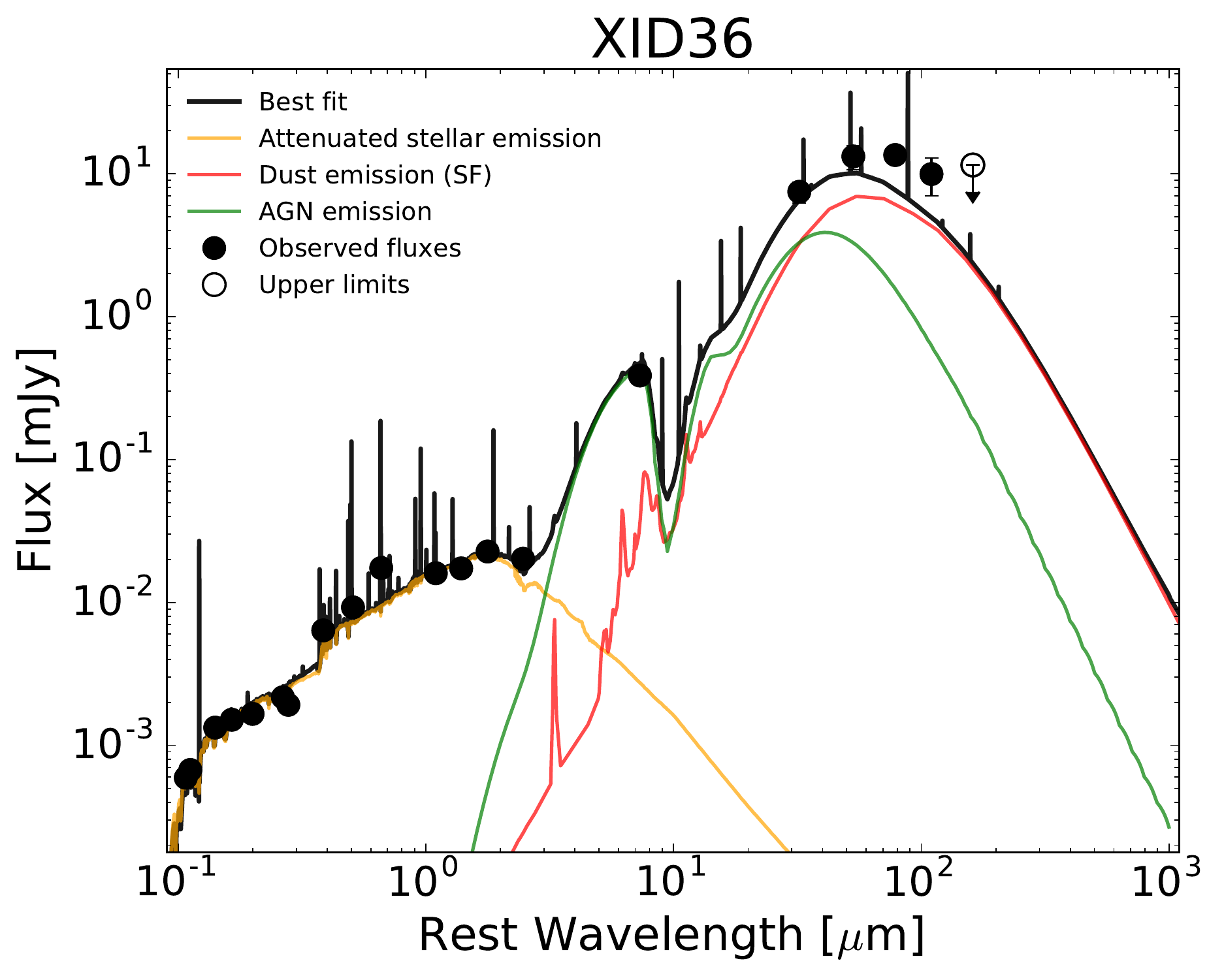}
 	\hspace{2mm}
    \includegraphics[width=7.8cm]{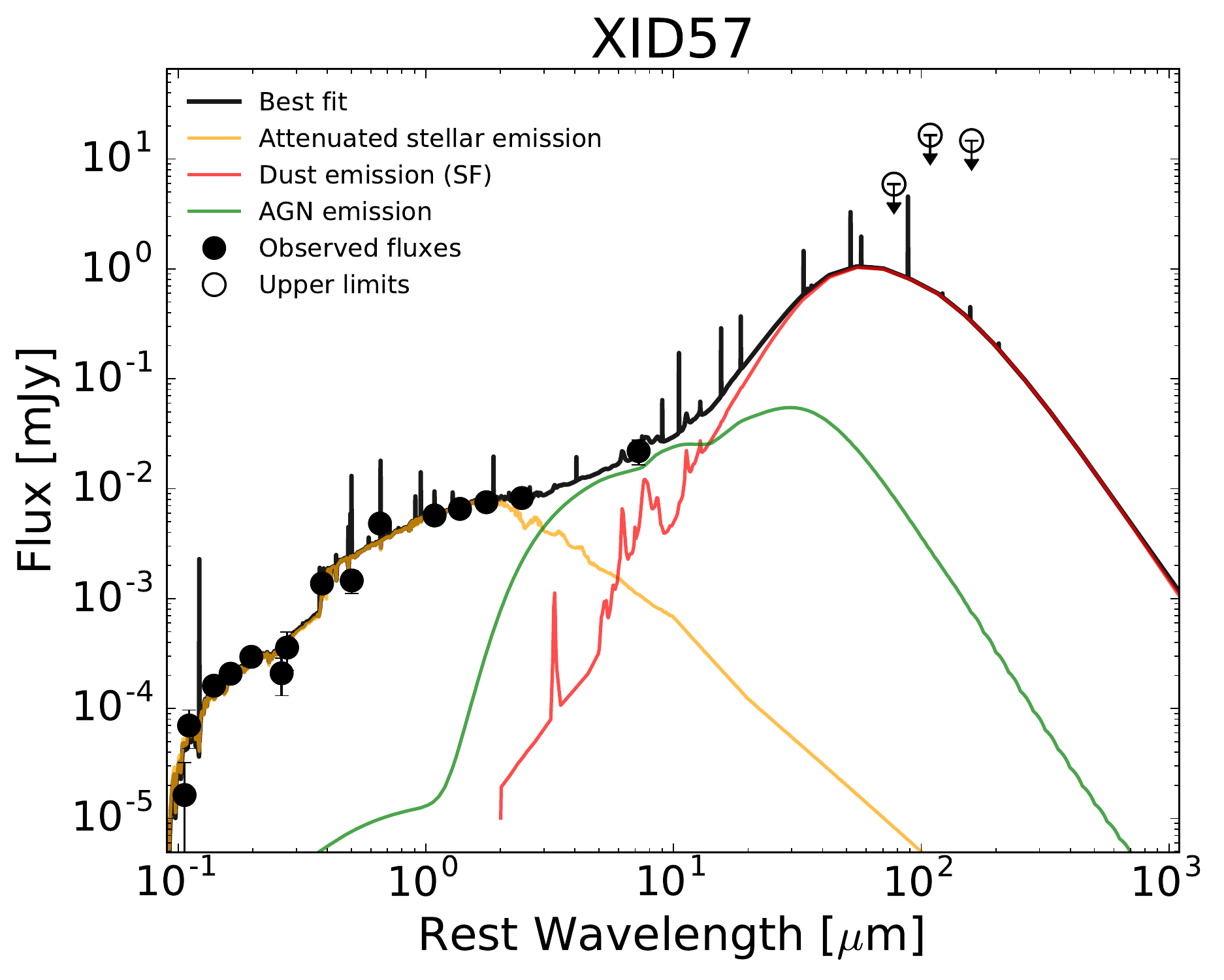}
  	\hspace{2mm}
  	\includegraphics[width=7.8cm]{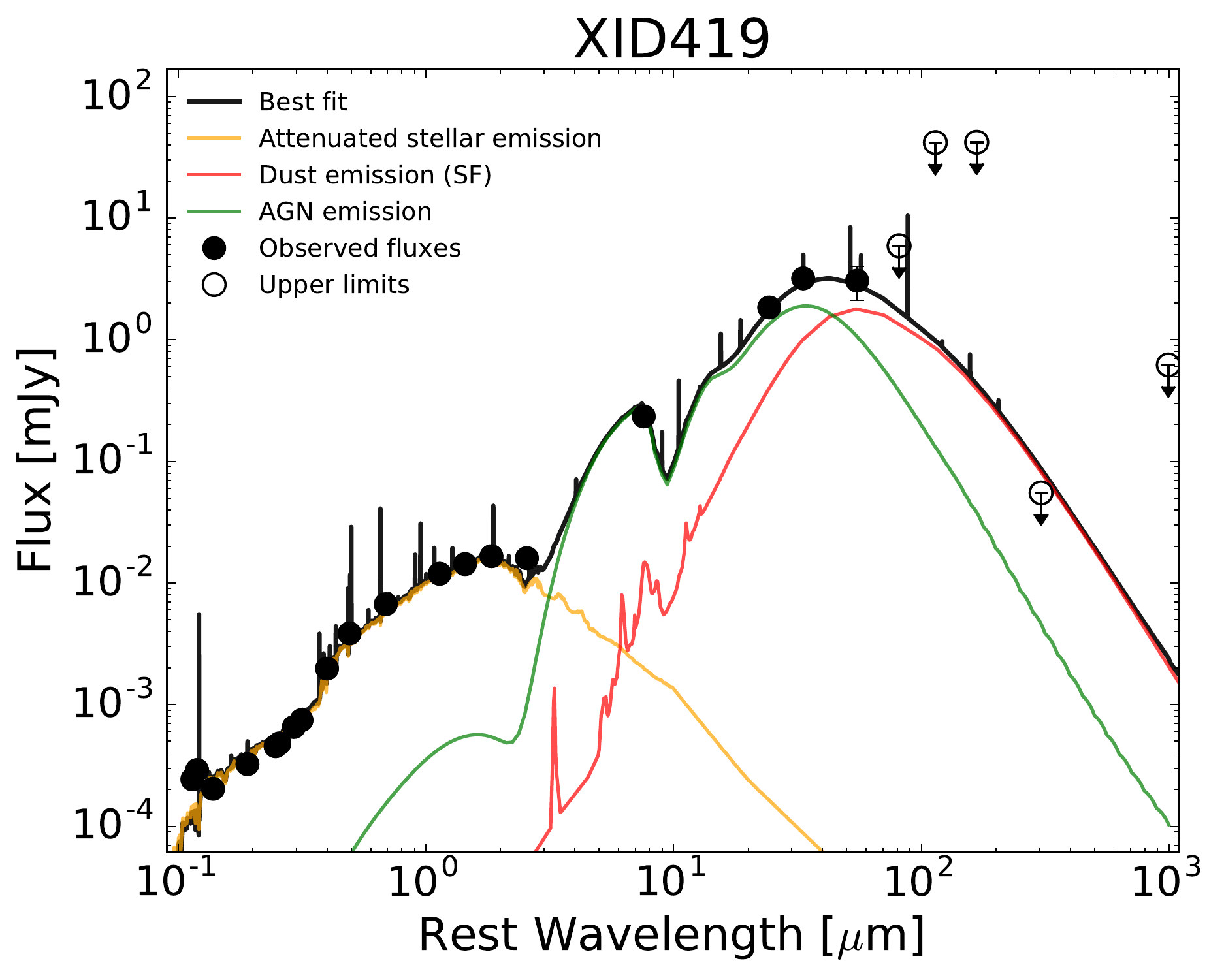}
  	\hspace{2mm}
  	\includegraphics[width=7.8cm]{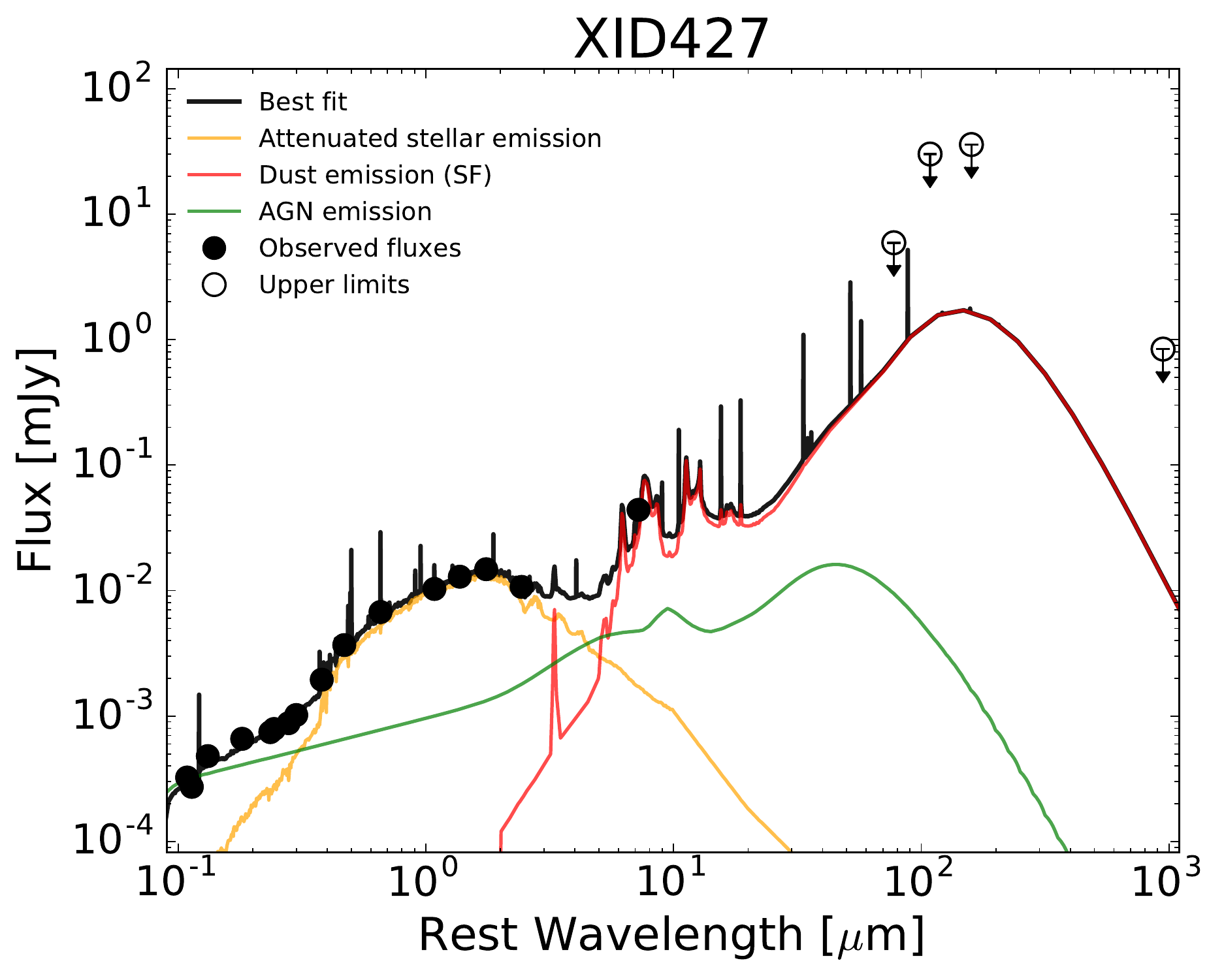}
\end{figure*}

\begin{figure*}
	\centering
  	\includegraphics[width=7.8cm]{C522sed} 
 	\hspace{2mm}
 	\includegraphics[width=7.8cm]{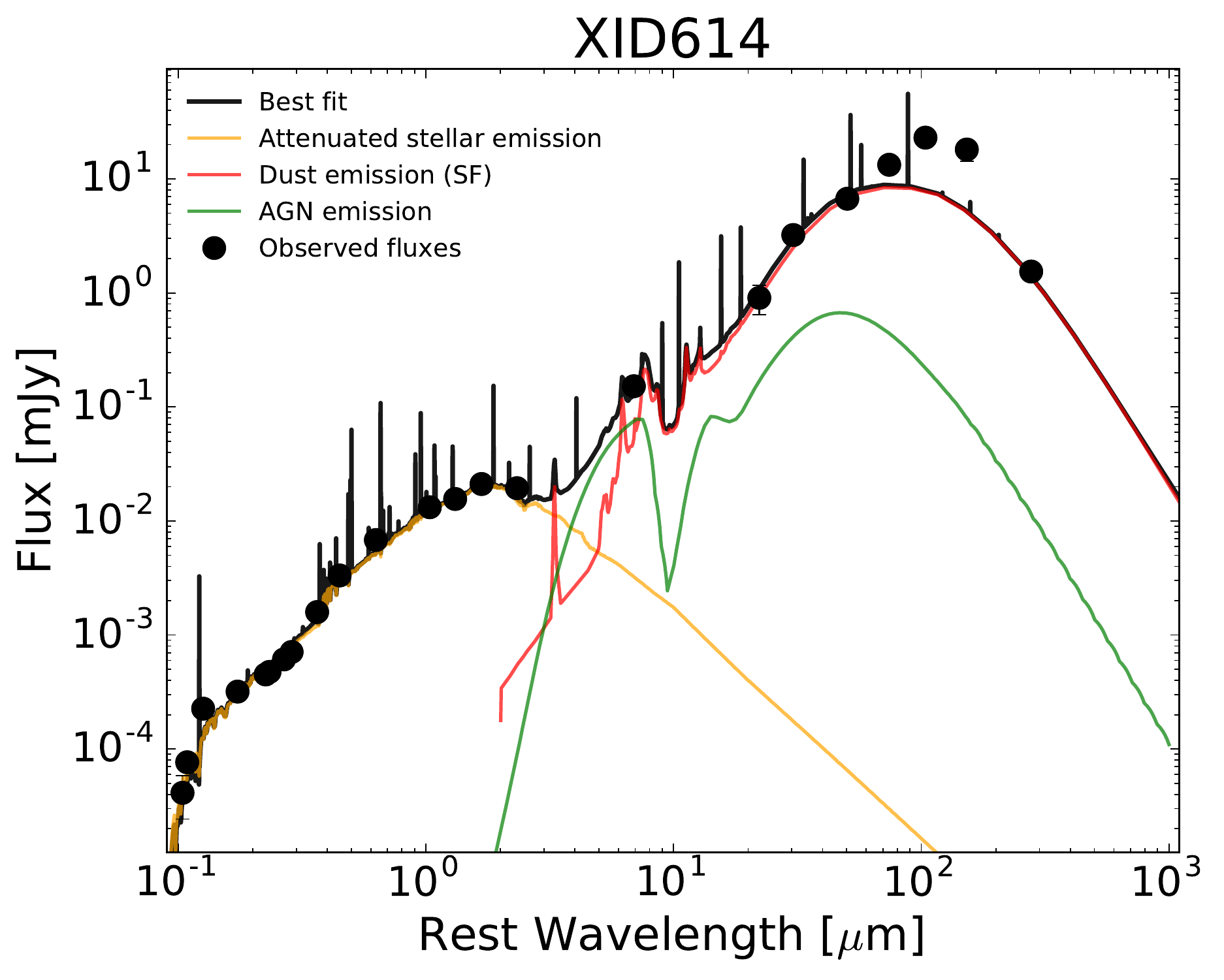}
  	\hspace{2mm}
  	\includegraphics[width=7.8cm]{cid_166sed}
    \hspace{2mm}
  	\includegraphics[width=7.8cm]{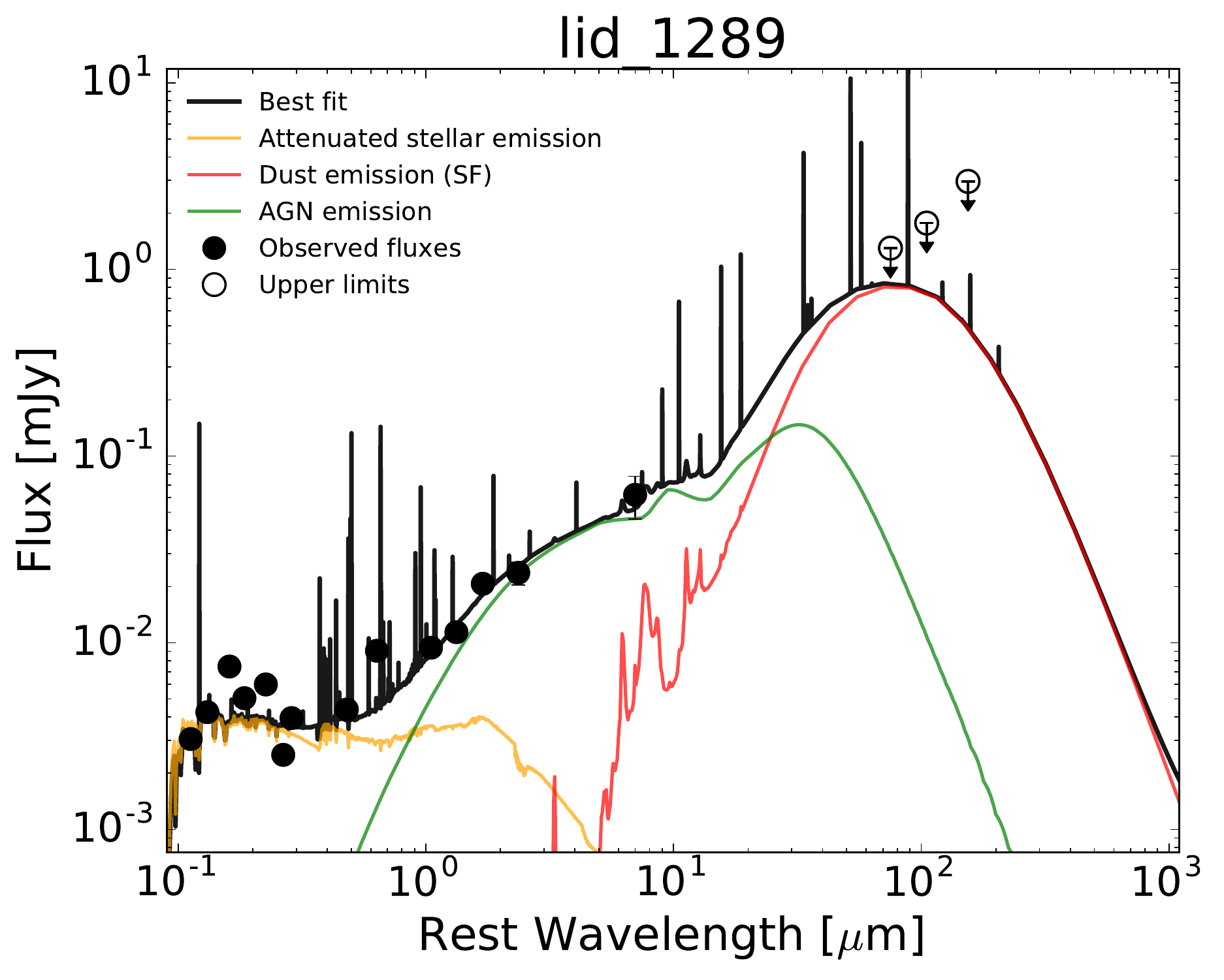}
  	\hspace{2mm}
    \includegraphics[width=7.8cm]{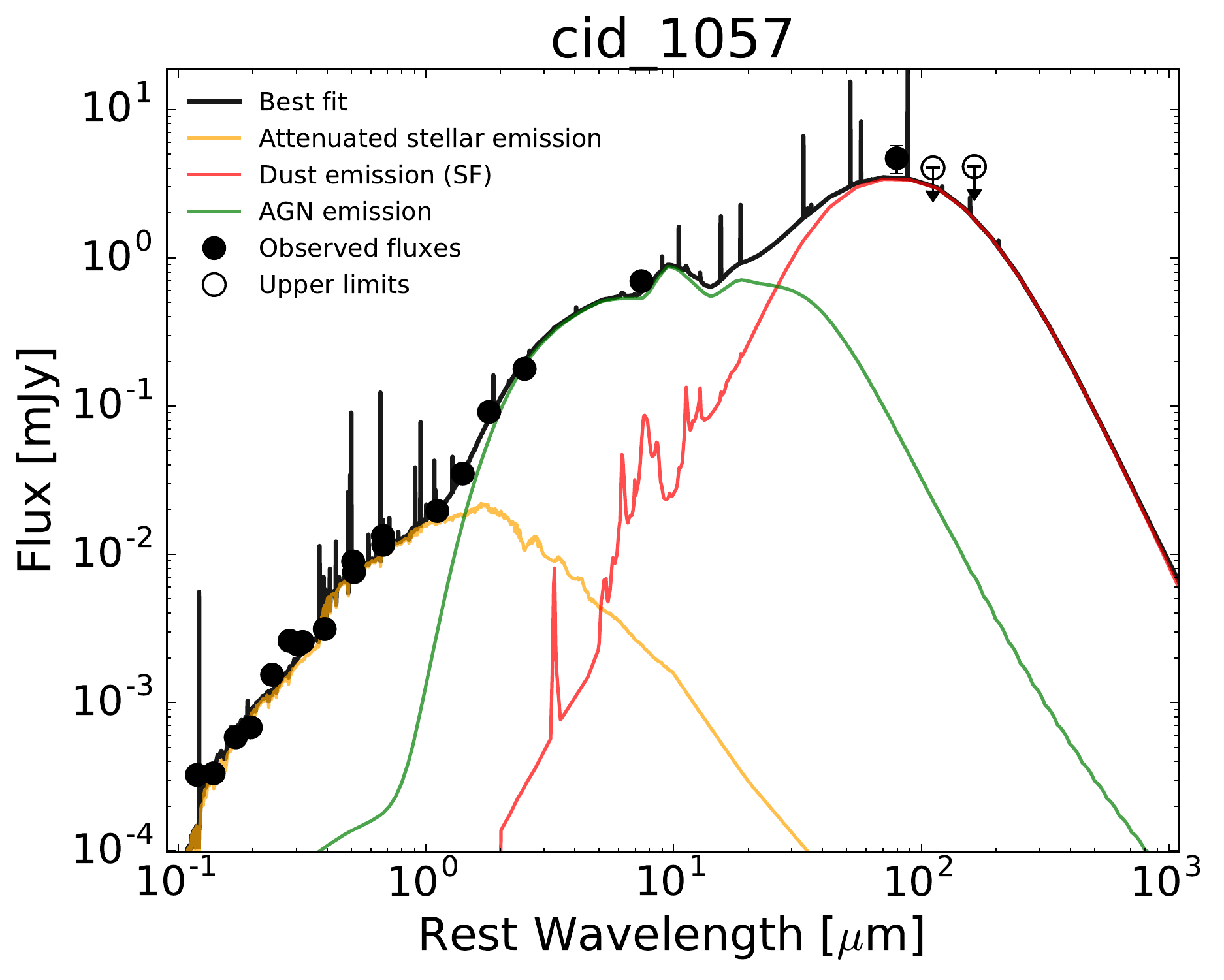}
    \hspace{2mm}
  	\includegraphics[width=7.8cm]{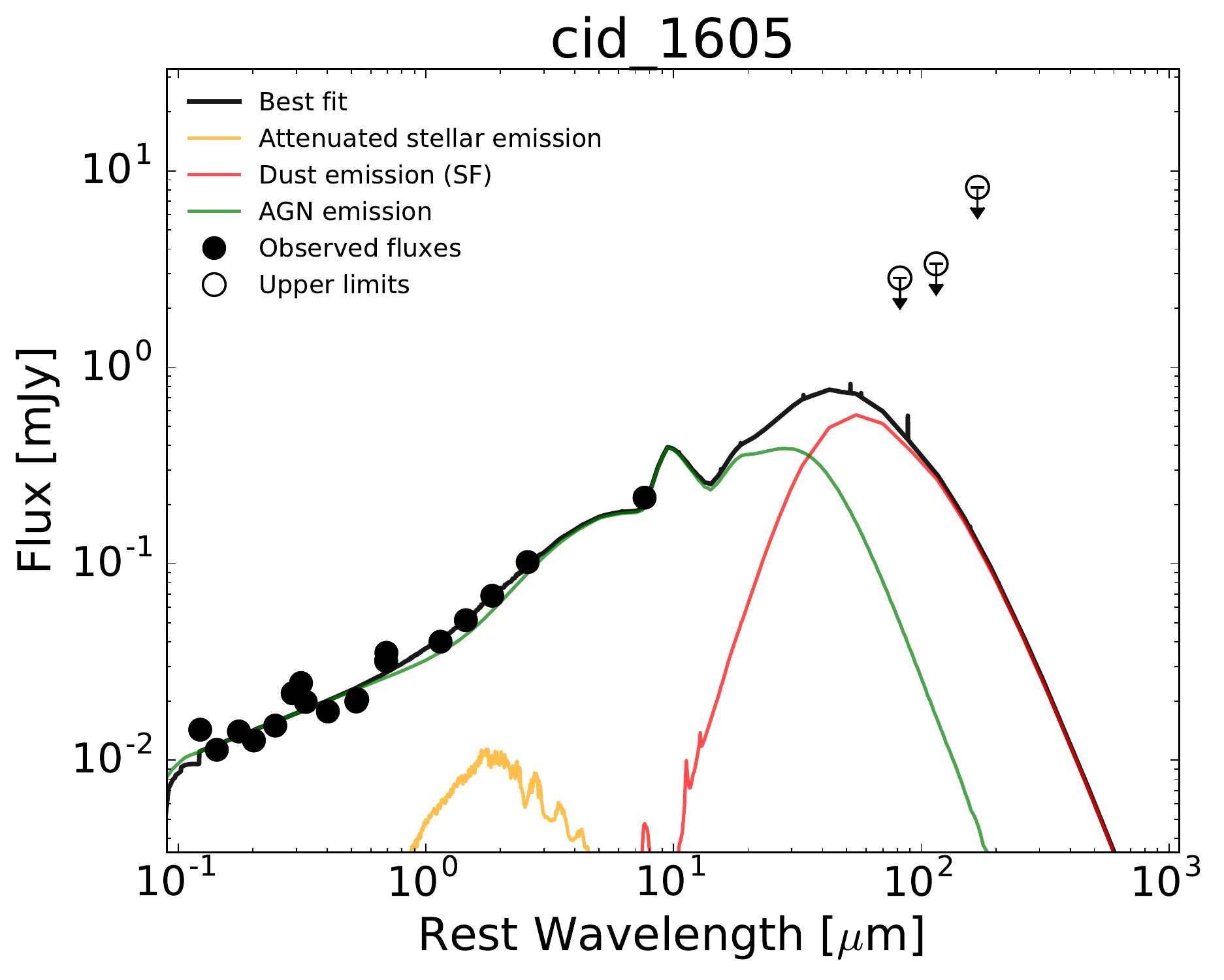}
    \hspace{2mm}
  	\includegraphics[width=7.8cm]{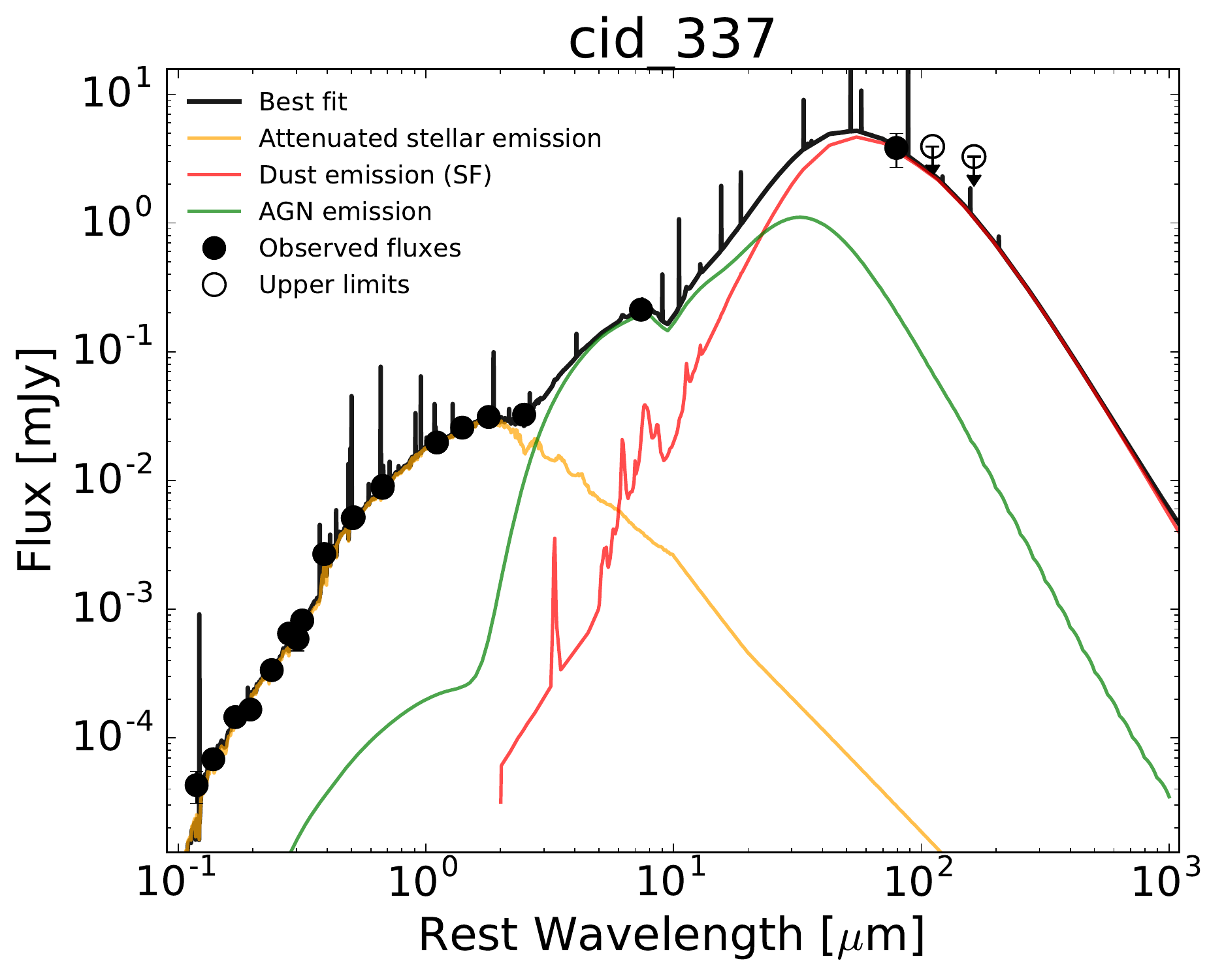}
  	\hspace{2mm}
  	\includegraphics[width=7.8cm]{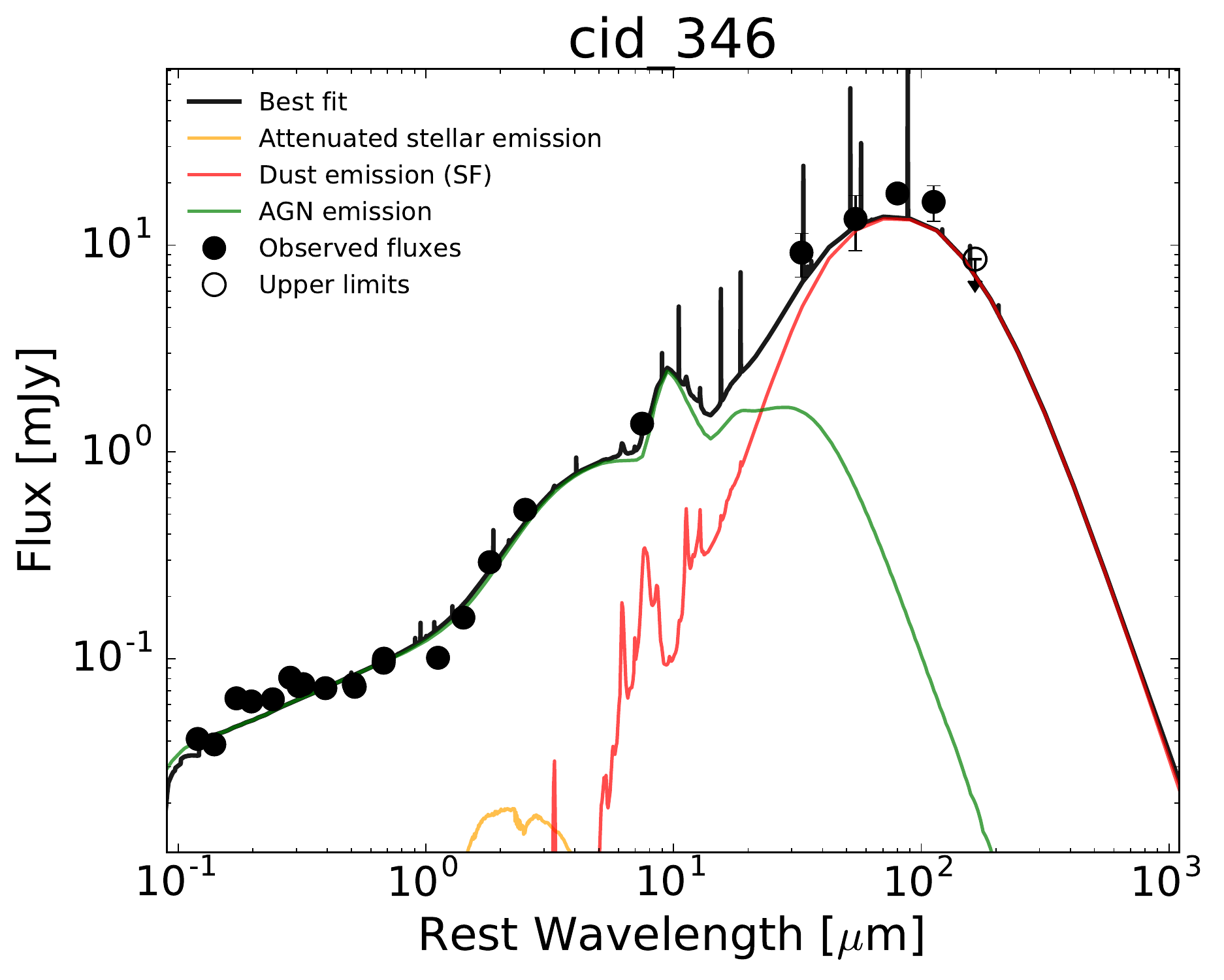}
\end{figure*}

\begin{figure*}
	\centering
  	\includegraphics[width=7.8cm]{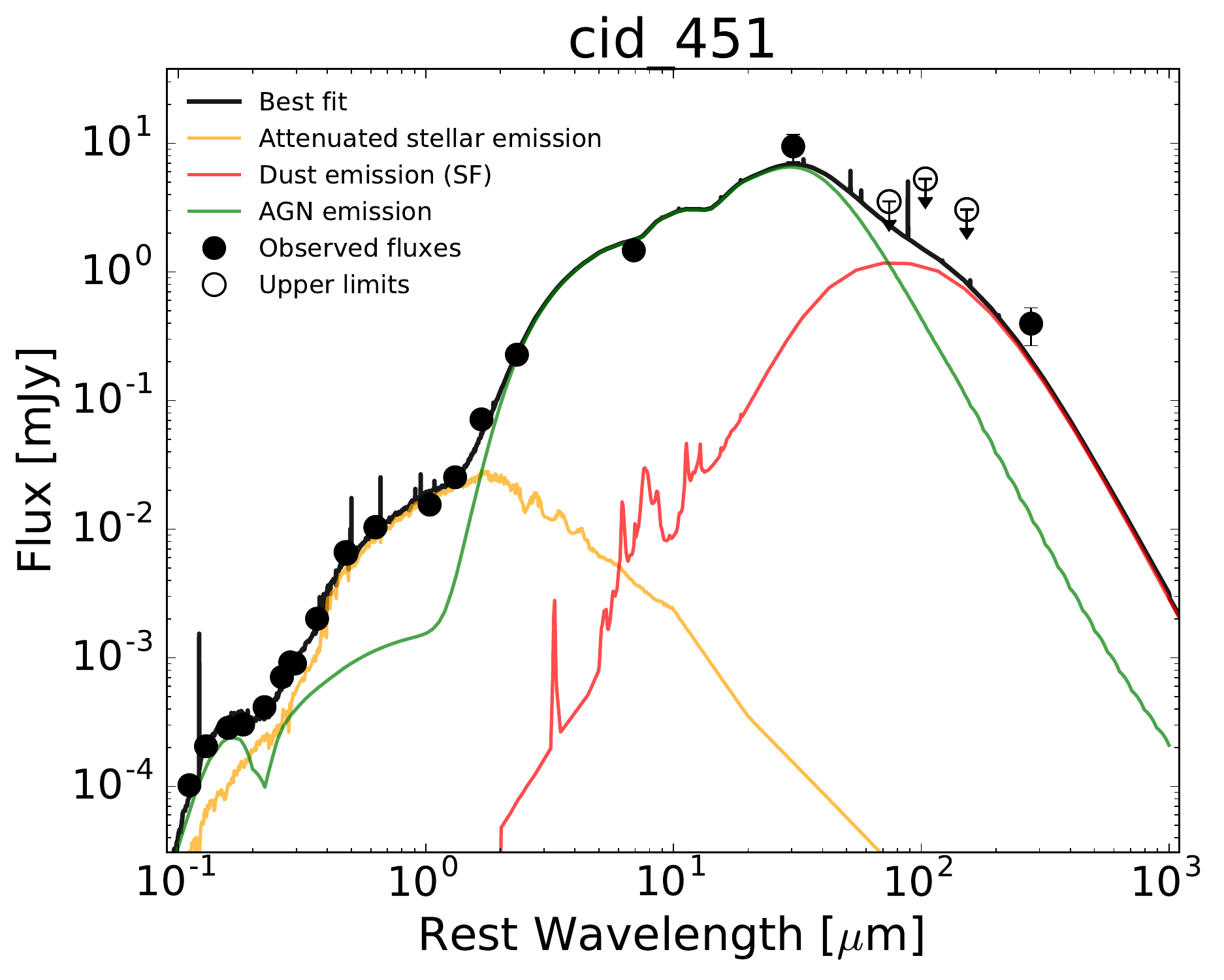} 
  	\hspace{2mm}
  	\includegraphics[width=7.8cm]{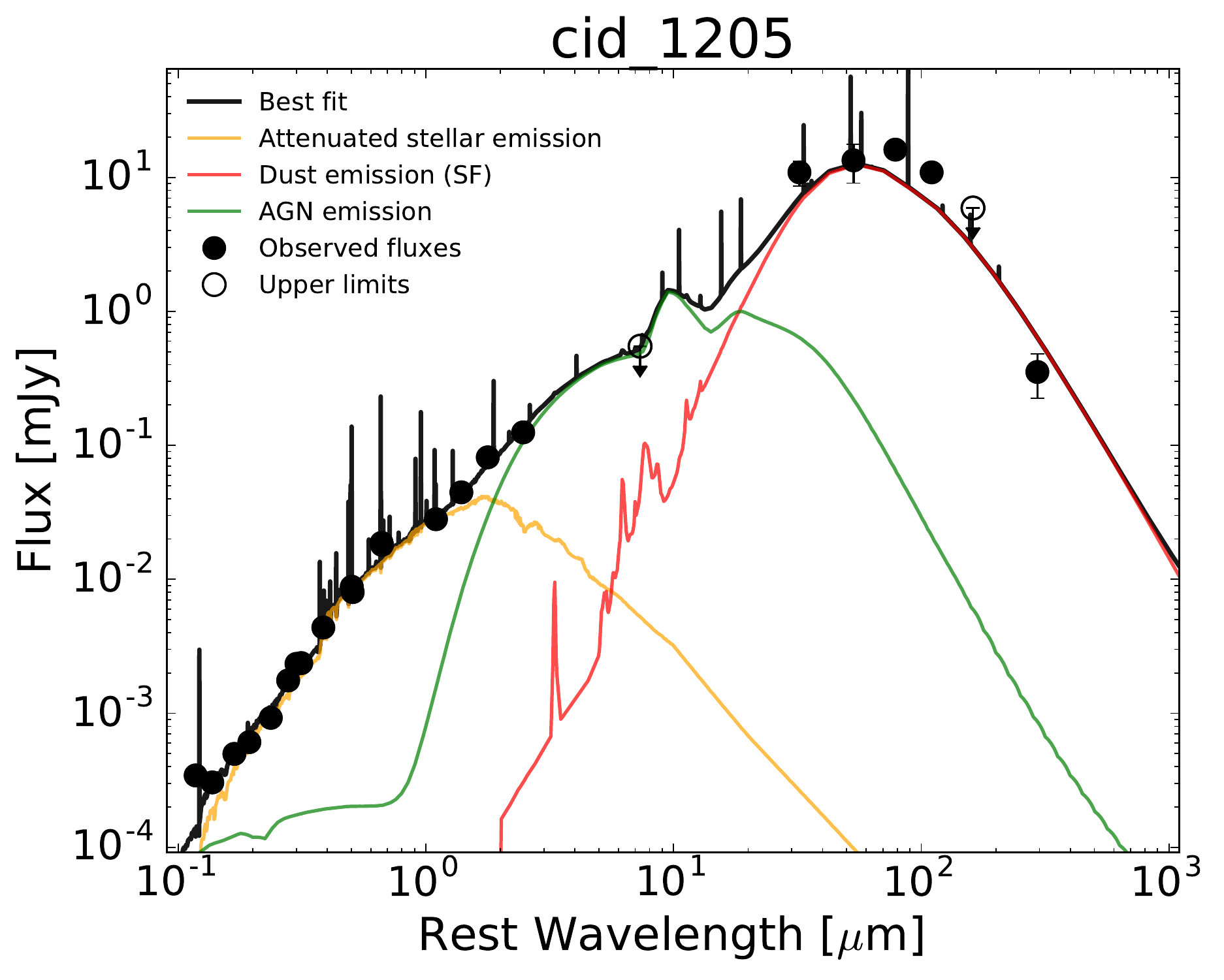}
  	\hspace{2mm}
  	\includegraphics[width=7.8cm]{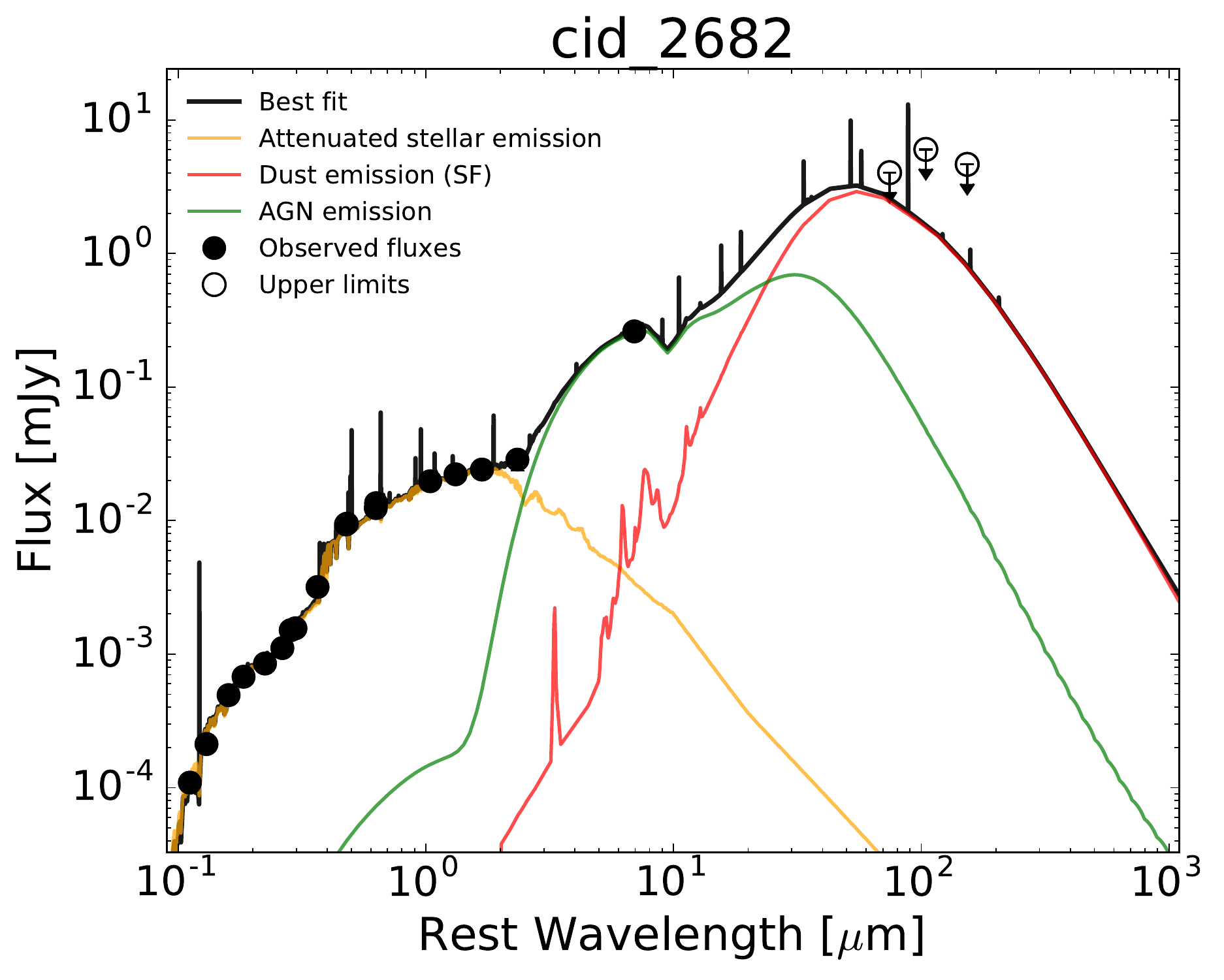}
  	\hspace{2mm}
  	\includegraphics[width=7.8cm]{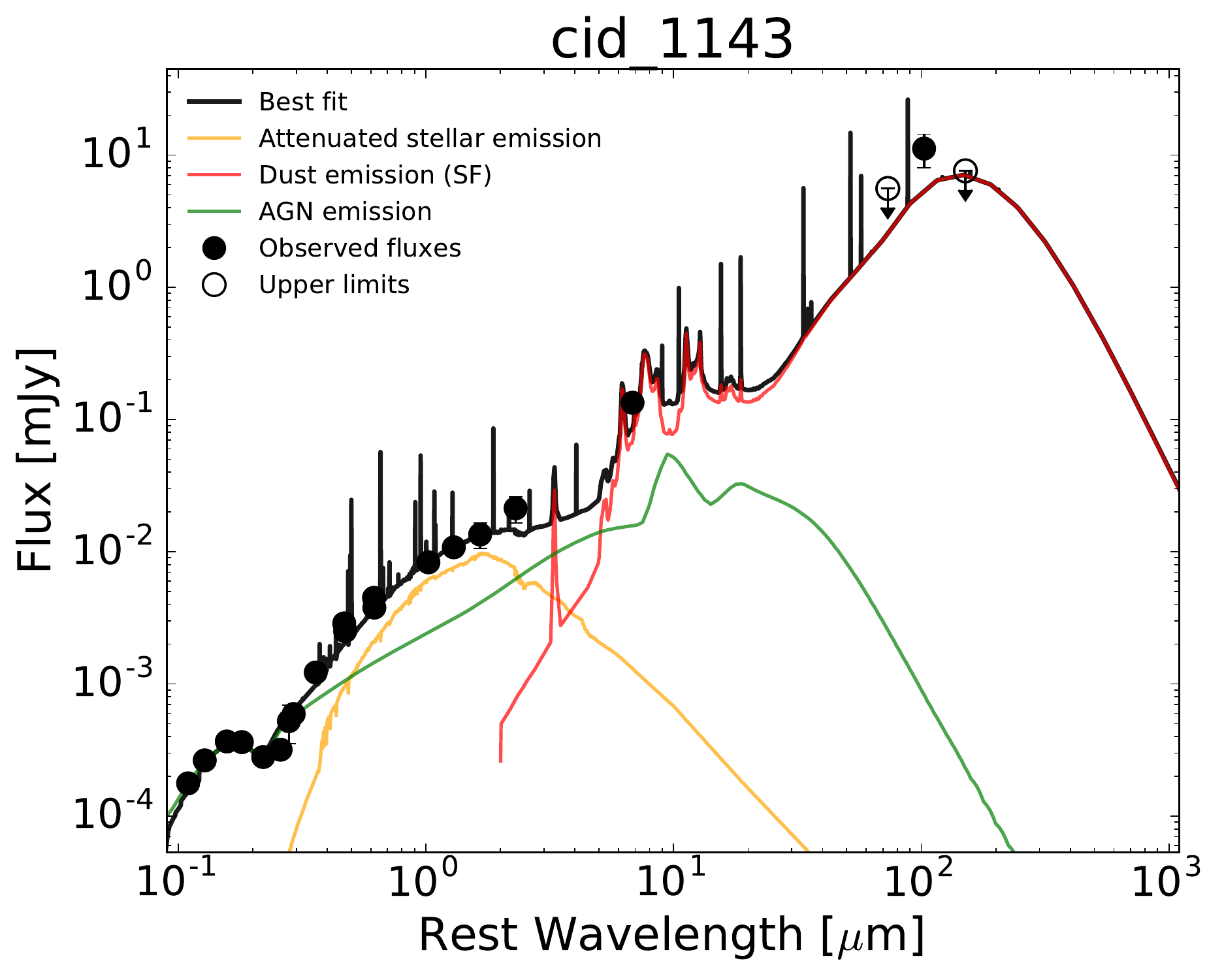}
 	\hspace{2mm}
  	\includegraphics[width=7.8cm]{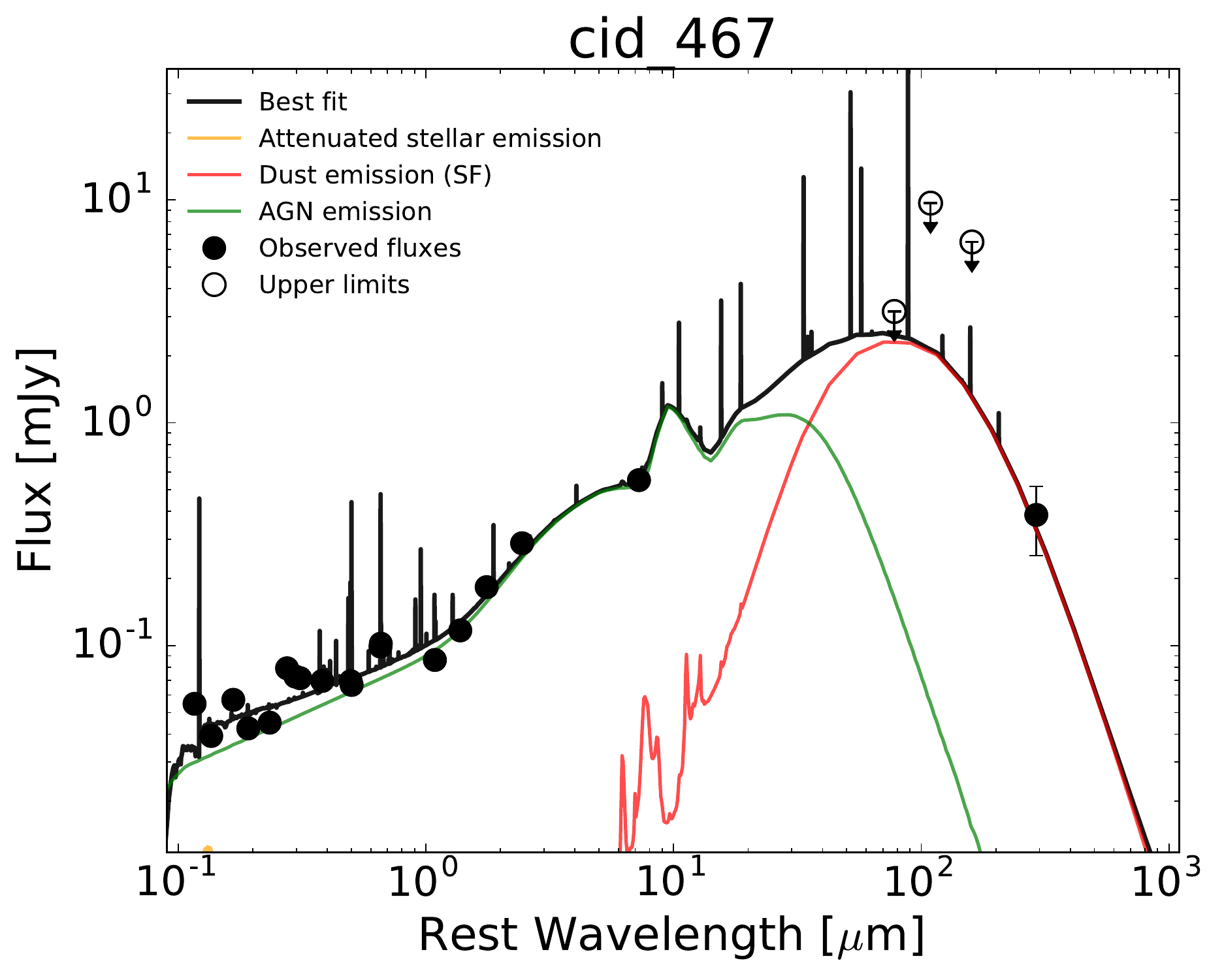}
    \hspace{2mm}
  	\includegraphics[width=7.8cm]{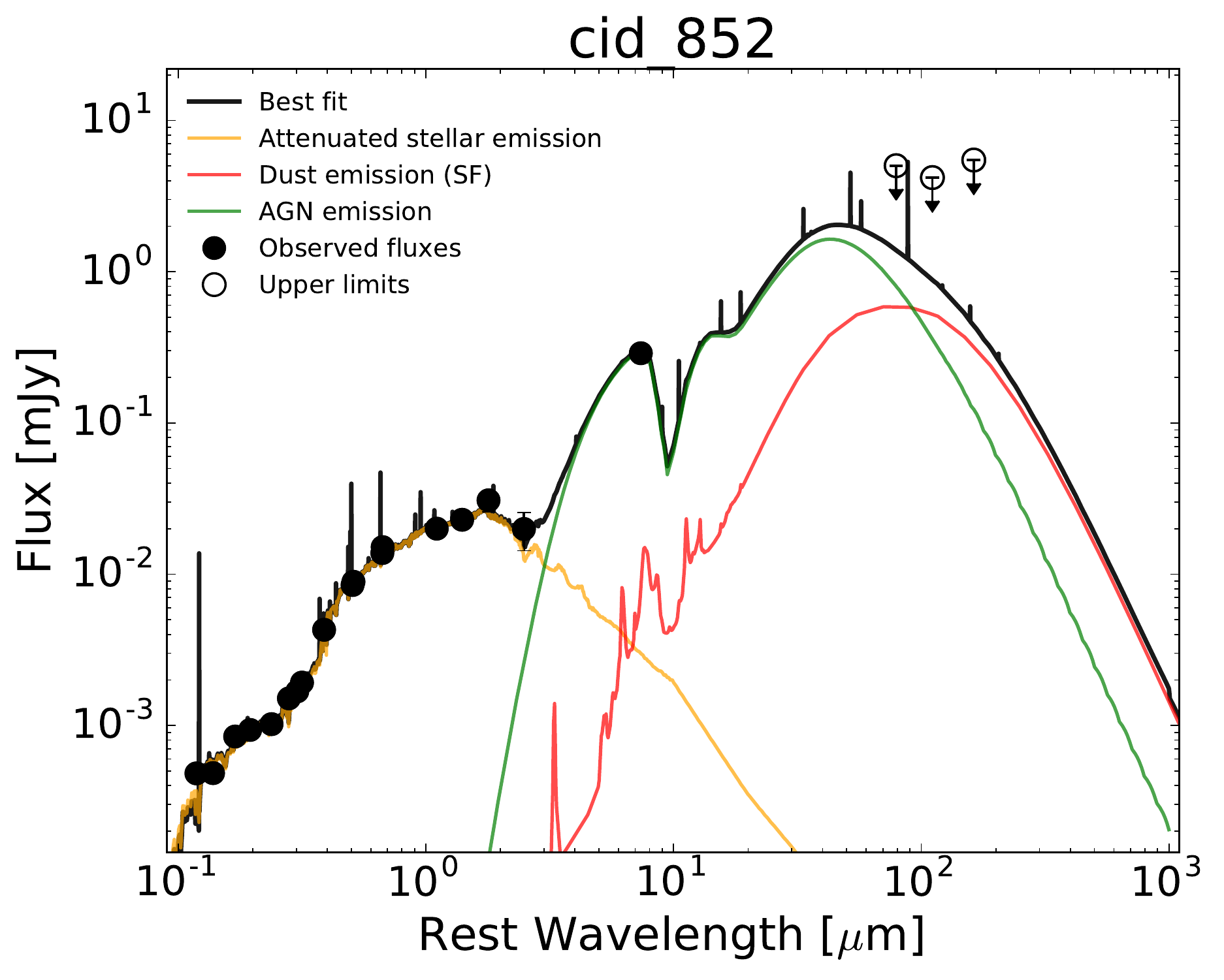}
  	\hspace{2mm}
  	\includegraphics[width=7.8cm]{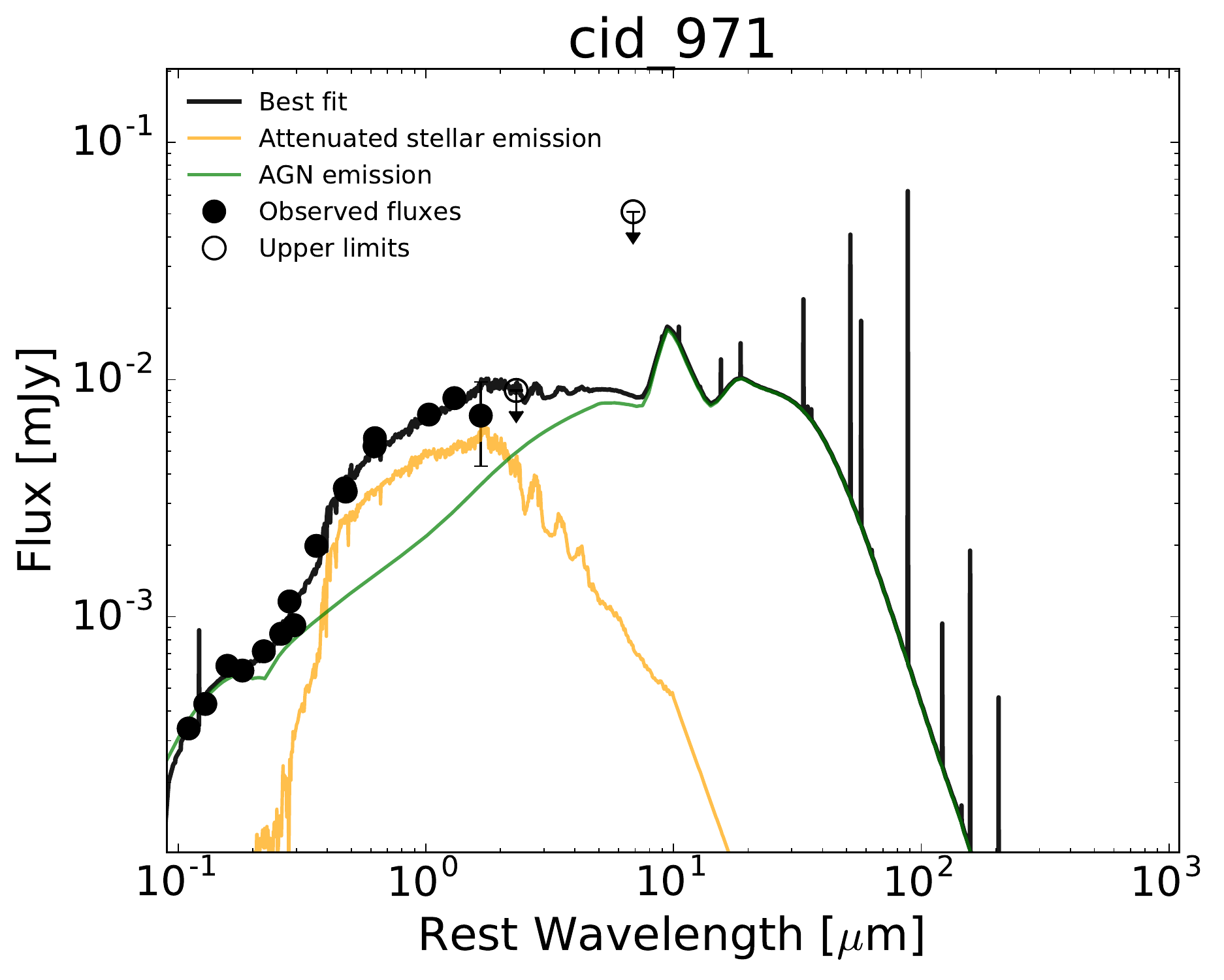}
    \hspace{2mm}
  	\includegraphics[width=7.8cm]{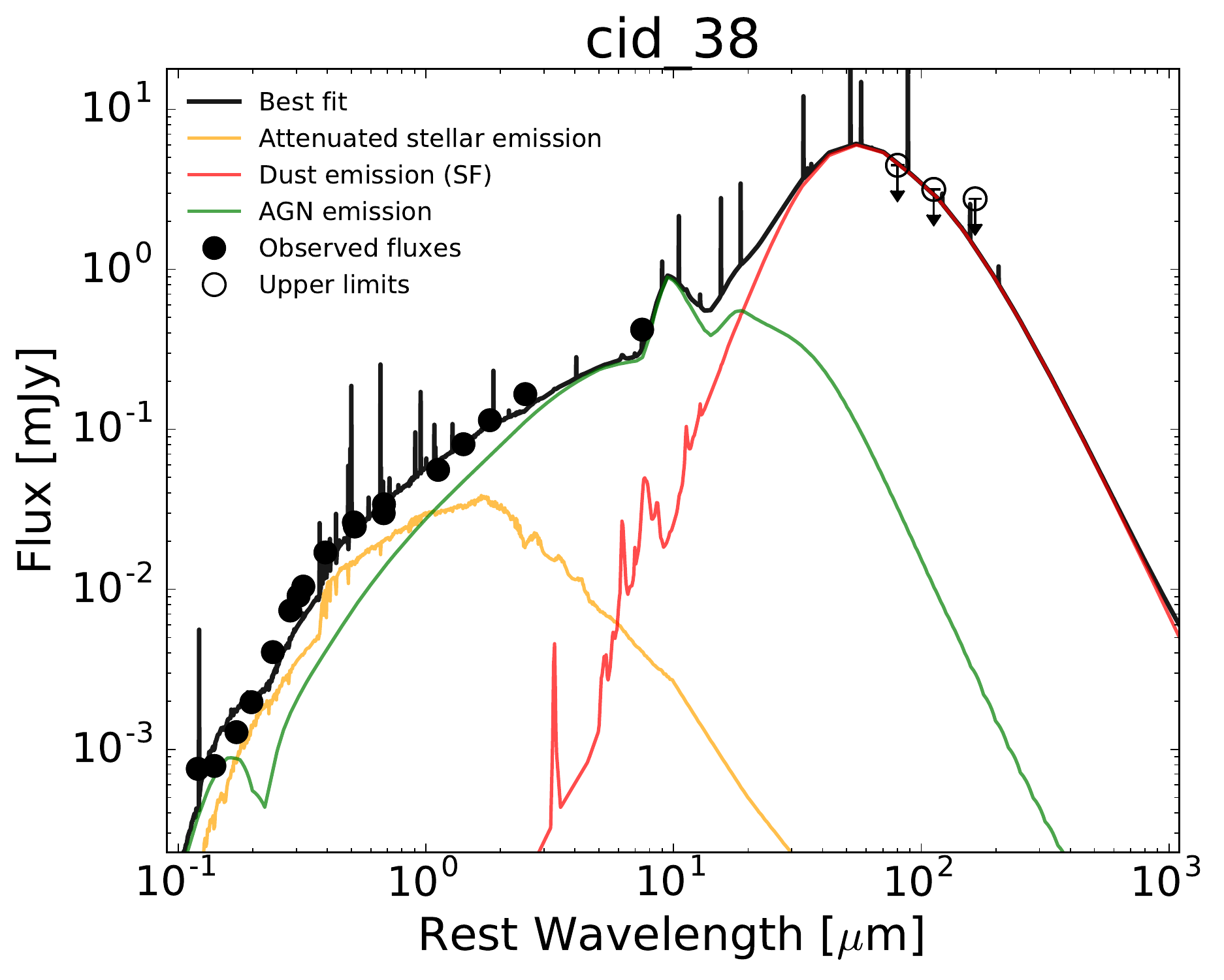}
\end{figure*}

\begin{figure*}
	\centering
 	\includegraphics[width=7.7cm]{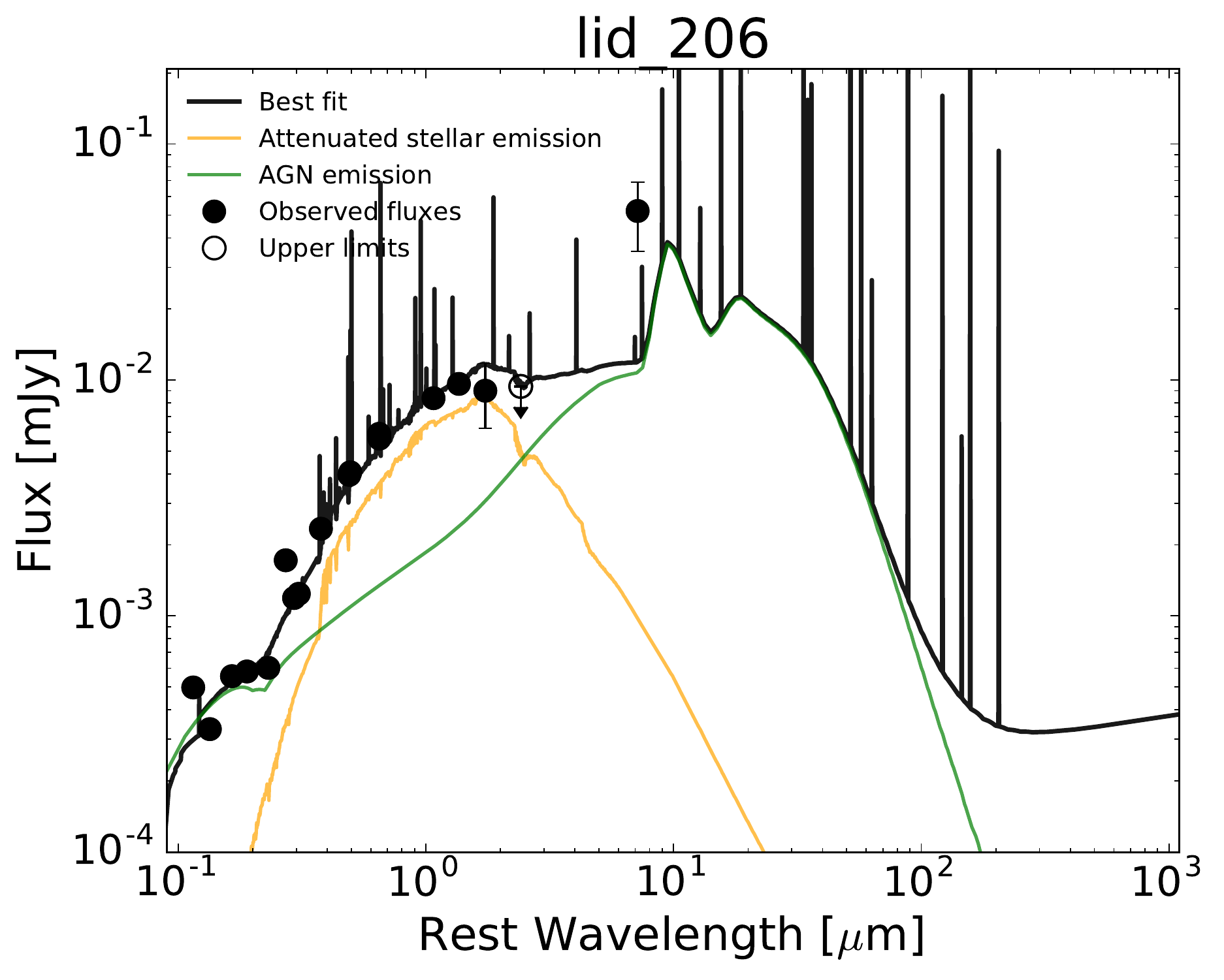} 
  	\hspace{2mm}
  	\includegraphics[width=7.7cm]{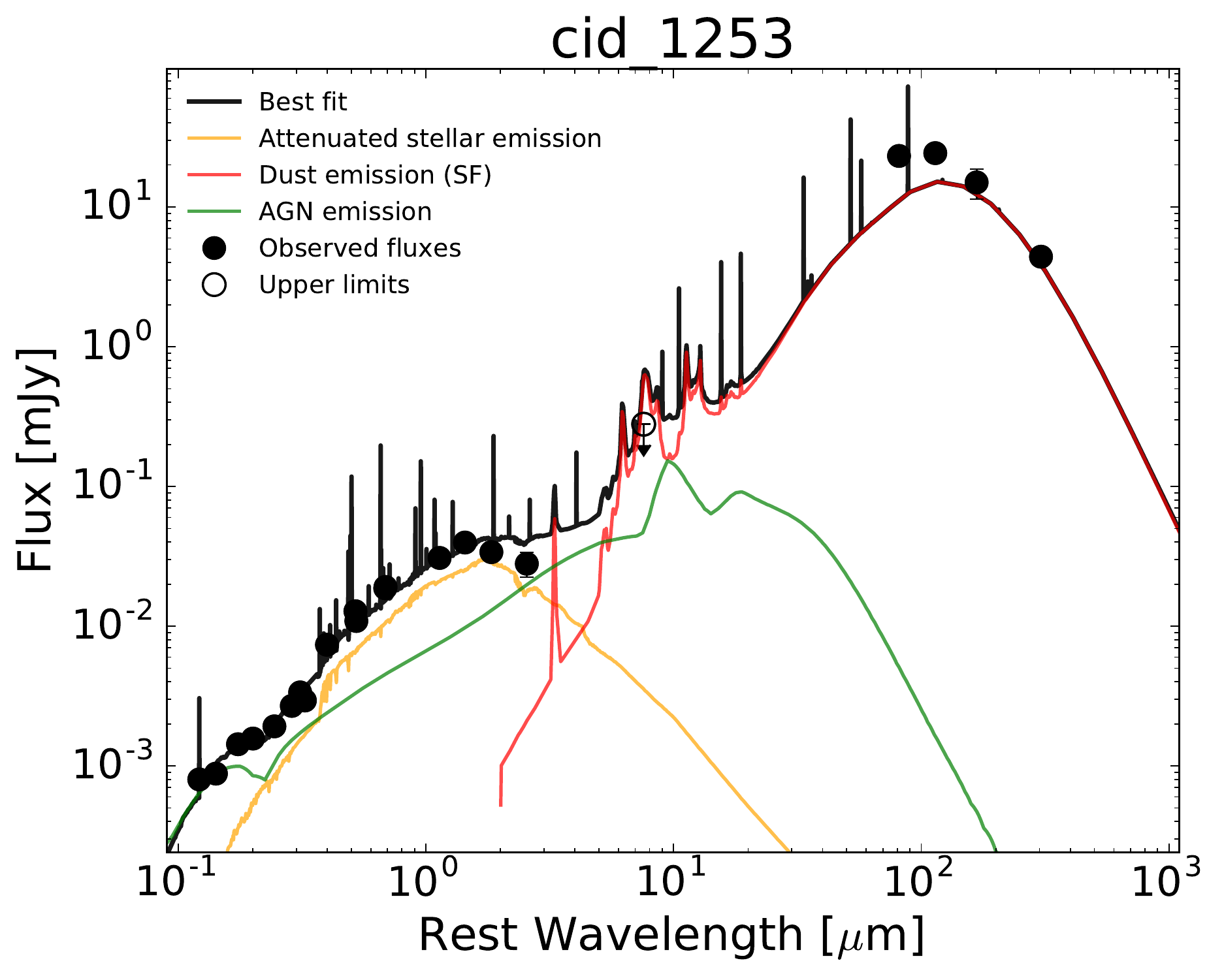}
  	\hspace{2mm}
  	\includegraphics[width=7.7cm]{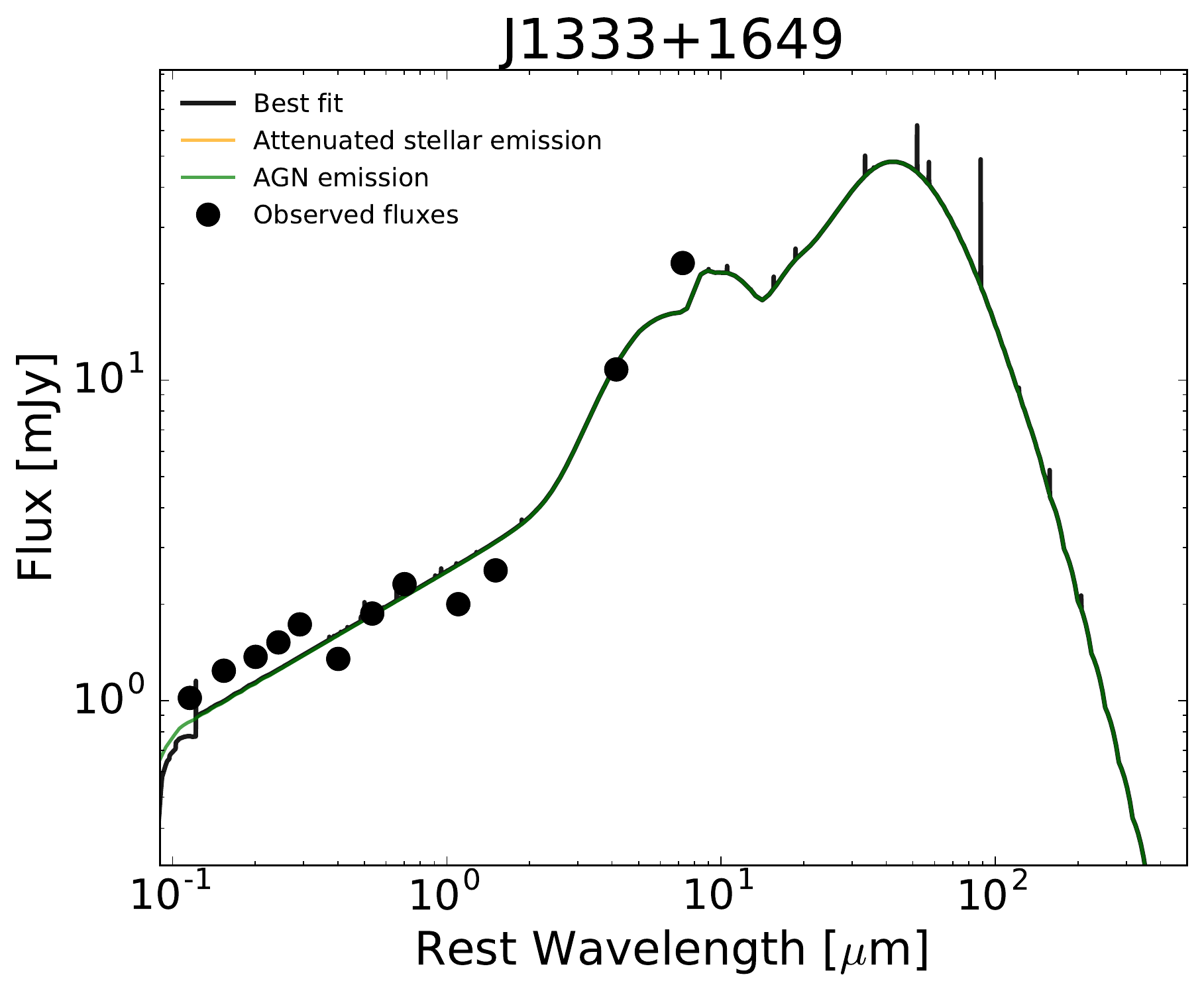}
  	\hspace{2mm}
  	\includegraphics[width=7.7cm]{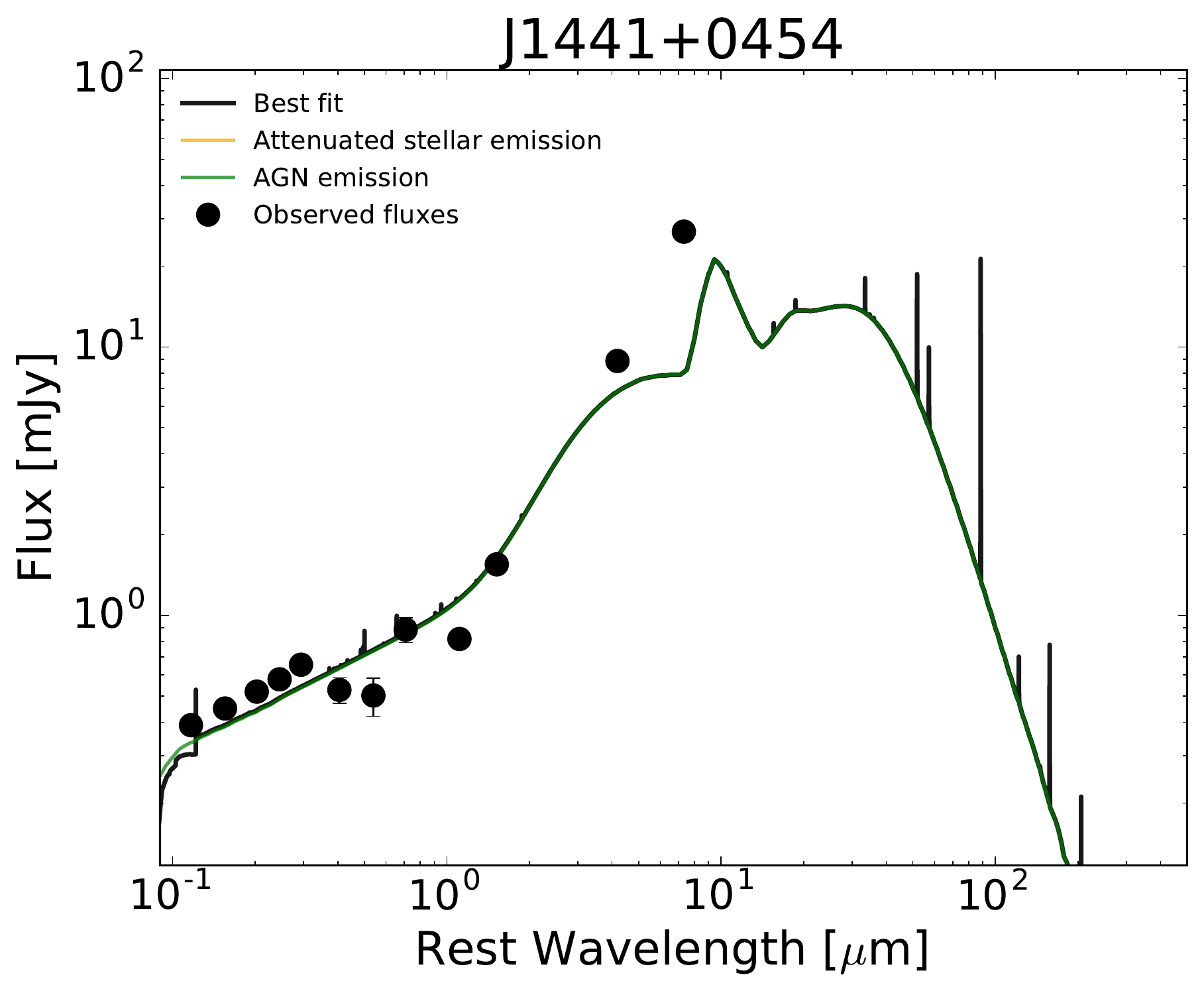}
  	\hspace{2mm}
  	\includegraphics[width=7.7cm]{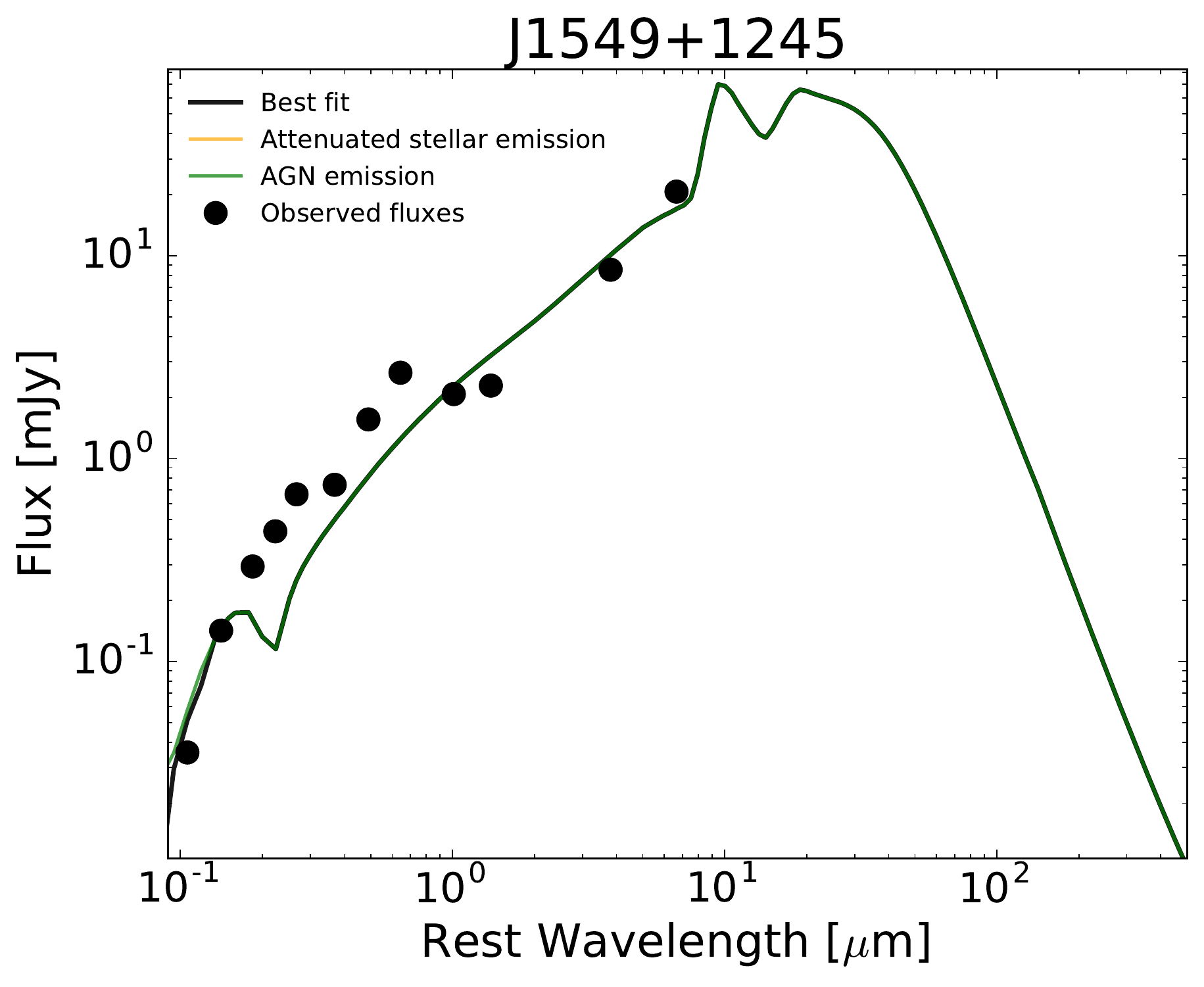}
  	\hspace{2mm}
  	\includegraphics[width=7.7cm]{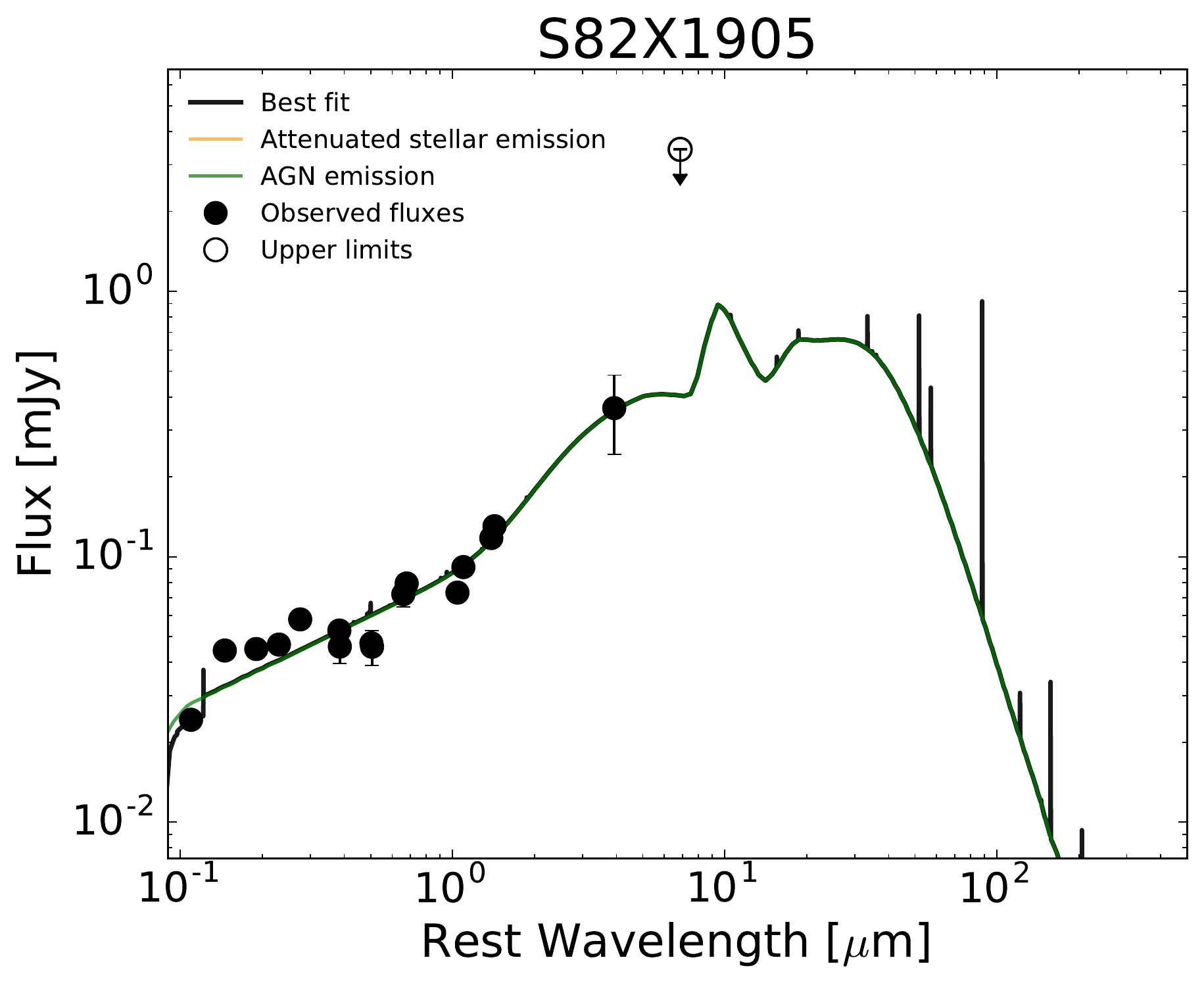}
  	\hspace{2mm}
  	\includegraphics[width=7.7cm]{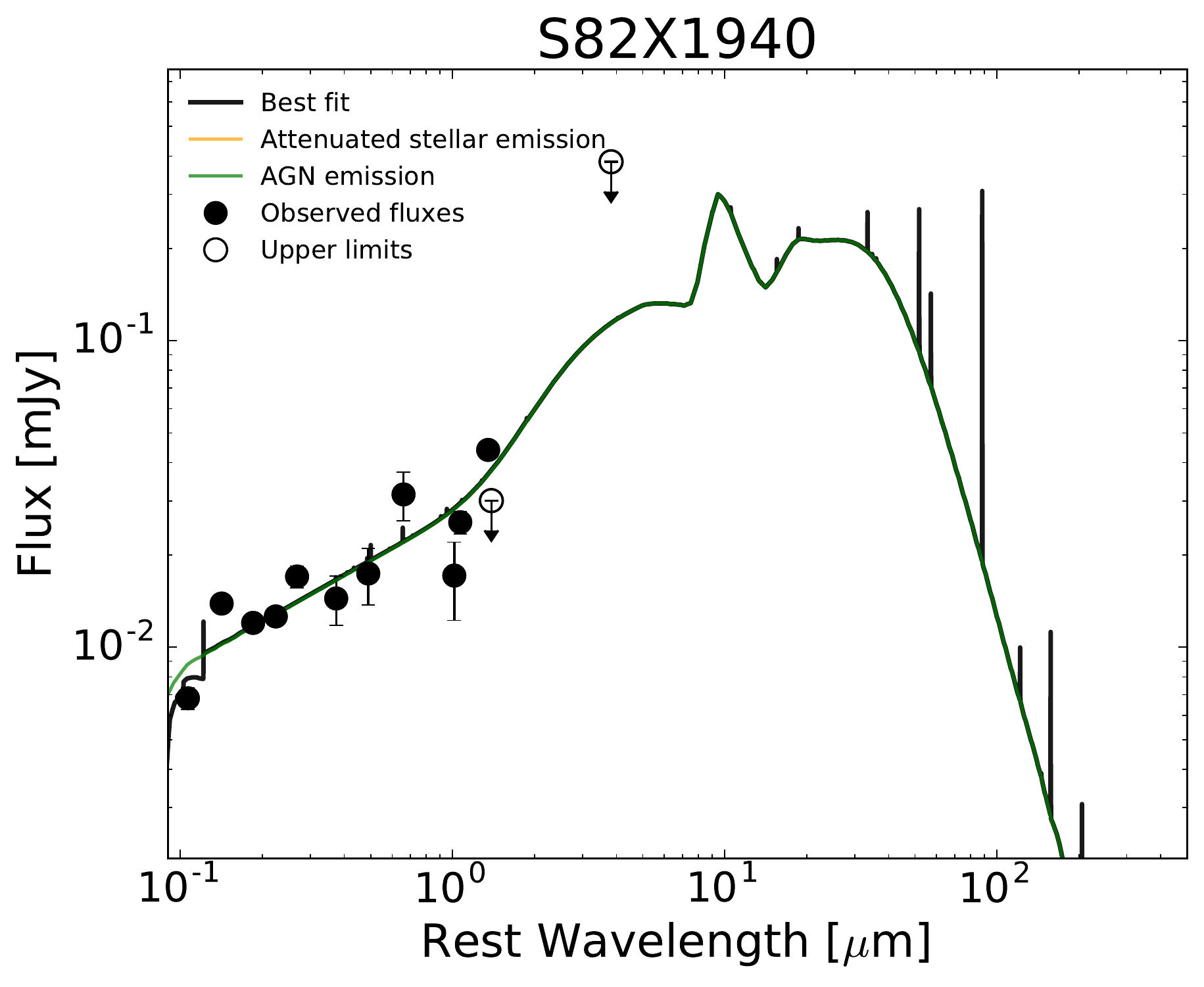}
    \hspace{2mm}
  	\includegraphics[width=7.6cm]{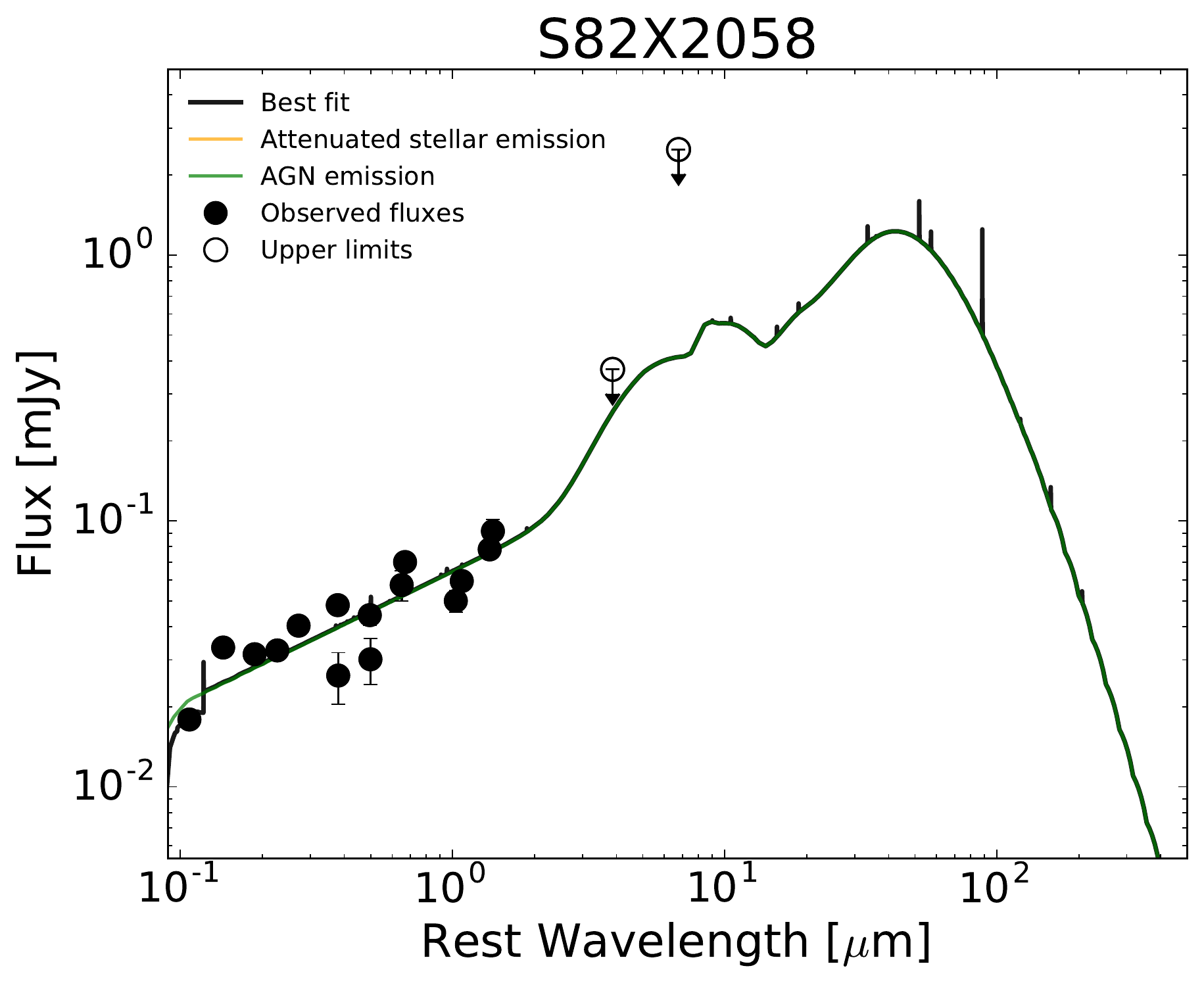} 
\end{figure*}

\begin{figure*}
	\centering
  	\includegraphics[width=7.8cm]{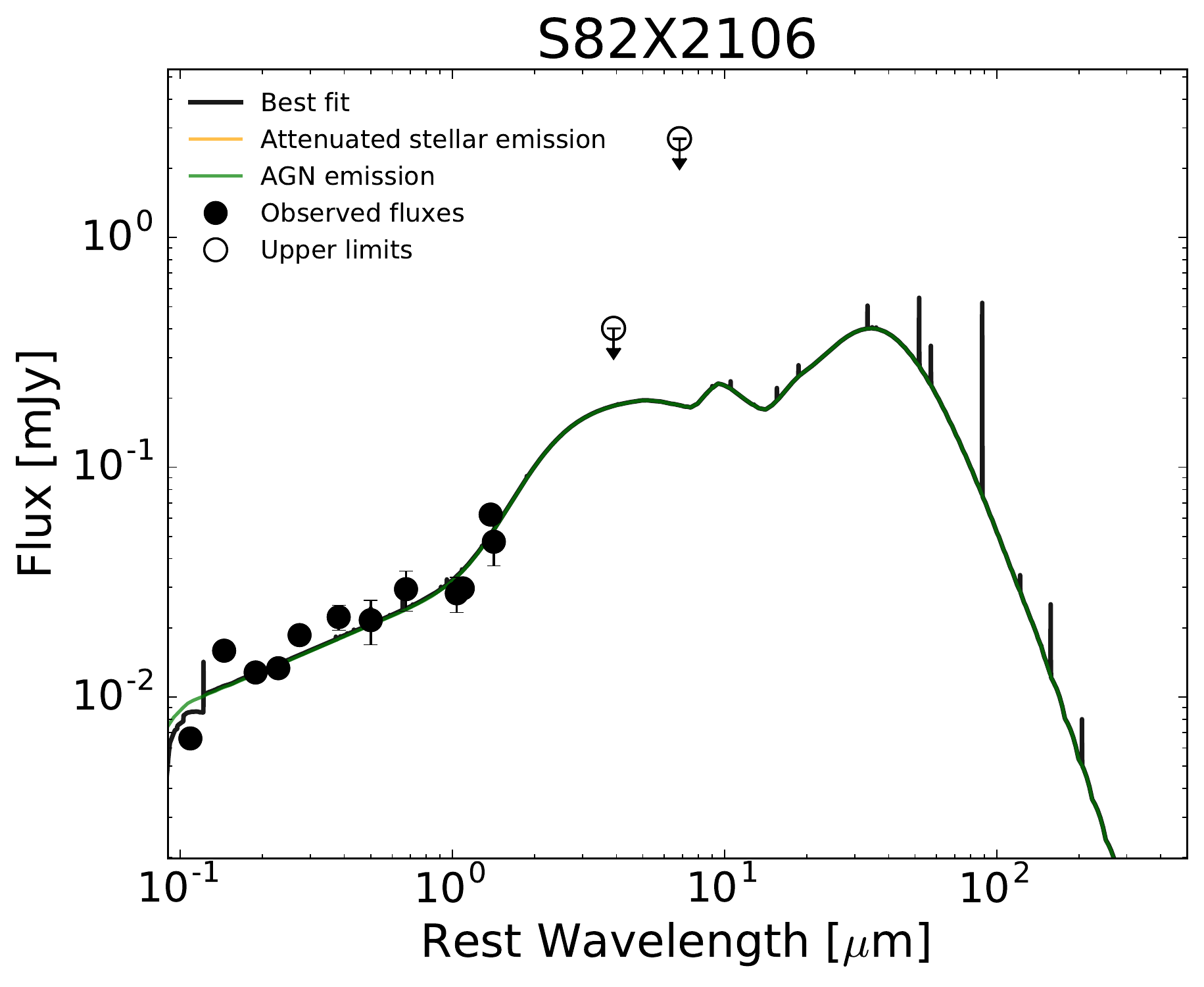} 
\end{figure*}
\end{appendix}

\end{document}